\documentclass[12pt]{article}
\usepackage{amsfonts}
\usepackage[dvips]{color}
\usepackage{xcolor}

\usepackage{amssymb}
\usepackage{graphicx}
\usepackage{amsmath}
\usepackage{natbib}
\usepackage{rotating}
\usepackage{afterpage}
\usepackage{cancel}

\usepackage{setspace}
\newcommand{\bc}{\begin{center}}
\newcommand{\ec}{\end{center}}
\newcommand{\be}{\begin{equation}}
\newcommand{\ee}{\end{equation}}
\newcommand{\bea}{\begin{eqnarray}}
\newcommand{\eea}{\end{eqnarray}}
\newcommand{\bean}{\begin{eqnarray*}}
\newcommand{\eean}{\end{eqnarray*}}
\newcommand{\bt}{\begin{tabular}}
\newcommand{\et}{\end{tabular}}

\usepackage{html}   %
\usepackage{url}       %
\usepackage{xcolor}
\usepackage{graphicx}

\newtheorem{theorem}{Theorem}

\numberwithin{theorem}{section}

\newcounter{pkt}
\newenvironment{tlist}%
{\begin{list}{(\roman{pkt})}{\usecounter{pkt}\parskip0ex\parsep0ex\itemsep0ex\topsep0ex}}%
        {\end{list}}

\newcounter{saveeqn}

\usepackage{hyperref}

\begin{document}

\title{\bf Forecasting with a Panel Tobit Model}

\author{
        Laura Liu\\ {\footnotesize {\em Indiana University}}
        \and
        Hyungsik Roger Moon\\ {\footnotesize {\em University of Southern California}} \\[-2ex]
        {\footnotesize{\em and Yonsei}}
        \and
        Frank Schorfheide\thanks{Correspondence:
                L. Liu: Department of Economics, Indiana University, 100 S. Woodlawn Ave, Bloomington, IN 47405. Email:  lauraliu@iu.edu. 
                H.R. Moon: Department of Economics, University of Southern California, KAP 300, Los Angeles, CA
                90089. E-mail: moonr@usc.edu. F. Schorfheide: Department of Economics, 133 S. 36th Street, University of Pennsylvania, Philadelphia, PA 19104-6297. Email:
                schorf@ssc.upenn.edu. We thank Mitchell Berlin, Siddhartha Chib, Tim Armstrong, and participants at various seminars and conferences for helpful comments and suggestions. Moon and Schorfheide gratefully acknowledge financial support from the National Science Foundation under Grants SES 1625586 and SES 1424843, respectively.} \\
        {\footnotesize{\em University of Pennsylvania}} \\[-2ex] {\footnotesize{\em CEPR, NBER, and PIER}} }

\date{This Version:  May 26, 2022}
\maketitle


\begin{abstract}
We use a dynamic panel Tobit model with heteroskedasticity to generate forecasts for a large cross-section of short time series of censored observations. Our fully Bayesian approach allows us to flexibly estimate the cross-sectional distribution of heterogeneous coefficients and then implicitly use this distribution as prior to construct Bayes forecasts for the individual time series. In addition to density forecasts, we construct set forecasts that explicitly target the average coverage probability for the cross-section. We present a novel application in which we forecast bank-level loan charge-off rates for small banks.
\end{abstract}

\noindent JEL CLASSIFICATION: C11, C14, C23, C53, G21

\noindent KEY\ WORDS: Bayesian inference, density forecasts, loan charge-offs, panel data,  set forecasts, Tobit model.

\thispagestyle{empty}
\setcounter{page}{0}
\newpage

\section{\color{black} Introduction}
\label{sec:introduction}

This paper considers the problem of forecasting a large collection of short time series with censored observations. In the empirical application we forecast charge-off rates on loans for a panel of small banks. A charge-off occurs if a loan is deemed unlikely to be collected because the borrower has become delinquent. The prediction of charge-off rates is interesting to banks, regulators, and investors because they are losses on loan portfolios. If charge-off rates are large, the bank may be entering a period of distress and require additional capital. Due to mergers and acquisitions, changing business models, and changes in regulatory environments the time series dimension that is useful for forecasting is often short. The general methods developed in this paper are not tied to the charge-off rate application and can be used in any setting in which a researcher would like to analyze a panel of censored data with a large cross-sectional and a short time-series dimension.

In a panel data setting, cross-sectional heterogeneity in the data is modeled through unit-specific parameters. The more precisely they are estimated, the more accurate the forecasts are. The challenge in forecasting panels with a short time dimension is that the data set does not contain a lot of information about the heterogeneous parameters. A natural way of adding information to the estimation of these parameters is the use of prior distributions. 
The key insight in panel data applications is that one can extract information from the cross section and equate the prior distribution with the cross-sectional distribution of unit-specific coefficients. 
An empirical Bayes implementation of this idea creates a point estimate of the cross-sectional distribution of the heterogeneous coefficients and then conditions the subsequent posterior calculations on the estimated prior distribution. The classic James-Stein estimator for a vector of means can be interpreted as an empirical Bayes estimator.\footnote{Empirical Bayes methods have a long history in the statistics literature going back to \cite{Robbins1955}; see \cite{Robert1994} for a textbook treatment.}  

Rather than pursuing an empirical Bayes approach, we conduct a full Bayesian analysis by specifying a hyperprior for the distribution of heterogeneous coefficients and constructing a joint posterior for the coefficients of this hyperprior as well as the actual unit-specific coefficients. This approach can in principle handle quite general nonlinearities and generate predictions under a wide variety of loss functions. It is preferable for interval and density forecasts, because it captures all sources of uncertainty. 

The contributions of our paper are threefold. First, we extend the full Bayesian estimation and prediction with a linear panel data model in \cite{liu2018density} to a dynamic panel Tobit model with heteroskedastic innovations and correlated random effects. {\color{black}We hereby build on work on the Bayesian estimation of static, dynamic, and panel Tobit models by \cite{Chib1992}, \cite{Wei1999}, \cite{baranchuk2008assessing}, and \cite{LiZheng2008}.} 

Second, we construct interval forecasts that target average posterior coverage probability across all units in our panel instead of pointwise coverage probability for each unit. We show that it is optimal to generate these forecasts as highest posterior density sets that use the same threshold for each unit instead of unit-specific thresholds. Because the predictive distributions associated with the Tobit models are mixtures of discrete and continuous distributions, ``interval'' forecasts may take the form of the union of one or more intervals and the value zero, and thus we refer to them as set forecasts subsequently. {\color{black}We prove that the empirical coverage frequency converges to the average nominal coverage frequency of the sets as the cross-sectional dimension of the panel tends to infinity.} This result is connected to similar findings in the literature on nonparametric function estimation and dates back to \cite{Wahba1983} and \cite{Nychka1988}. {\color{black}The underlying insights also have been recently used in concurrent research by \cite{armstrong2020robust} to construct empirical Bayes confidence intervals for vectors of means that are valid for multiple priors.} In the Monte Carlo study and the empirical application the proposed Bayesian set forecasts have good finite sample frequentist coverage properties in the cross-section. 

Third, we present a novel application in which we forecast bank-level loan charge-off rates. Our empirical analysis is based on more than 100 short panel data sets with a time dimension of $T=10$. These panel data sets include predominantly credit card (CC) and residential real estate (RRE) loans and cover various (overlapping) time periods. We also include local economic conditions as bank-specific regressors with homogeneous coefficients. For each data set, we document the density forecasting performance of several model specifications. We find that allowing for heteroskedasticity is important for good density and set forecasting performance. Overall, a specification with flexibly modeled correlated random effects and heteroskedasticity performs well in terms of density forecasting and is used in the subsequent analysis. In addition, we generate maps that compare the spatial distribution of predicted loan losses during and after the Great Recession and plot cross-sectional distribution of set forecasts. We document how set forecasts change as we move from targeting pointwise coverage probability to targeting average coverage probability. The latter approach smooths out differences among the lengths of the set forecasts and overall improves the forecasts with respect to both coverage probability and average length. 

                

The heterogeneous intercepts in our model can be interpreted as estimates of the quality of the banks' loan portfolios. Loan quality is potentially determined by many factors: the risk taking behavior of the bank, the potential customer base, and its ability to efficiently screen borrowers. In regressing heterogeneous coefficient estimates on bank characteristics we find that bank size as measured in total assets is positively related to inverse quality of the loan portfolio. A favorable interpretation of this finding is that larger banks are able to take higher risks on loans because they are better diversified or have a higher tolerance for risk. However, overall bank characteristics explain only a very small fraction of the estimated heterogeneity. 

Because the Tobit model is nonlinear, the effect of a change in local economic conditions that enter the model with homogeneous coefficients depends on the heterogeneous intercept and is thereby bank specific. We are able to compute a posterior distribution of the ``treatment'' effect for each bank and decompose it into an extensive-margin effect (a bank switches from no charge-offs to positive charge-offs during an economic downturn) and an intensive-margin effect (a bank increases its positive charge-offs during a downturn). 
We find that the variation in charge-off rates generated by local economic conditions is very small compared to the variation due to the heterogeneous intercept estimates. 


Our paper relates to several branches of the literature. {\color{black}We build on the Bayesian literature on the estimation of censored regression models.\footnote{A general survey of the literature on Bayesian estimation of univariate and multivariate censored regression models can be found, for instance, in the handbook chapter by \cite{LiTobias2011}.} The approach of using data augmentation for limited-dependent variable models that impute the latent uncensored variables dates back to \cite{Chib1992} and \cite{albert1993bayesian}. To sample the latent  observations we rely on an algorithm tailored toward dynamic Tobit models by \cite{Wei1999}. Sampling from Truncated Normal distributions is implemented with a recent algorithm of \cite{Botev2017}. Bayesian panel Tobit models have been estimated by \cite{baranchuk2008assessing} and~\cite{LiZheng2008}. Our flexible benchmark model is most closely related to the semiparametric model of \cite{LiZheng2008} which we generalize by introducing heteroskedasticity through a latent unit-specific error variance and allowing for a more flexible form of correlated random effects. As mentioned previously, the former is very important for the density and set forecast performance.\footnote{\cite{baranchuk2008assessing} report some results on point forecasts of the probability of zeros versus non-zeros, whereas we focus on set and density forecasts.}} 

We model the unknown distribution of the heterogeneous coefficients (intercepts and innovation variances) as Dirichlet process mixtures (DPM) of Normals. Even though we do not emphasize the nonparametric aspect of this modeling approach (due to a truncation, our mixtures are strictly speaking finite and in that sense parametric), our paper is related to the literature on nonparametric density modeling using DPM.\footnote{\cite{KeaneStavrunova} introduce a smooth mixture of Tobits to model a cross-section of healthcare expenditures. Our model is related, but different in that we are using a DPM to average across different intercept values and innovation variances.} Examples of econometrics papers that use DPMs in the panel data context are \citet{Hirano2002}, \citet{burda2013panel}, \citet{10.2307/j.ctt5hhrfp},
and \citet{jensen2015mutual}. The implementation of our Gibbs sampler relies on \cite{IshwaranJames2001,IshwaranJames2002}. 

{\color{black}As an alternative to a full Bayesian analysis, recent papers by \cite{GuKoenkerJAE2016,GuKoenker2014} and \cite{LiuMoonSchorfheide2015} have pursued an empirical Bayes strategy to generate predictions based on linear panel data models with heterogeneous coefficients. Forecasts from empirical Bayes and full Bayesian estimation approaches have desirable optimality properties as the cross-sectional dimension of the data set gets large. \cite{LiuMoonSchorfheide2015} generalize optimality results for the estimation of a vector of means in \cite{BrownGreenshtein2009} to a linear dynamic panel data forecasting setting. \cite{liu2018density} shows that the predictive density obtained from the full Bayesian analysis of a linear panel data model converges to the predictive density derived from the true cross-sectional distribution of the heterogeneous coefficients as the cross-section gets large.}

There also exists a literature on estimating the determinants of loan losses. This literature often uses nonperforming loans (loans that have not been serviced for more than 90 days) and tends to ignore the censoring which is reasonable if one uses an average across banks but can be problematic if one uses bank-level data. The two papers most closely related to our work are \cite{Ghosh2015,Ghosh2017}. We base our choice of bank-characteristic regressors on these papers.

The remainder of our paper is organized as follows. Section~\ref{sec:model} presents the specification of our dynamic panel Tobit model, a characterization of the posterior predictive distribution for future observations, and discusses the construction and evaluation of density and set forecasts. Section~\ref{sec:details} provides details on how we model the correlated random effects distribution and heteroskedasticity. It also presents the prior distributions for the parametric and flexible components of the model, and outlines a posterior sampler. We conduct a Monte Carlo experiment in Section~\ref{sec:montecarlo} to examine the performance of the proposed techniques in a controlled environment. The empirical application in which we forecast charge-off rates on various types of loans for a panel of banks is presented in Section~\ref{sec:empirics}. Finally, Section~\ref{sec:conclusion} concludes. Detailed derivations and proofs, a description of the data sets, and additional simulation and empirical results are relegated to the Online Appendix.

\section{Model Specification and Forecast Evaluation}
\label{sec:model}

Throughout this paper we consider the following dynamic panel Tobit model with heterogeneous intercepts and innovation variances:
\begin{eqnarray}
   && y_{it} = y_{it}^* \mathbb{I} \{ y_{it}^* \ge 0 \}, \label{eq:paneltobit} \\  
   && y_{it}^* | ({\color{black}Y_{1:N,0:t-1}^*, X_{1:N,-1:t-1}, \lambda_{1:N},\sigma^2_{1:N},\rho,\beta,\xi}) \stackrel{\rm indep}{\sim} N \big(\lambda_i + \rho y_{it-1}^* + \beta'x_{it-1}, \, \sigma_i^2 \big), \nonumber
\end{eqnarray}
%
where $i=1,\ldots,N$, $t=1,\ldots,T$, and $\mathbb{I}\{ y \ge a \}$ is the indicator function that is equal to one if $y \ge a$ and equal to zero otherwise. {\color{black} Throughout the paper, we abbreviate sequences of the form $(a_1,\ldots,a_n)$ by $a_{1:n}$. For instance,  
$Y^*_{1:N,0:t-1} = \big\{ (y_{10}^*,\ldots,y_{N0}^*),\ldots,(y_{1t-1}^*,\ldots, y_{Nt-1}^*) \big\}$, and $\lambda_{1:N} = (\lambda_1,\ldots,\lambda_N)$. 
The $n_x \times 1$ vector $x_{it}$ comprises a set of sequentially exogenous regressors. $\xi$ is a vector of hyperparameters defined in (\ref{eq:panelCRE}) below that does not affect the conditional distribution of $y_{it}^*$. It is assumed that conditional on the parameters and the regressors $x_{it-1}$, the observations $y_{it}$ are cross-sectionally independent.} {\color{black}The distributional assumption in (\ref{eq:paneltobit}) implies that we can write
\be
  p(y_{it}^* | Y_{1:N,0:t-1}^*, X_{1:N,-1:t-1},\lambda_{1:N},\sigma^2_{1:N},\rho,\beta,\xi) =
    p(y_{it}^* | y_{it-1}^*, x_{it-1},\lambda_i,\sigma^2_i,\rho,\beta),
    \label{eq:paneltobit.2}
\ee
which we will use subsequently to simplify formulas.}
Our specification uses the lagged latent variable $y_{it-1}^*$ on the right-hand side because it is more plausible for our empirical application. The Bayesian computations described in Section~\ref{subsec:details.posterior} below can be easily adapted to the alternative model, in which the lagged censored variable $y_{it-1}$ appears on the right-hand side.

We model the heterogeneous parameters as correlated random effects (CRE) with density
\be
  p( \lambda_i,y_{i0}^*,\sigma_i^2 |x_{i,-1},\xi ), \label{eq:panelCRE}
\ee
assuming cross-sectional independence of the heterogeneous coefficients.\footnote{We consider period $t=-1$ for $x$ in the conditioning set because of the timing assumption that charge-off rates can only respond with a one-period lag to changes in local economic conditions so as to accommodate possible sequentially exogenous regressors. See Section~\ref{subsec:model.simultaneity} for more details.} Here $\xi$ is a hyperparameter vector that indexes a family of CRE distributions. For instance, the candidate distribution of $( \lambda_i,y_{i0}^*,\ln\sigma_i^2 )$ could be jointly Normal with a mean that is a linear function of $x_{i,-1}$. In this case $\xi$ would include the parameters of the conditional mean function and the non-redundant parameters of the covariance matrix. To achieve a flexible representation of the distribution of $( \lambda_i,y_{i0}^*,\sigma_i^2 )$ we  consider a family of mixtures of Normal distributions in Section~\ref{sec:details}. {\color{black}We define the homogeneous parameter $\theta = [\rho,\beta']'$ and complete the model with the specification of a prior distribution for $\big( \theta,\xi \big)$.}

{\color{black}Our model is closely related to the panel Tobit models of \cite{baranchuk2008assessing} and \cite{LiZheng2008}, henceforth BC and LZ, respectively. However, the modeling approaches differ with respect to the treatment of coefficient heterogeneity and heteroskedasticity.\footnote{\color{black} As in the panel Probit model of \cite{chib2006inference}, one could allow for additional lags of $y_{it}^*$.} As in LZ, we restrict regression coefficient heterogeneity to the intercept. We also follow LZ in modeling the CRE distribution in (\ref{eq:panelCRE}) nonparametrically, albeit the details are slightly different. Because the regressors $x_{it}$ in our application are not assumed to be strictly exogenous, we condition the distribution of $(\lambda_i,y_{i0}^*)$ only on the initial values $x_{i,-1}$ and not on other $x_{it}$s. The most important difference between our specification and that of LZ is that we allow for heterogeneous innovation variances $\sigma^2_i$, whereas LZ set $\sigma^2_i = \sigma^2$ for all $i$. As documented in  Section~\ref{subsec:empirics.density},  $\sigma^2_i$ heterogeneity is very important for the construction of accurate set and density forecasts in our empirical application.
        
BC restrict the distribution of the heterogeneous coefficients to be Normal, but they do allow regression coefficients other than the intercept to be heterogeneous.\footnote{\color{black} Our framework can be easily extended to accommodate heterogeneous slope coefficients (see \cite{LiuMoonSchorfheide2015} and \cite{liu2018density}).} Rather than linking the heterogeneity to the regressors $x_{it}$, they let the mean of the distribution depend on additional unit-specific covariates. Instead of embedding additional covariates (such as bank characteristics) {\em ex ante} into (\ref{eq:panelCRE}), we run {\em ex post} regressions of estimates of the ratio $\widehat{\lambda_i/\sigma_i}$ on additional unit-specific covariates to explore potential relationships. The reasons for conducting an {\em ex post} analysis in our application are threefold: (i) it is not clear {\em ex ante} which bank characteristics are relevant, (ii) the relationship between bank characteristics and cross-sectional heterogeneity could be nonlinear, and (iii) bank characteristics may only explain a small fraction of the cross-sectional heterogeneity.

BC's interaction between regressors and the Normal CRE distribution generates heteroskedasticity in what could be interpreted as composite error term that consists of a homoskedastic innovation in the regression equation for $y_{it}^*$ and the randomness in the heterogeneous coefficients scaled by the regressors. In our model specification, the heteroskedasticity is unrelated to the regressors $x_{it}$ because we are treating the $\sigma^2_i$ as random effects. A relationship to the regressors could be generated through a CRE specification for $\sigma^2_i$, but we did not pursue this extension because in our application the regressors, local unemployment and house price growth, cannot explain the dispersion in $\sigma^2_i$. }

In the remainder of this section, we {\color{black}discuss our assumptions about the simultaneous determination of outcomes $y_{it}$ and regressors $x_{it}$ in Section~\ref{subsec:model.simultaneity},} the derivation of the posterior predictive density in Section~\ref{subsec:model.prediction}, the density forecast evaluation criteria in Section~\ref{subsec:model.pfcst.dfcst}, and the construction and evaluation of set forecasts in Section~\ref{subsec:model.ivfcst}.

\subsection{Simultaneity and Timing Assumptions}
\label{subsec:model.simultaneity}

{\color{black}In our application $y_{it}$ corresponds to bank-level loan charge-off rates and the regressors $x_{it}$ measure local economic conditions, such as unemployment and house prices, in the state in which the bank operates.\footnote{We consider a sample of small banks that conduct most of their business locally.} In this context it is plausible to assume that there is feedback from the bank charge-offs, which affect profitability and overall health of the banking sector, to the local economic conditions. 

The key assumption that we are making throughout the paper is that charge-off rates are only affected by lagged economic conditions and not by contemporaneous economic conditions. For concreteness, suppose that $x_{it}$ corresponds to economic conditions in the state in which bank $i$ operates. We assume that the state-level conditions in period $t=0,\cdots,T$ are described by the conditional density 
\begin{eqnarray}
p(X_{1:N,t}|Y_{1:N,0:t},Y_{1:N,0:t}^*,X_{1:N,-1:t-1},\theta_x, \lambda_{1:N},\sigma^2_{1:N}, \theta,\xi )
=
p(X_{1:N,t}|Y_{1:N,0:t},X_{1:N,-1:t-1},\theta_x).
\label{eq:zt} 
\end{eqnarray}
Thus, we allow current charge-offs to affect current state-level conditions.  
However, we assume that $X_{1:N,t}$ does not separately depend on the latent 
variables $Y_{1:N,0:t}^*$ and the heterogeneous coefficients 
$(\lambda_i,\sigma_i^2)$. In our application only actual charge-off rates are assumed to 
matter for economic outcomes. $\theta_x$ is a vector of 
parameters determining the law of motion for the state-level conditions.

Timing restrictions such as the one above have traditionally been widely used 
in the macroeconometric literature on structural vector autoregressions; see, 
for instance, the survey by \cite{Ramey2016}. Here we are assuming that a 
deterioration of macroeconomic conditions affects banks' decisions to write off 
loans with a one period delay, where the length of a period is a quarter in our 
application.\footnote{\color{black} Relaxing this assumption is beyond the scope 
of this paper.} Combining (\ref{eq:paneltobit}), (\ref{eq:paneltobit.2}), and (\ref{eq:zt}), we can write 
\begin{eqnarray}
	\lefteqn{
	p(Y_{1:N,1:T},Y_{1:N,1:T}^*,X_{1:N,0:T}|Y_{1:N,0},Y_{1:N,0}^*,X_{1:N,-1},\lambda_{1:N},\sigma^2_{1:N},\theta,\xi,\theta_x )} \label{eq:factorize.pypx}\\
    &=& \prod_{t=1}^T \left\{ p(X_{1:N,t}|Y_{1:N,0:t},X_{1:N,-1:t-1},\theta_x) 
        \times \left[ \prod_{i=1}^N  p(y_{it}|y_{it}^*) p(y_{it}^*|y_{it-1}^*,x_{it-1},\lambda_i,\sigma^2_i,\theta) \right] \right\}  \nonumber \\
&& \times  p(X_{1:N,0}|Y_{1:N,0},X_{1:N,-1},\theta_x) \nonumber \\[1ex]
&=& \left\{ \prod_{i=1}^N \left[ \prod_{t=1}^T  p(y_{it}|y_{it}^*) p(y_{it}^*|y_{it-1}^*,x_{it-1},\lambda_i,\sigma^2_i,\theta)        \right] \right\}  \prod_{t=0}^T p(X_{1:N,t}|Y_{1:N,0:t},X_{1:N,-1:t-1},\theta_x). \nonumber 
\end{eqnarray}
In slight abuse of notation $p(y_{i0}|y_{i0}^*)$ represents the censoring.
The distribution of $y_{it} | y_{it}^*$ is a unit point mass that is located at 0 if $y_{it}^* \le 0$ or at $y_{it}^*$ if $y_{it}^* > 0$.
Because the system is triangular, the panel Tobit component in (\ref{eq:paneltobit}) can be estimated independently of (\ref{eq:zt}) and without the use of 
instrumental variables.}

\subsection{Posterior Predictive Densities} 
\label{subsec:model.prediction}
 
Our goal is to generate forecasts of $Y_{1:N,T+h}$ conditional on the observations $(Y_{1:N,0:T},X_{1:N,-1:T})$. In the empirical analysis in Section~\ref{sec:empirics} we focus on $h=1$-step-ahead forecasts  which require the predictor $x_{iT}$, which is known at the forecast origin $t=T$. The extension to multi-step forecasts is discussed in Section~\ref{subsec:details.multistep}. {\color{black}Because in a Bayesian framework uncertainty with respect to parameters, latent variables, and future shocks is treated identically through the use of random variables, it is conceptually straightforward to construct a predictive distribution of $Y_{1:N,T+1}$ conditional on $(Y_{1:N,0:T},X_{1:N,-1:T})$ by integrating out all sources of uncertainty. The general approach is summarized, for instance, in \cite{GewekeWhiteman2006}. We subsequently describe the integration steps required for our panel Tobit model.}

According to (\ref{eq:panelCRE}) the distribution of $(Y_{1:N,0},Y_{1:N,0}^*)$ conditional on $X_{1:N,-1}$ does not depend on $\theta_x$. Using the factorization in (\ref{eq:factorize.pypx}), the CRE density (\ref{eq:panelCRE}), and the prior $p(\theta,\xi) = p(\theta)p(\xi)$, we can write the posterior distribution of the parameters and time-$T$ latent variables as
\begin{eqnarray}
 \lefteqn{p \big(Y_{1:N,T}^*, \lambda_{1:N},\sigma^2_{1:N},\theta,\xi| Y_{1:N,0:T}, X_{1:N,-1:T} \big)} \label{eq:paneltobit.posterior}\\
 & \propto & \bigg[ \prod_{i=1}^N \int \left(\prod_{t=1}^T  p(y_{it}|y_{it}^*) p(y_{it}^*|y_{it-1}^*,x_{it-1},\lambda_i,\sigma^2_i,\theta) \right)
                    \nonumber \\
                    && \times p(y_{i0}|y_{i0}^*) p\big( \lambda_i,y_{i0}^*,\sigma_i^2 |x_{i,-1},\xi \big) dY_{i,0:T-1}^* \bigg] p(\theta) p(\xi) \nonumber ,
\end{eqnarray}
where $\propto$ denotes proportionality. The posterior predictive distribution for units $i=1,\ldots,N$
is given by
\begin{eqnarray}
    \lefteqn{p(Y_{1:N,T+1}|Y_{1:N,0:T},X_{1:N,-1:T}) } \label{eq:basic_oracleposteriorpred} \\
    &=& \int \prod_{i=1}^N \bigg[ \int \int p(y_{iT+1}|y_{iT+1}^*) 
             p\big(y_{iT+1}^*|y_{iT}^*,x_{iT}, \lambda_i,\sigma_i^2,\theta\big)  \nonumber \\
    && \times         p\big(y_{iT}^*, \lambda_{i},\sigma^2_{i}|\theta,\xi, Y_{1:N,0:T}, X_{1:N,-1:T}\big) d y_{iT}^* d (\lambda_i,\sigma_i^2) \bigg] p(\theta,\xi|Y_{1:N,0:T}, X_{1:N,-1:T}) d(\theta,\xi). \nonumber
\end{eqnarray}
Draws from $p(Y_{1:N,T+1}|Y_{1:N,0:T},X_{1:N,-1:T})$ can be generated by sampling $(Y_{1:N,T}^*, \lambda_{1:N},\sigma^2_{1:N},\theta,\xi)$ from the posterior (\ref{eq:paneltobit.posterior}) and then evaluating the autoregressive law of motion for $y_{it}^*$ in (\ref{eq:paneltobit}) for $t=T+1$. 

To simplify the notation, we drop $X_{1:N,-1:T}$ from the conditioning set in the remainder of this section. Moreover, we denote the forecast horizon by $h$ again with the understanding that the discussion of multi-step forecasts is deferred to Section~\ref{subsec:details.multistep}.
We denote expectations and probabilities under the posterior predictive distribution by $\mathbb{E}_{Y_{1:N,0:T}}^{y_{iT+h}}[\cdot]$ and $\mathbb{P}_{Y_{1:N,0:T}}^{y_{iT+h}}\{\cdot\}$, respectively. More generally, we use subscripts to indicate the conditioning set and superscripts to denote the random variables over which the operators integrate. 
The predictive distribution is a mixture of a point mass at zero and a continuous distribution for realizations of $y_{iT+h}$ that are greater than zero:
\be
   p(y_{iT+h}|Y_{1:N,0:T}) = \mathbb{P}_{Y_{1:N,0:T}}^{y_{iT+h}}\{y_{iT+h}=0\} \delta_0(y_{iT+h}) + p_c(y_{iT+h}|Y_{1:N,0:T}) \mathbb{I}\{ y_{iT+h} \ge 0\}. \label{eq:model.preddensity.decomposition}
\ee
Here $\delta_0(y)$ is the Dirac function with the property $\delta_0(y)=0$ for $y \not=0$ and $\int \delta_0(y)dy = 1$. The density $p_c(y_{iT+h}|Y_{1:N,0:T})$ represents the continuous part of the predictive distribution.

\subsection{Evaluating Density Forecasts}
\label{subsec:model.pfcst.dfcst}


To compare the density forecast performance of various model specifications $M$ we report the average log predictive scores
\begin{eqnarray}
LPS_h(M) &=& \frac{1}{N} \sum_{i=1}^N \ln \big( \mathbb{I}\{ y_{iT+h} =0 \} \cdot \mathbb{P}_{Y_{1:N,0:T}}^{y_{iT+h}} \{ y_{iT+h} =0 | M \} \\
& & + \mathbb{I}\{ y_{iT+h}  > 0 \}   p(y_{iT+h}|Y_{1:N,0:T}) \big) \nonumber
\end{eqnarray}
and continuous ranked probability scores (CRPSs). The CRPS measures the $L_2$ distance between the cumulative distribution function $F_{Y_{1:N,0:T}}^{y_{iT+h}}(y|M)$ associated with $p(y_{iT+1}|Y_{1:N,0:T})$ and a ``perfect'' density forecasts which assigns probability one to the realized $y_{iT+h}$. Then,
\be
CRPS_h(M) = \frac{1}{N} \sum_{i=1}^N \int_{0}^\infty \big( F^{y_{iT+h}}_{Y_{1:N,0:T}}(y|M) - \mathbb{I}\{ y_{iT+h} \le y \} \big)^2 dy.
\ee
Both LPS and CRPS are proper scoring rules, meaning that it is optimal for the forecaster to truthfully reveal her predictive density \citep{GneitingRaftery}.

\subsection{Constructing and Evaluating Set Forecasts}
\label{subsec:model.ivfcst}

We construct set forecasts from the posterior predictive distribution $p(y_{iT+h}|Y_{1:N,0:T})$ in (\ref{eq:basic_oracleposteriorpred}) of the form:
\be
C_{iT+h|T}(Y_{1:N,0:T}) =   \{0\} \cup \left( \bigcup_{k=1}^{K_i} [a_{ik}, \, b_{ik}] \right) \label{eq:interval.forecast.generic}
\ee
with the understanding that (i) $C_i = \{0\}$ if $K_i=0$, (ii) $a_{i1}$ may be equal to zero, and (iii) 
\[
a_{i1} < b_{i1} < a_{i2} < b_{i2} < \ldots < a_{iK_i} < b_{iK_i}.
\]
The $\{0\}$ value arises from the discrete portion of the predictive density, whereas the interval components are obtained from the continuous portion of the predictive density; see the decomposition in (\ref{eq:model.preddensity.decomposition}).\footnote{Because in our model the support of the posterior predictive distribution of $y_{iT+h}^*$ includes $y < 0$, the probability of censoring is strictly positive and the set that includes $\{0\}$ is strictly shorter than the one without zero.} The disjoint interval segments may arise if the continuous part of the predictive density is multimodal. If we target an average coverage probability in the cross section, then for some units $i$ we might obtain the empty set, i.e., $C_{iT+h|T}(Y_{1:N,0:T}) = \emptyset$.

\noindent {\bf Constructing Set Forecasts.} To generate the set forecasts, we adopt a Bayesian approach and require that the probability of $\{y_{iT+h} \in  C_{iT+h|T}(Y_{1:N,0:T})\}$ conditional on having observed $Y_{1:N,0:T}$ reaches a pre-specified level. Given that the estimation of the Tobit model is executed with Bayesian techniques, the use of posterior predictive credible sets is natural. We distinguish between forecasts that are constructed to satisfy the coverage probability constraint pointwise, that is,
\be
\mathbb{P}_{Y_{1:N,0:T}}^{y_{iT+h}} \big\{ y_{iT+h} \in C_{iT+h|T}(Y_{1:N,0:T}) \big\} \ge 1-\alpha \quad \mbox{for all } i,
\label{eq:model.coverage.pointwise}
\ee
and sets that are constructed to satisfy the constraint on average:
\be
\frac{1}{N} \sum_{i=1}^N \mathbb{P}_{Y_{1:N,0:T}}^{y_{iT+h}} \big\{ y_{iT+h} \in C_{iT+h|T}(Y_{1:N,0:T}) \big\} \ge 1-\alpha.
\label{eq:model.coverage.average}
\ee     
The latter approach allows the sets $C_{iT+h|T}(Y_{1:N,0:T})$ for some units $i$ to be ``shortened'' in the sense that their posterior credible level drops below $1-\alpha$, whereas sets for other units are ``lengthened.''

It is well known that the shortest credible sets take the form of highest posterior density sets. Suppose that we require to satisfy the coverage constraint for each $i$ individually. If $\mathbb{P}_{Y_{1:N,0:T}}^{y_{iT+h}} \{ y_{iT+h} = 0\} \ge 1-\alpha$, then $C_{iT+h|T}(Y_{1:N,0:T}) = \{0\}$. Otherwise, 
the set takes the form 
\be
  C_{iT+h|T}(Y_{1:N,0:T}) = \{0\} \cup \big\{ y_{iT+h} \; \big| \; p_c(y_{iT+h}|Y_{1:N,0:T})\mathbb{I}\{ y_{iT+h} \ge 0\} \ge \kappa_{i}
  \big\},
  \label{eq:model.hpd.i}
\ee
where the threshold $\kappa_i$ is chosen such that
\[
   \int_{y_{iT+h} \in C} p_c(y_{iT+h}|Y_{1:N,0:T})\mathbb{I}\{ y_{iT+h} \ge 0\}dy_{iT+h} = 1 - \alpha - \mathbb{P}_{Y_{1:N,0:T}}^{y_{iT+h}} \{ y_{iT+h} = 0\}.
\]
Because $p_c(y|\cdot)$ is a continuous density, the HPD set can be represented as a collection of disjoint intervals as in (\ref{eq:interval.forecast.generic}).

If the objective is to minimize average length across $i$ conditional on the constraint on coverage probability holding only on average, then the unit-specific thresholds $\kappa_i$ in (\ref{eq:model.hpd.i}) are replaced by a common threshold $\kappa$ that applies to all units $i$. One can establish the optimality of the common threshold as follows. Suppose that one lowers the threshold for unit $i$ ($\kappa_i < \kappa$) and raises it for unit $j$ ($\kappa_j > \kappa)$. This lengthens the set for unit $i$ by $\delta_i > 0$ and shortens the set for unit $j$ by $\delta_j < 0$. The increase in coverage probability for unit $i$, $\Delta \pi_i > 0$, is less than $\delta_i \kappa$, whereas the decrease in coverage probability for unit $j$, $\Delta \pi_j < 0$, is less than $\delta_j \kappa$. Because we are holding the overall coverage probability constant, we obtain:
\[
  \delta_i \kappa > \Delta \pi_i = - \Delta \pi_j > - \delta_j \kappa.
\]
Thus, $\delta_i > - \delta_j$, which means that the overall average length increases and the uniform threshold of $\kappa$ dominates.

\noindent {\bf Evaluation of Set Forecasts.} The assessment of the set forecasts in our simulation study and the empirical application is based on the cross-sectional coverage frequency
\be
\frac{1}{N} \sum_{i=1}^N \mathbb{I} \big\{ y_{iT+h} \in C_{iT+h|T}(Y_{1:N,0:T}) \big\}
\label{eq:model.covfreq}
\ee
and the average length of the sets $C_{iT+h|T}(Y_{1:N,0:T})$
\be
\frac{1}{N} \sum_{i=1}^N \sum_{k=1}^{K_i} (b_{ik} - a_{ik}).
\label{eq:model.setlength}
\ee
Rather than trading off average length against deviations of average coverage frequency from the nominal coverage probability in a single criterion, we simply report both.\footnote{For various approaches to rank set forecasts see \cite{AskanaziEtAl2018}.}

The relationship between the nominal credible level of the set forecasts and the empirical coverage frequency is delicate. In {\color{black}Theorem~\ref{prop:post.emp.cov}} below we provide high-level regularity conditions under which 
\be
  \frac{1}{N} \sum_{i=1}^N \mathbb{I} \big\{ y_{iT+h} \in C_{iT+h|T}(Y_{1:N,0:T}) \big\} \stackrel{p}{\longrightarrow} 1-\alpha
  \label{eq:convergence}
\ee
in $\mathbb{P}^{Y_{1:N,0:T},Y_{1:N,T+h}}$ probability as $N \longrightarrow \infty$. {\color{black}Underlying this results is the well-known insight -- see, for instance, the textbook by \cite{Robert1994} -- that, for a generic parameter $\varsigma$ and data set $Y$, the following relationship between credible sets and confidence sets holds: 
\[
   \mathbb{P}^{\varsigma,Y} \{ \varsigma \in C(Y)\} = \int_Y \mathbb{P}_Y^\varsigma \{ \varsigma \in C(Y) \} d\mathbb{P}^Y= \int_\varsigma \mathbb{P}_\varsigma^Y \{ \varsigma \in C(Y)\} d \mathbb{P}^\varsigma. 
\]
Thus, $1-\alpha$ Bayesian credible sets have on average $1-\alpha$ frequentist coverage probability, but not pointwise for each $\varsigma$. In our framework the cross-sectional averaging across $i$ approximates the integration under the prior distribution.} The basic insight has previously been used in the literature on nonparametric function estimation, dating back to \cite{Wahba1983} and \cite{Nychka1988}, to obtain results that link average coverage probabilities to Bayesian credible levels. {\color{black}More recently, \cite{armstrong2020robust} constructed empirical Bayes confidence intervals for vectors of means that are valid for multiple priors.} 

{\color{black} Let $\vartheta = (\theta,\xi)$.} To state the theorem we define the following probability associated with the interval $[a_{ik,N},\, b_{ik,N}]$ conditional on $(Y_{i,0:T},\vartheta)$:
\be
F_{ik,N}(\vartheta) = \int_{a_{ik,N}}^{b_{ik,N}} p(y_{iT+h}^*|Y_{i,0:T},\vartheta) d y_{iT+h}^*.
\label{eq:theory.def.FiN}
\ee

\begin{theorem} \label{prop:post.emp.cov} Suppose the following assumptions hold: 
        \begin{tlist}
                    \item {\color{black} The future observations are sampled from the predictive density $p(y_{1:N,T+h}|Y_{1:N,0:T})$.}
                \item {\color{black}The posterior distribution $p(\vartheta|Y_{1:N,0:T})$ has the unique mode $\bar{\vartheta}_N$.} There exists a sequence of shrinking neighborhoods ${\cal N}_N(\bar{\vartheta}_N)$ with complements ${\cal N}^c_N(\bar{\vartheta}_N)$ and a sequence $\delta_N$, such that $\|\vartheta - \bar{\vartheta}_N \| \le \delta_N$ for all $\vartheta \in {\cal N}_N(\bar{\vartheta}_N)$ and  
                \[
                \mathbb{P}^\vartheta_{Y_{1:N,0:T}}  \big\{ \vartheta \in {\cal N}^c_N(\bar{\vartheta}_N) \big\} \stackrel{p}{\longrightarrow} 0, \quad
                \delta_N \stackrel{p}{\longrightarrow} 0
                \]      
                in $\mathbb{P}^{Y_{1:N,0:T}}$ probability as $N\longrightarrow \infty$.
                \item The functions $F_{ik,N}(\vartheta)$ defined in (\ref{eq:theory.def.FiN}) are locally Lipschitz in any compact neighborhood ${\cal N}_N(\vartheta)$ with Lipschitz constants $M_{ik,N}({\cal N}_N(\vartheta))$. 
                \item For some $M< \infty$ independent of $N$, the Lipschitz constants satisfy
                \[
                \mathbb{P}^{Y_{1:N,0:T}} \left\{\frac{1}{N} \sum_{i=1}^N \sum_{k=1}^{K_i} M_{ik,N}({\cal N}_N(\bar{\vartheta}_N)) > M \right\}
                \longrightarrow 0.
                \]
                \item The Bayesian coverage probability constraint, see (\ref{eq:model.coverage.pointwise}) or (\ref{eq:model.coverage.average}), holds with equality.
        \end{tlist}
        Then the empirical coverage frequency converges to the Bayesian credible level in the sense of (\ref{eq:convergence}). 
\end{theorem}

A proof of this theorem is provided in the Online Appendix. {\color{black}Assumption (i) states that the future observations are generated from the ``true'' predictive density $p(Y_{1:N,T+h}|Y_{1:N,0:T})$.} In Assumption (ii) we require the posterior distribution of $\vartheta$ to concentrate. Throughout the paper, we represent the CRE distribution through finite-dimensional mixtures; see Section \ref{subsec:details.specifications}. Thus, $\vartheta$ is finite-dimensional and the concentration results can be obtained from the literature on the consistency and asymptotic Normality of posterior distributions; see \cite{Hartigan1983}, \cite{vanderVaart1998}, \cite{GoshRamamoorthi2003}, or \cite{GhosalVaart2017} for textbook treatments. The only difference to many of the results stated in the literature is that we assume that the convergence in probability to occur under the marginal distribution of $Y_{1:N,0:T}$ rather than its distribution conditional on a ``true'' parameter which imposes some restrictions on the prior for $\vartheta$. Assumptions~(iii) and (iv) require the probabilities $F_{ik,N}$ to be smooth functions of $\vartheta$. In our model the probabilities are computed from finite-dimensional mixtures of Normal distributions, which are smooth functions of the underlying parameters. However, the Lipschitz constants are generally sample dependent and one needs to require that their average across $i$ and $k$ is stochastically bounded. In the Online Appendix we verify the conditions for a simple model without censoring.

\section{Correlated Random Effects, Priors, and Posteriors}
\label{sec:details}

We provide a characterization of the CRE distribution $p(\lambda_i,y^*_{i0},\sigma^2_i|x_{i,-1},\xi)$ and a specification of the prior distribution for $(\theta,\xi)$ in Section~\ref{subsec:details.specifications}. Section~\ref{subsec:details.posterior} contains a description of the posterior sampler, and Section~\ref{subsec:details.multistep} outlines multi-step forecasting approaches.

\subsection{(Correlated) Random Effects and Prior Distributions}
\label{subsec:details.specifications}

We now describe the prior distribution for $\theta$, the parametrization of the  distribution of $(\lambda_i,y_{i0}^*)$, and the prior distribution for the hyperparameter vector $\xi$. We begin with a homoskedastic random effects (RE) setup in which $\lambda_i$ and $y_{i0}^*$ are independent of each other and of $x_{i,-1}$. We then introduce heteroskedasticity and finally extend the model specification to CRE. The prior distribution involves a small number of tuning constants, denoted by $\tau$, that allow the researcher to scale the prior in various dimensions. 

The subsequent exposition involves various parametric probability distributions in addition to the Normal distribution that appeared in (\ref{eq:paneltobit}). We use $B(a,b)$, $G(a,b)$, and $IG(a,b)$ to denote the Beta, Gamma, and Inverse Gamma distributions, respectively. The pair $(\theta,\sigma^2)$ has a Normal-Inverse-Gamma distribution $NIG(m,v,a,b)$ if $\sigma^2 \sim IG(a,b)$ and $\theta|\sigma^2 \sim N(m,\sigma^2 v)$. Finally, the pair $(\Phi,\Sigma)$ has a matricvariate Normal-Inverse-Wishart distribution $MNIW(M,V,\nu,S)$ if $\Sigma \sim IW(\nu,S)$ has an inverse Wishart distribution and $\mbox{vec}(\Phi) |\Sigma \sim N(\mbox{vec}(M), \Sigma \otimes V )$.

\noindent {\bf Prior for $\theta$.}  We standardize the regressors $x_{it}$ to have zero mean and unit variance and use the following Normal prior for the regression coefficients $\theta$: 
\be
\theta \sim N(0,\tau_\theta I_{n_x+1} ),
\label{eq:details.theta.distribution}
\ee
where $\tau_\theta$ is a tuning constant that controls the prior variance. 

\noindent {\bf Flexible RE with homoskedasticity.} Under RE, the distribution of $\lambda_i$ and $y_{i0}^*$ does not depend on $x_{i,-1}$. Moreover, we assume that $\lambda_i$ and $y_{i0}^*$ are independent. Thus,
\[
p( \lambda_i,y^*_{i0} | x_{i,-1},\xi ) = p(\lambda_i|\xi)p(y^*_{i0}|\xi). 
\]
We consider a mixture representation for $p(\lambda_i|\xi)$ while assuming that the initial values $y_{i0}^*$ are normally distributed:
\begin{eqnarray}
\lambda_i|\xi      &\stackrel{iid}{\sim}&  N(\phi_{\lambda,k},\Sigma_{\lambda,k}) \; \mbox{with prob.} \; \pi_{\lambda,k}, \quad k=1,\ldots,K \label{eq:details.RE} \\
y_{i0}^* |\xi      &\stackrel{iid}{\sim}&  N(\phi_{y},\Sigma_{y}) \nonumber .
\end{eqnarray} 
The maximum number of mixture components $K$ is assumed to be pre-specified.\footnote{We use $K=20$ in the simulation exercise and the empirical analysis. This leads to the following uniform bound on the approximation error (see Theorem 2 of \cite{IshwaranJames2001}):
        $
        \Vert f^{\lambda,K}-f^{\lambda}\Vert \sim4N\exp[-(K-1)/\alpha]\le2.24\times10^{-5},
        $
        at the prior mean of $\alpha$ ($=1$) and a cross-sectional
        sample size $N=1000$.}

A prior over the RE distributions is induced through a prior $p(\xi)$ for the hyperparameter vector
\[
\xi = \big[ \phi_{\lambda,1},\Sigma_{\lambda,1},\pi_{\lambda,1}, \ldots, \phi_{\lambda,K},\Sigma_{\lambda,K},\pi_{\lambda,K}, \phi_y,\Sigma_y \big]'.
\]
During the Bayesian inference stage, the prior is updated in view of the data and we obtain a posterior distribution for $\xi$ and hence a posterior distribution for the RE distribution. 
The priors for the coefficients of the Normal distributions are
\be
(\phi_{\lambda,k},\Sigma_{\lambda,k}) \stackrel{iid}{\sim} NIG(0,\tau_\phi,3,2 \tau_{\sigma}^\lambda), \quad (\phi_{y},\Sigma_{y}) \sim NIG(0,\tau_\phi^y,3,2 \tau_{\sigma}^y).
\label{eq:details.RE.xi}
\ee
We parameterized the IG distribution such that the variances $\Sigma_{\lambda,k}$ and $\Sigma_{y}$ have a prior distribution with mean $\tau_\sigma$ and variance $\tau_\sigma^2$ (omitting the superscripts).\footnote{Under our parametrization of the $X \sim IG(a,b)$ distribution, $\mathbb{E}[X] = b/(a-1)$ for $a>1$, and $\mathbb{V}[X] = (\mathbb{E}[X])^2/(a-2)$ for $a>2$.} Conditional on $\Sigma$, the mean parameter $\phi$ has a $N(0,\tau_\phi \Sigma)$ distribution (omitting the subscripts). 
The marginal distribution of $y_{i0}^*$ implied by (\ref{eq:details.RE}) and (\ref{eq:details.RE.xi}) is a Student-$t$ distribution, whereas the distribution of $\lambda_i$ is a mixture of Student-$t$ distributions. The tuning constants can be used to control the spread of the means of the mixture components as well as the magnitude and variation of the variances of the mixture components. 

The prior for the probabilities $\pi_{\lambda,1:K}$ is generated by a mixture of truncated stick breaking processes $TSB(1,\alpha_\lambda,K)$ of the form
\be
\pi_{\lambda,1:K}|(\alpha_\lambda,K)
\sim \begin{cases}
        \zeta_{1}, & k=1,\\
        \prod_{j=1}^{k-1}\left(1-\zeta_{j}\right)\zeta_{k}, & k=2,\ldots,K-1,\\
        1-\sum_{j=1}^{K-1}p_{j}, & k=K,
\end{cases}, \quad \zeta_{k}\sim B(1,\alpha_\lambda), \quad \alpha_\lambda \sim G(2,2).
\label{eq:details.kappa.distribution}
\ee
Note that the $B(1,\alpha_\lambda)$ prior has a density $p(\zeta_k) \propto (1-\zeta_k)^{(\alpha_\lambda-1)}$. If $\alpha_\lambda$ is close to zero, then a lot of the mass of the distribution is concentrated near $\zeta_k=1$. This means that the first mixture component has a probability that is close to one, whereas the remaining mixture components have very small probabilities. If $\alpha_\lambda$ is close to two, then most of the mass of the distribution of $\zeta_k$ is concentrated on values of $\zeta_k$ that are close to zero. In turn, a larger number of mixture components receive non-trivial probabilities. The $G(2,2)$ distribution is recommended by Ishwaran and James (2002). It has a mean of one and draws fall with 95\% probability into the interval $[0.12,\, 2.75]$ which means that the prior covers both mixtures dominated by few components and mixtures with many non-trivial components. 


In the homoskedastic specification, we use the conjugate prior for $\sigma^2$ that arises in the context of a linear regression model:
\be
\sigma^2 \sim IG\big(3,2 \tau_v V^* \big).
\label{eq:details.homoskedasticity.sigma.distribution}
\ee
The IG distribution is parameterized in a similar way as the IG distributions in (\ref{eq:details.RE.xi}). $V^*=\frac{1}{N} \sum_{i=1}^N \widehat{\mathbb{V}}_{i}(y_{it})$ is the cross-sectional average of the time-series variances of $y_{it}$ and the tuning constant $\tau_v$ provides additional flexibility to scale the prior for $\sigma^2$.

\noindent {\bf Heteroskedasticity.} To generate heteroskedasticity one could simply replace (\ref{eq:details.homoskedasticity.sigma.distribution}) by $\sigma^2_i \sim IG\big(3,2 \tau_v V^* \big)$. However, to make the distribution a bit more flexible, we augment the hyperparameter vector $\xi$ and also represent the distribution of $\ln \sigma_i^2$ as a mixture of Normals:\footnote{\color{black}In an earlier version of the paper we used a mixture of IG distributions. We switched to a mixture of Normals for $\ln \sigma^2_i$ for a more symmetric treatment of $\lambda_i$ and $\sigma^2_i$. Alternatively, \cite{chib2002semiparametric} used Dirichlet process prior with an IG base measure to generate scale mixtures of Normals.} 
\be
\ln \sigma_i^2 |\xi \sim N \big( \psi_k, \omega_k^2 \big) \; \mbox{with prob.} \; \pi_{\sigma,k}, \quad k=1,\ldots,K.
\label{eq:details.sigma.distribution}
\ee
A straightforward change-of-variables yields the distribution $p(\sigma^2_i|\xi)$. As for the RE distribution, the coefficients $\psi_k$ and $\omega_k$ have $NIG$ priors:
\be
(\psi_k,\omega_k^2) \stackrel{iid}{\sim} NIG \big( \ln \left(\tau_v V^*\right)-\ln(2)/2, 1, 3, 2\ln 2 \big), \quad \quad k=1,\ldots,K.
\label{eq:details.psi.omega.distribution}
\ee
The parametrization is chosen so that the implied prior mean $\mathbb{E}[\sigma_i^2]$ and prior variance $\mathbb{V}[\sigma_i^2]$ for each mixture component $k$ matches the one implied by the prior used in the homoskedastic version of the Tobit model; see (\ref{eq:details.homoskedasticity.sigma.distribution}).\footnote{The marginal IG distribution implies $\mathbb{E}[\omega_k^2] =\ln 2$. Conditional on $\omega_k^2 = \ln 2$, the transformed parameter $\exp(\psi_k)$ has a Lognormal distribution with mean $ \tau_v V_*$ and variance $( \tau_v V_*)^2$.} Moreover, we verified by simulation that the marginal density of $\sigma_i^2$ under this prior is very similar to the $IG(3,(3-1) \tau_v V^*)$ distribution used for the homoskedastic specification. It does, however, have fatter tails as it is a mixture of log $t$ distributions. 

\noindent {\bf Flexible CRE with heteroskedasticity.} We extend the RE specification in two directions: first, we allow for correlation of $\lambda_i$ and $y_{i0}^*$ with $x_{i,-1}$. Second, we allow $\lambda_i$ and $y_{i0}^*$ to be correlated with each other conditional on $x_{i,-1}$. The CRE distribution is given by the following location and scale mixture of Normal distributions: 
\be
\big[ \lambda_i, \, y_{i0}^* \big] \,
\big| \, (x_{i,-1},\xi) \stackrel{iid}{\sim} N \big( [1,\, x_{i,-1}'] \Phi_k ,\Sigma_k \big) \; \mbox{with prob.} \; \pi_{\lambda,k}, \quad k=1,\ldots,K,
\label{eq:details.lambda.y.distribution}
\ee
where $\Phi_k$ is an $(n_x+1) \times 2 $ matrix and $\Sigma_k$ is a $2\times 2$ matrix. 
The hyperparameter vector $\xi$ is now defined to include the non-redundant elements of $(\Phi_k,\Sigma_k,\pi_{\lambda,k})$.

For the mixture probabilities $\pi_{\lambda,1:K}$ we use the same prior distribution as in (\ref{eq:details.kappa.distribution}). 
The prior distribution for the coefficient matrices $\Phi_k$ and $\Sigma_k$ is a multivariate generalization of the RE distribution. We assume:
\be
(\Phi_k,\Sigma_k)  \stackrel{iid}{\sim} MNIW \big( 0, \tau_\phi I_{n_x+1}, 7, 4D(\tau_\sigma)), \quad  k=1,\ldots,K, \quad D(\tau_\sigma)=\begin{bmatrix}\tau_\sigma^\lambda & 0 \\
0 & \tau_\sigma^y \\
\end{bmatrix}.
\label{eq:details.Phi.Sigma.distribution}
\ee
Under this parametrization the marginal IW distribution of the $2 \times 2$ matrix $\Sigma_k$ has mean $D(\tau_\sigma)$.
The conditional distribution of $\Phi_k|\Sigma_k$ is $ MN(0,\tau_\phi \Sigma_k \otimes I_{n_z+1})$, where $\tau_\phi$ scales the variance of the Normal distribution. The dimension of $\Sigma_k$ is $2 \times 2$  and, hence, the marginal distribution of $\lambda_i$ is identical to the RE case.\footnote{The marginal distribution of the (1,1) element of the $IW(7,4 D(\tau_\Sigma))$ distribution is $IW(6,4 D_{11}(\tau_\Sigma))$. Converted into the parametrization of the Gamma distribution, this corresponds to an $IG(3,2 D_{11}(\tau_\Sigma))=IG(3,2\tau_\sigma^\lambda)$ distribution.}

\noindent {\bf Tuning of the Prior.} The scale of the prior distribution is controlled by a vector of tuning constants:
\[
   \tau = \big[ \tau_\theta, \tau_\phi, \tau_\sigma^\lambda, \tau_\sigma^y, \tau_v \big]'.
\]
While these tuning constants could in principle be determined in a data-driven way, using a marginal data density criterion (see the approach used in the Bayesian vector autoregression (VAR) literature, for instance, \cite{DelNegro2007b} and \cite{GiannoneLenzaPrimiceri2015}), we do not pursue that route in this paper. Instead we choose $\tau$ \textit{ex ante} in an informal calibration step. While $\tau_\theta$ has a straightforward interpretation after the regressors have been normalized, the implications of the remaining constants are less transparent because they control priors that are specified over a set of distributions. We recommend the researcher makes an initial choice and then samples from the prior. We found it useful to examine plots of moments or number of modes associated with the distributions. Similar plots can be generated based on the posterior. If a researcher finds that the posterior is located in an area that has essentially no prior mass, then the scaling of the prior can be adjusted to examine whether the initial prior unduly biases the posterior estimates. An example in the context of our empirical application is provided in the Online Appendix.

\subsection{Posterior Sampling}
\label{subsec:details.posterior}

Draws from the posterior distribution can be obtained with a Gibbs sampling algorithm. We subsequently describe the conditional distributions over which the Gibbs sampler iterates. We focus on the flexible CRE specification with heteroskedasticity, which is the most complicated specification. A key feature of the Gibbs sampler is that it uses data augmentation by sampling the sequences of latent variables $Y_{i,0:T}^*$, $i=1,\ldots,N$. {\color{black} In this regard we are building on  \cite{TannerWong1987} (data augmentation for a general state-space model), \cite{Chib1992} (static Tobit model), \cite{albert1993bayesian} (Probit model), \cite{CarterKohn1994} (linear state space model), and \cite{Wei1999} (dynamic Probit model). The general blocking of parameters in the Gibbs sampler is related to \cite{baranchuk2008assessing} and \cite{LiZheng2008}. The sampler for the flexible mixture representation of the CRE distribution is based on \cite{IshwaranJames2001,IshwaranJames2002}.} In terms of the actual implementation, the computations for the Tobit model are very similar to the ones for the linear model studied in \cite{liu2018density}. The only exception is the treatment of the latent variables $Y_{i,0:T}^*$ which closely follows \cite{Wei1999}.  
        
In order to characterize the conditional posterior distributions for the Gibbs sampler, we introduce some additional notation. Because $p(\lambda_i,y^*_{i0}|x_{i,-1},\xi)$ and $p(\sigma_i^2|\xi)$ are mixture distributions, {\em ex post} each $(\lambda_i,y_{i0}^*)$ and $\sigma_i^2$ is associated with one of the $K$ mixture components, respectively. We denote the component membership indicators by $\gamma_{i,\lambda}$ and $\gamma_{i,\sigma} \in \{1,\ldots,K\}$, respectively.

\noindent {\bf Step~1: Drawing from $Y^*_{i,0:T} | ( Y_{1:N,0:T}, X_{1:N,-1:T}, \lambda_{1:N},\sigma_{1:N}^2,\gamma_{1:N,y},\gamma_{1:N,\sigma},\theta,\xi)$.} To fix ideas, consider the following sequence of observations
$y_{i0},\ldots,y_{iT}$:
\[
y_{i0}^*, \; y_{i1}^*, \; 0, \; 0, \; 0, \; y_{i5}^*, \; y_{i6}^*, \; 0, \; 0, \; 0, \; y_{i10}^*.
\]
Our model implies that whenever $y_{it} > 0$ we can deduce that $y_{it}^*=y_{it}$. Thus, we can focus our attention on periods in which $y_{it}=0$. In the hypothetical sample we observe two strings of censored observations: $(y_{i2}, y_{i3}, y_{i4})$ and $(y_{i7}, y_{i8}, y_{i9})$.
We use $t_1$ for the start date of a string of censored observations and $t_2$ for the end date. In the example we have two such 
strings, we write $t_1^{(1)}=2$, $t_2^{(1)}=4$, $t_1^{(2)}=7$, $t_2^{(2)}=9$. The goal is to characterize
$p(Y^*_{i,t_1^{(1)}:t_2^{(1)}},Y^*_{i,t_1^{(2)}:t_2^{(2)}} |Y_{i,0:T},\ldots)$. 
Because of the AR(1) structure, observations in periods $t<t_1-1$ and $t>t_2+1$ contain no additional information about $y^*_{it_1},\ldots,y^*_{it_2}$. Thus, we obtain
\begin{eqnarray*}
        \lefteqn{p(Y^*_{i,t_1^{(1)}:t_2^{(1)}},Y^*_{i,t_1^{(2)}:t_2^{(2)}} |Y_{i,0:T},\ldots)} \\
        &=& p(Y^*_{i,t_1^{(1)}:t_2^{(1)}}|Y_{i,t_1^{(1)}-1:t_2^{(1)}+1},\ldots) 
        p(Y^*_{i,t_1^{(2)}:t_2^{(2)}}|Y_{i,t_1^{(2)}-1:t_2^{(2)}+1},\ldots),
\end{eqnarray*}
which implies that we can sample each string of latent observations independently.

Let $s=t_2-t_1+2$ be the length of the segment that includes the string of censored observations as well as the adjacent
uncensored observations.
Iterating the AR(1) law of motion for $y_{it}$ forward from period $t_1-1$ we deduce that 
the vector of random variables $[Y_{i,t_1:t_2}^*,y_{it_2+1}]'$ conditional on $y_{it_1-1}$ is multivariate Normal with mean
\be
M_{1:s|0} = [ \mu_1, \ldots, \mu_s ]', \quad
\mu_1 = \lambda_i + \rho y_{it_1-1} + \beta'x_{it_1-1}, \quad \mu_\tau = \lambda_i + \rho \mu_{\tau-1} + \beta'x_{i\tau-1} \; \mbox{for} \; \tau =2,\ldots,s.
\label{eq:details.oracle_M1sg0}
\ee
The covariance matrix takes the form
\be
\Sigma_{1:s|0} = \sigma^2_i \left[ 
\begin{array}{ccc}
        \rho_{1,1|0} & \cdots & \rho_{1,s|0} \\
        \vdots & \ddots & \vdots \\
        \rho_{s,1|0} & \cdots & \rho_{s,s|0} 
\end{array}    
\right], \quad
\rho_{i,j|0} = \rho_{j,i|0} = \rho^{j-i} \sum_{l=0}^{i-1} \rho^{2l} \; \mbox{for} \; j \ge i.
\label{eq:details.oracle_Sigma1sg0}
\ee
We can now use the formula for the conditional mean and variance of a multivariate Normal distribution
\begin{eqnarray}
M_{1:s-1|0,s} &=& M_{1:s-1|0} - \Sigma_{1:s-1,s|0} \Sigma_{ss|0}^{-1} (y_{it_2+1} - \mu_s)  \label{eq:details.oracle_MSigma1s1g0s} \\
\Sigma_{1:s-1,1:s-1|0,s} &=&
\Sigma_{1:s-1,1:s-1|0} - \Sigma_{1:s-1,s|0} 
\Sigma_{ss|0}^{-1} \Sigma_{s,1:s-1|0} \nonumber
\end{eqnarray}
to deduce that 
\be
Y^*_{i,t_1:t_2} \sim TN_- \big( M_{1:s-1|0,s}, \Sigma_{1:s-1,1:s-1|0,s} \big).
\label{eq:details.oracle_post_Ystar}
\ee
Here we use $TN_-(\mu,\Sigma)$ to denote a Normal distribution that is truncated to satisfy $y \le 0$. 
Draws from this Truncated Normal distribution can be efficiently generated using the algorithm recently proposed by \cite{Botev2017}.

There are two important special cases. First, suppose that $t_2=T$, meaning that the last observation in the sample is censored. Then the mean vector and the covariance matrix of the Truncated Normal distribution are given by (\ref{eq:details.oracle_M1sg0}) and (\ref{eq:details.oracle_Sigma1sg0}) with the understanding that $s=t_2-t_1+1$. Second, suppose that $t_1=0$, meaning that the initial 
observation in the sample $y_{i0}=0$. Because in this case the observation $y_{it_1-1} = y_{i,-1}$ is missing, we need to modify
the expressions in (\ref{eq:details.oracle_M1sg0}) and (\ref{eq:details.oracle_Sigma1sg0}). According to (\ref{eq:details.lambda.y.distribution}), the joint distribution of $(\lambda_i,y_{i0}^*)$ is a mixture of Normals. Using the mixture component membership indicator $\gamma_{i,\lambda}$, we can express 
$y^*_{i0}|(\lambda_i ,x_{i,-1}) \sim N(\mu_*(\lambda_i,x_{i,-1}),\sigma^2_*)$. This leads to the mean vector
\be
M_{1:s} = [ \mu_1, \ldots, \mu_s ], \quad
\mu_1 = \mu_*(\lambda_i,x_{i,-1}), \quad \mu_\tau = \lambda_i + \rho \mu_{\tau-1} + \beta'x_{i\tau-1} \; \mbox{for} \; \tau =2,\ldots,s
\ee
and the covariance matrix 
\be
\Sigma_{1:s} = \sigma^2_i \left[ 
\begin{array}{cccc}
        0 & 0 & \cdots & 0 \\
        0 & \rho_{1,1} & \cdots & \rho_{1,s-1} \\
        \vdots & \vdots & \ddots & \vdots \\
        0 & \rho_{s-1,1}& \cdots & \rho_{s-1,s-1} 
\end{array}    
\right]
+ \sigma_*^2 \left[ 
\begin{array}{ccc}
        \rho^{0+0} & \cdots & \rho^{0+(s-1)} \\
        \vdots     & \ddots & \vdots \\
        \rho^{(s-1)+0} & \cdots & \rho^{(s-1)+(s-1)} 
\end{array}    
\right],
\ee
where the definition of $\rho_{i,j}$ is identical to the definition of $\rho_{i,j|0}$ in (\ref{eq:details.oracle_Sigma1sg0}).
One can then use the formulas in (\ref{eq:details.oracle_MSigma1s1g0s}) to obtain the mean and covariance parameters of the Truncated 
Normal distribution. 

\noindent {\bf Step~2: Drawing from $\lambda_i| (Y_{1:N,0:T}, Y^*_{1:N,0:T},X_{1:N,-1:T},\sigma_{1:N}^2,\gamma_{1:N,y},\gamma_{1:N,\sigma},\theta,\xi)$}. Posterior inference with respect to $\lambda_i$ becomes ``standard'' once we condition on the latent variables $Y^*_{i,0:T}$ and the component membership $\gamma_{i,\lambda}$. It is based on the Normal location-shift model
\be
y_{it}^* - \rho y_{it-1}^* - \beta'x_{it-1} = \lambda_i + u_{it}, \quad u_{it} \stackrel{iid}{\sim} N(0,\sigma_i^2), \quad t=1,\ldots,T.
\label{eq:details.oracle_lambdalh}
\ee
Because the conditional prior distribution $\lambda_i|(y_{i0}^*,x_{i,-1},\gamma_{i,\lambda})$ is Normal, the posterior of $\lambda_i$ is also Normal and direct sampling is possible.

\noindent {\bf Step~3: Drawing from $\sigma_i^2| (Y_{1:N,0:T}, Y^*_{1:N,0:T},,X_{1:N,-1:T},\lambda_{1:N},\gamma_{1:N,y},\gamma_{1:N,\sigma},\theta,\xi)$}. Posterior inference with respect to $\sigma_i^2$ is based on the Normal scale model
\be
y_{it}^* - \rho y_{it-1}^* - \beta'x_{it-1} - \lambda_i = u_{it}, \quad u_{it} \stackrel{iid}{\sim} N(0,\sigma_i^2), \quad t=1,\ldots,T.
\label{eq:details.draw.sigma.i}
\ee
However, even conditional on the mixture component membership indicator $\gamma_{i,\sigma}$, the prior for $\sigma_i^2$ in (\ref{eq:details.sigma.distribution}) is not conjugate and direct sampling is not possible. Instead, we sample from this non-standard posterior via an adaptive random walk Metropolis-Hastings (RWMH) step.\footnote{We use an adaptive procedure based on \cite{AtchadeRosenthalothers2005}, which adaptively adjusts the random walk step size to keep acceptance rates around 30\%.}

\noindent {\bf Step~4: Drawing from $\theta| (Y_{1:N,0:T}, Y^*_{1:N,0:T},,X_{1:N,-1:T},\lambda_{1:N},\sigma_{1:N}^2,\gamma_{1:N,\lambda},\gamma_{1:N,\sigma},\xi)$}. Conditional on the latent variables $Y^*_{i,0:T}$ and the heterogeneous coefficients $\lambda_i,\sigma_i^2$, we can express our model as
\be
    y_{it}^* - \lambda_i  =  \rho y_{it-1}^* + \beta'x_{it-1} + u_{it}, \quad u_{it} \stackrel{iid}{\sim} N(0,\sigma_i^2), \quad i=1,\ldots,N, \quad t=1,\ldots,T.
\ee
The temporal and spatial independence of the $u_{it}$'s allows us to pool observations across $i$ and $t$.
Under the Normal prior in (\ref{eq:details.theta.distribution}), the posterior distribution of $\theta = [\rho,\beta']'$ is also Normal and we can obtain draws by direct sampling. 

\noindent {\bf Step~5: Drawing from $(\gamma_{i,\lambda},\gamma_{i,\sigma})| (Y_{1:N,0:T}, Y^*_{1:N,0:T},,X_{1:N,-1:T},\lambda_{1:N},\sigma_{1:N}^2,\theta,\xi)$.} We describe how to draw the component membership indicator $\gamma_{i,\lambda}$. Straightforward modifications lead to a sampler for $\gamma_{i,\sigma}$. Note that $\xi$ contains the elements of $\Phi_{1:K}$, $\Sigma_{1:K}$, and $\pi_{\lambda,1:K}$. The prior probability that unit $i$ is a member of component $k$ is given by $\pi_{\lambda,k}$. Let $\bar{\pi}_{i,\lambda,k}$ denote the posterior probability of unit $i$ belonging to component $k$ conditional on the set of means $\Phi_{1:K}$ and variances $\Sigma_{1:K}$ as well as $\lambda_i$. The $\bar{\pi}_{i,\lambda,k}$'s are given by
\be
\bar{\pi}_{i,\lambda,k}  = \frac{\pi_{\lambda,k} p_N\big(\lambda_i| y_{i0}^*, x_{i,-1}, \Phi_k,\Sigma_k \big)}{ \sum_{k=1}^K \pi_{\lambda,k} p_N\big(\lambda_i| y_{i0}^*, x_{i,-1}, \Phi_k,\Sigma_k \big)}.
\ee
Note that the conditional distribution $\lambda_i| (y_{i0}^*, x_{i,-1}, \Phi_k,\Sigma_k )$ is Normal, indicated by the notation $p_N(\cdot)$, and can be derived from the joint Normal distributions of the mixture components in (\ref{eq:details.lambda.y.distribution}). 
Thus,
\be
\gamma_{i,\lambda} | (\Phi_{1:k}, \Sigma_{1:K}, \lambda_i ) = k \; \mbox{with prob.} \; \bar{\pi}_{i,\lambda,k}.
\label{eq:details.pinonparametric_post_gammai}
\ee

\noindent {\bf Step~6: Drawing from $\xi |(Y_{1:N,0:T}, Y^*_{1:N,0:T},X_{1:N,-1:T},\lambda_{1:N},\sigma_{1:N}^2,\gamma_{1:N,\lambda},\gamma_{1:N,\sigma},\theta)$.} Sampling from the conditional posterior of $\Phi_{1:K}$, $\Sigma_{1:K}$, and $\pi_{\lambda,1:K}$ can be implemented as follows.  Let $n_{\lambda,k}$ be the number of units and $J_{\lambda,k}$ the set of units that are members of component $k$. Both $n_{\lambda,k}$ and $J_{\lambda,k}$ can be determined based on $\gamma_{1:N,\lambda}$. The conditional posterior of the component probabilities takes the form of a generalized truncated stick breaking process
\be
     \pi_{\lambda,1:K} | (n_{\lambda,1:K},\alpha,K) \sim TSB \left( \{1+n_{\lambda,k} \}_{k=1}^K,\left\{ \alpha_\lambda+\sum_{j=k+1}^K n_{\lambda,j} \right\}_{k=1}^K,K \right),
\label{eq:details.pinonparametric_post_pik}
\ee
meaning that the $\zeta_k$'s in (\ref{eq:details.kappa.distribution}) have a $B\big(1+n_{\lambda,k},\alpha_\lambda+\sum_{j=k+1}^K n_{\lambda,j}\big)$ distribution. Conditional on $\pi_{\lambda,1:K}$ the hyperparameter $\alpha_\lambda$ has a Gamma posterior distribution of the form
\be
    \alpha_\lambda | \pi_{\lambda,1:K} \sim G(2+K-1,2 - \ln \pi_{\lambda,K}).
\label{eq:details.pinonparametric_post_alpha}
\ee
The conditional posterior for $(\Phi_k,\Sigma_k)$ takes the form
\begin{eqnarray}
\lefteqn{p(\Phi_k,\Sigma_k|Y_{1:N,0:T},Y^*_{1:N,0:T},\lambda_{1:N},\sigma^2_{1:N},\gamma_{1:N,\lambda}, \gamma_{1:N,\sigma}, \theta)} \label{eq:details.pinonparametric_post_muomk} \\ 
&&\propto
p(\Phi_k,\Sigma_k ) \prod_{i \in J_{\lambda,k}}  p(\lambda_i,y_{i0}^*|x_{i,-1},\Phi_k,\Sigma_k). \phantom{hallo hallo hallo}\nonumber
\end{eqnarray}
Because here the prior $p(\Phi_k,\Sigma_k )$ is MNIW and the likelihood $\prod_{i \in J_{\lambda,k}}  p(\lambda_i,y_{i0}^*|x_{i,-1},\Phi_k,\Sigma_k)$ is derived from a multivariate Normal linear regression model, the conditional posterior of $(\Phi_k,\Sigma_k)$ is also MNIW. All three conditional posteriors allow direct sampling. The derivations can be modified to obtain the conditional posterior of $\psi_{1:K}$, $\omega_{1:K}$, and $\pi_{\sigma,1:K}$.

\noindent {\bf Step~7: Drawing from the predictive density.} Conditional on $(y_{iT}^*,\lambda_i,\sigma_i^2,\theta)$ and $x_{i,T:T+h-1}$, paths from the predictive distribution for $y_{i,T+1:T+h}$ can be easily generated by simulating (\ref{eq:paneltobit}) forward; see Section~\ref{subsec:details.multistep} for further details. 

\noindent {\bf Modifications for the simplified model specifications.} If the CRE distribution is modeled parametrically instead of flexibly, then the drawing of the component membership indicators $(\gamma_{i,\lambda},\gamma_{i,\sigma})$ in Step~5 and the drawing of $\pi_{\cdot,1:K}$ and $\alpha$ in Step 6 are unnecessary. One only has to sample from the MNIW posterior of $(\Phi_1,\Sigma_1)$ and the NIG posterior of $(\psi_1,\omega_1)$. 
Under homoskedasticity, i.e., $\sigma_i^2=\sigma^2$ for all $i$, we can pool (\ref{eq:details.draw.sigma.i}) in Step~3 across $t$ and $i$. In combination with the prior in (\ref{eq:details.homoskedasticity.sigma.distribution}) this leads to an IG posterior for $\sigma^2$ from which one can sample directly.
The RE specification requires modifications to Step~1, because the distribution of $y_{i0}$ is now simplified to $y_{i0}^* \sim N(\phi_y,\Sigma_y)$, to Step~2 because the prior distribution of $\lambda_i$ is different, and to Step~6 because the pairs of VAR coefficients $(\Phi_k,\Sigma_k)$ are replaced by $(\phi_{\lambda,k},\Sigma_{\lambda,k})$ and $(\phi_y,\Sigma_{y})$, which leads to NIG posteriors.  

\subsection{Multi-Step Forecasting}
\label{subsec:details.multistep}

In general, there are two ways of extending one-step-ahead to multi-step-ahead forecasting: an iterated approach and a direct approach.

First, iterating the law of motion of $y_{it}^*$ in (\ref{eq:paneltobit}) forward by $h$ periods, starting from period $t=T$, yields 
\be
y_{iT+h}^* = \lambda_i \left( \sum_{s=0}^{h-1} \rho^s \right) + \rho^h y_{iT}^*
+ \beta' \left( \sum_{s=0}^{h-1} \rho^s x_{iT+h-1-s}\right) +  \sum_{s=0}^{h-1} \rho^s u_{iT+h-s}.
\label{eq:paneltobit.hstep}
\ee
{\color{black}Thus, forecasting $y_{iT+h}^*$ iteratively requires the path $x_{i,T:T+h-1}$. We can distinguish the following scenarios: (i) the path is given at time $T$. For instance, in a stress-testing application of our framework the path of the exogenous variables would be specified by the regulator as part of the stressed macroeconomic scenario.  (ii) $x_{it}$ is strictly exogenous. In this case the user has to specify a separate model for  $x_{it}$ to simulate future trajectories along which (\ref{eq:paneltobit.hstep}) is evaluated. Because of the exogeneity, this simulation can be conducted independently of the simulation of (\ref{eq:paneltobit.hstep}). Suppose one has draws $(\lambda_i^{(j)},\rho^{(j)}, \beta^{(j)},\sigma^{2(j)}_i)$ and draws $x_{i,T:T+h-1}^{(j)}$ from the posterior predictive distribution of the exogenous regressors, then one can define
        \begin{eqnarray*}
                \mu_{iT+h|T}^{(j)} &=& \lambda_i^{(j)} \left( \sum_{s=0}^{h-1} (\rho^{(j)})^s \right) + (\rho^{(j)})^h y_{iT}^{*(j)}
                + \beta^{(j)} \left( \sum_{s=0}^{h-1} (\rho^{(j)})^s x_{iT+h-1-s}^{(j)}\right) \\
                \sigma^{2(j)}_{iT+h|T} &=& \left( \sum_{s=0}^{h-1} (\rho^{(j)})^{2s}  \right) \sigma_i^{2(j)}.
        \end{eqnarray*}
One can sample $y_{iT+h}^{*(j)}$ from a $N(\mu_{iT+h|T}^{(j)},\sigma^{2(j)}_{iT+h|T})$ and apply the censoring to obtain a draw $y_{iT+h}^{(j)}$. 
(iii) The $x_{it}$s are endogenous and interact with the $y_{it}$s, which is the case in our application. To capture the feedback from the dependent variables to the regressors, one has to simulate $(Y_{1:N,T+1:T+h},Y^*_{1:N,T+1:T+h},X_{1:N,T+1:T+h-1})$ jointly; see (\ref{eq:factorize.pypx})}.

Second, rather than generating $h$-step ahead forecasts iteratively, in practice forecasters often engage in direct estimation of an $h$-step-ahead prediction function. In our framework, this approach amounts to estimating a model of the form
\[
   y_{it}^* = \lambda_i + \rho y_{it-h}^* + \beta'x_{it-h} + u_{it}
\]
with the understanding that the serial correlation in $u_{it}$ implied by our original model (\ref{eq:paneltobit}) is ignored. A discussion of the disadvantages and advantages of multi-step estimation in the context of VARs can be found in \cite{Schorfheide2005}.

\section{Monte Carlo Experiment}
\label{sec:montecarlo}

{\color{black} We conduct a Monte Carlo experiment to illustrate the performance of the set and density forecasts from the dynamic panel Tobit model in (\ref{eq:paneltobit}) under ideal conditions. We also discuss the estimation of the heterogeneous coefficients. We simplify the model} by omitting the additional predictors $x_{it}$ and using the RE specification. We endow the forecaster with knowledge of the true $p(y^*_{i0})$ and factorize $p(\lambda_i,y^*_{i0},\ln \sigma_i^2|\xi)$ as $p(\lambda_i|\xi) p(y^*_{i0}) p(\ln \sigma_i|\xi)$.
The data generating process (DGP) is summarized in Table~\ref{t_MC1design}. We set the autocorrelation parameter to $\rho=0.8$ and consider skewed random effects distributions for $\lambda_i$ and $\ln \sigma_i^2$ that are generated as mixtures of Normals. 

\begin{table}[t!]
        \caption{Monte Carlo Design}
        \label{t_MC1design}
        \begin{center}
                \scalebox{0.9}{
                        \begin{tabular}{l} \hline \hline
                                Law of Motion: $y^*_{it} = \lambda_i + \rho y^*_{it-1} + u_{it}$ where $u_{it} \sim N(0,\sigma^2_i)$ and $\rho = 0.8$ \\
                                Initial Observations: $y_{i0}^* \sim N(0,1)$ \\
                                Skewed Random Effects Distributions:\\
                                \hspace{1em} $p(\lambda_i|y_{i0}^*) = \frac{1}{9}p_N\left(\lambda_i|\frac{5}{2},\frac{1}{2}\right)+\frac{8}{9}p_N\left(\lambda_i|\frac{1}{4},\frac{1}{2}\right)$ \\
                                \hspace{1em} $p(\ln\sigma^2_i|y_{i0}^*) = \frac{1}{9}p_N\left(\ln\sigma^2_i-c|\frac{5}{2},\frac{1}{2}\right)+\frac{8}{9}p_N\left(\ln\sigma^2_i-c|\frac{1}{4},\frac{1}{2}\right)$, $c$ is chosen such that $\mathbb E[\sigma^2_i] = 1$\\
                                Sample Size: $N=1,000$, $T=10$ \\
                                Number of Monte Carlo Repetitions: $N_{sim}=100$ \\ 
                                Fraction of Zeros: 45\%, Fraction of All-Zeros: 15\%\\ \hline
                        \end{tabular}
                }
        \end{center}
\end{table}

The simulated panel data sets consist of $N=1,000$ cross-sectional units and the number of time periods in the estimation sample is $T=10$. We generate one-step-ahead forecasts for period $t=T+1$. The fraction of zeros across all samples is 45\% and for roughly 15\% of the cross-sectional units the sample consists of $T=10$ zeros (``all zeros'').\footnote{In the Online Appendix we report additional results for Monte Carlo designs with 60\% and 75\% zeros, respectively. The overall message from the baseline Monte Carlo design is preserved under the alternative specifications.}
The measures of forecast accuracy discussed in Sections~\ref{subsec:model.pfcst.dfcst} and~\ref{subsec:model.ivfcst} are first computed for the cross section $i=1,\ldots,N=1,000$ and we then average the performance statistics over the $n_{sim}=100$ Monte Carlo repetitions.

\noindent {\bf Model Specifications and Predictors.} We  compare the performance of six predictors described below: four Bayes predictors derived from different versions of the dynamic panel Tobit model, a predictor derived from a Tobit model with homogeneous coefficients, and a predictor from a linear model with homogeneous coefficients that ignores the censoring. The prior distributions used for the estimation of the various models were described in Section~\ref{subsec:details.specifications} and are summarized in Table~\ref{t_MC1priors}. Further implementation details are provided in the Online Appendix.

\begin{sidewaystable}
        \caption{Summary of Prior Distributions}
    \label{t_MC1priors}
    \begin{center}
        \scalebox{0.95}{
        \begin{tabular}{lllll} \hline \hline \\[-1ex]
                Specification  & $\lambda$ & $p(\lambda|\xi)$ & $\sigma^2$ & $p(\sigma^2|\xi)$ \\[1ex] \hline \\[-1ex]
                Flexible RE \& Heterosk. & $\lambda \sim N(\phi_{\lambda,k},\Sigma_{\lambda,k}) $ & $(\phi_{\lambda,k},\Sigma_{\lambda,k}) \sim NIG(0,5,3,2)$ & $\ln\sigma^2 \sim N(\psi_{k},\omega_{k})$ & $(\psi_k,\omega_{k}) \sim NIG(\ln V^*-\ln(2)/2,$\\
                & $\mbox{w.p.} \; \pi_{\lambda,k}$ & $\pi_{\lambda,k} \sim TSB(1,\alpha_{\lambda},K)$  & $\mbox{w.p.} \; \pi_{\sigma,k}$ & \hspace*{2cm} $1,3,2\ln 2)$ \\
                  & & $\alpha_{\lambda} \sim G(2,2)$ & & $\pi_{\sigma,k} \sim TSB(1,\alpha_\sigma,K)$ \\
                & &  & & $\alpha_\sigma \sim G(2,2)$
\\[2ex]
                   Normal RE \& Heterosk. &
                        $\lambda \sim N(\phi_\lambda,\Sigma_\lambda)$ & $(\phi_{\lambda},\Sigma_{\lambda}) \sim NIG(0,5,3,2)$ & $\ln\sigma^2 \sim N(\psi,\omega)$ & $(\psi,\omega) \sim NIG(\ln V^*-\ln(2)/2,$\\
                &  & & & \hspace*{2cm} $1,3,2\ln 2)$
\\[2ex]
                    Flexible RE \& Homosk. & $\lambda \sim N(\phi_{\lambda,k},\Sigma_{\lambda,k})$ & $(\phi_{\lambda,k},\Sigma_{\lambda,k}) \sim NIG(0,5,3,2)$  & $\sigma^2\sim IG(3,2 V^*)$& N/A\\
                & $\mbox{w.p.} \; \pi_{\lambda,k}$ & $\pi_{\lambda,k} \sim TSB(1,\alpha_{\lambda},K)$ & & \\
                  & & $\alpha_{\lambda} \sim IG(2,2)$ & & 
\\[2ex]

                Normal RE \& Homosk. &
                        $\lambda \sim N(\phi_\lambda,\Sigma_\lambda)$ & $(\phi_{\lambda},\Sigma_{\lambda}) \sim NIG(0,5,3,2)$ & $\sigma^2\sim IG(3,2 V^*)$& N/A 
\\[2ex]
               
                Pooled Tobit / Linear & $\lambda \sim N(0,5)$ & N/A & $\sigma^2\sim G(3,2V^*)$& N/A \\[2ex]
                \hline \\[-1ex]
                Prior for $\rho$ &  $\rho \sim N(0,5)$ 
                \\[2ex]
                Prior for $y_{i0}^*$ & \multicolumn{4}{l}{$y_{i0}^* \sim N(\phi_y,\Sigma_y)$, $(\phi_y,\Sigma_y) \sim NIG(0,5,3,2)$}
                \\[1ex] \hline
        \end{tabular}
    }
    \end{center}
     {\footnotesize {\em Notes:} We set $V^* = \frac 1 N \sum_{i=1}^N \widehat{\mathbb{V}}_i(Y_{it})$, the cross-sectional average of the time-series variances of $y_{it}$.
}\setlength{\baselineskip}{4mm}
        
\end{sidewaystable}

We consider four versions of the dynamic panel Tobit model with random effects (see Section~\ref{subsec:details.specifications} for details): (i) flexible RE and heteroskedasticity; (ii) Normal RE and heteroskedasticity; (iii) flexible RE and homoskedasticity; and (iv) Normal RE and homoskedasticity. Versions (ii)-(iv) are misspecified in light of the DGP. 
The pooled Tobit specification ignores the heterogeneity in $\lambda_i$, setting $\lambda_i=\lambda$ for all $i$, and imposes homoskedasticity. Finally, the pooled linear specification imposes $\lambda_i=\lambda$, $\sigma_i=\sigma^2$ for all $i$, and, in addition, ignores the censoring of the observations during the estimation stage (and finally censors the forecasts at 0).

\begin{table}[t!]
        \caption{Monte Carlo Experiment: Forecast Performance and Parameter Estimates}
        \label{t_MC1results_ifcst_dfcst}
        \begin{center}
                \scalebox{0.97}{
                \begin{tabular}{lcccccccc} \\  \hline \hline
                        & \multicolumn{2}{c}{Density Fcst} &   \multicolumn{2}{c}{Set Forecast} &   \multicolumn{2}{c}{Set Forecast} &
                        \multicolumn{2}{c}{Estimates}
                         \\ 
                        & & &   \multicolumn{2}{c}{``Average"} &   \multicolumn{2}{c}{``Pointwise"} \\  
                        &       LPS         &       CRPS &      Cov.    &      Length  &      Cov.    &      Length  & Bias$(\hat{\rho})$        &       StdD$(\hat{\rho})$  \\ \hline
                        Flexible \& Heterosk. &  -0.757 &   0.277 &   0.910 &   1.260 &   0.933 &   1.503  &  -0.002 &   0.005   \\ 
                        Normal \& Heterosk.   &  -0.758 &   0.277 &   0.908 &   1.248 &   0.932 &   1.498 &  -0.006 &   0.005    \\ 
                        Flexible \& Homosk.   &  -0.902 &   0.294 &   0.929 &   1.506 &   0.942 &   1.698  &   0.007 &   0.008  \\ 
                        Normal \& Homosk.     &  -0.903 &   0.294 &   0.929 &   1.501 &   0.942 &   1.699 &   0.001 &   0.007   \\ 
                        Pooled Tobit       &  -0.935 &   0.313 &   0.935 &   1.705 &   0.947 &   1.911 &   0.252 &   0.004  \\ 
                        Pooled Linear      &  -1.243 &   0.357 &   0.923 &   1.925 &   0.933 &   1.951 &   0.229 &   0.005   \\
                        \hline
                \end{tabular}   
        }    
        \end{center}
        {\footnotesize {\em Notes:} The Monte Carlo design is summarized in Table~\ref{t_MC1design}. The true values for $\rho$ is 0.8. ``Cov.'' is coverage frequency and ``Length'' is an average across $i$.}\setlength{\baselineskip}{4mm}
\end{table}

\noindent {\bf Density and Set Forecasts.} To assess the density forecasts we compute LPS and CRPS; see Section~\ref{subsec:model.pfcst.dfcst}. The larger LPS and the smaller CRPS the better the forecast. The accuracy statistics are reported in columns 2 and 3 of Table~\ref{t_MC1results_ifcst_dfcst}. As expected, the flexible specification with heteroskedasticity that nests the DGP delivers the most accurate density forecasts. While replacing the flexible representations of the RE distributions with Normal distributions only leads to a marginal deterioration of forecast performance, imposing homoskedasticity generates a substantial drop in accuracy. The two ``pooled'' models that ignore the intercept heterogeneity perform the worst.  

We consider two types of set forecasts; see Section~\ref{subsec:model.ivfcst}.  The first type targets the average coverage probability in the cross-section (``average''), whereas the other type targets the correct coverage probability for each unit $i$ (``pointwise''). To assess the set forecasts we compute the coverage frequency and the average length of 90\% predictive sets. Results are presented in columns 4 to 7 of Table~\ref{t_MC1results_ifcst_dfcst}. 
The ``average'' sets constructed from the heteroskedastic specification have good frequentist coverage properties. They attain coverage frequencies of 91.0\% and 90.8\%, respectively. A comparison between the ``average'' and the ``pointwise'' set forecasts from the heteroskedastic models highlights that the average length of the ``average'' sets is indeed smaller. Moreover, the coverage frequency of the ``pointwise'' sets exceeds the nominal coverage level of 90\% by a larger amount. We observe a similar pattern also for the set forecasts from the homoskedastic model specifications.  Overall, the homoskedastic specifications generate worse set forecasts, in terms of coverage frequency {\em and} average length, than the heteroskedastic specifications. 

\noindent {\bf Parameter Estimates.} The last two columns of Table~\ref{t_MC1results_ifcst_dfcst} summarize the bias and standard deviation of the posterior mean estimator of the homogeneous parameter $\rho$. Under the correctly specified ``Flexible \& Heterosk.'' model the bias is close to zero and the standard deviation is small. Replacing the flexible RE specification by a Normal specification raises the bias by a factor of three. Replacing heteroskedasticity by homoskedasticity approximately increases the standard deviation by 50\% because of a loss of efficiency. Imposing intercept homogeneity (pooled Tobit and pooled linear specification) leads to a substantial increase in the bias.

\begin{figure}[t!]
        \caption{Posterior Means and Estimated RE Distributions for $\lambda_i$}
        \label{f_MC1results_phatlambda}
        \begin{center}
        \begin{tabular}{cc}
                Flexible \& Heteroskedastic & Normal \& Heteroskedastic \\
                \includegraphics[width=2.6in]{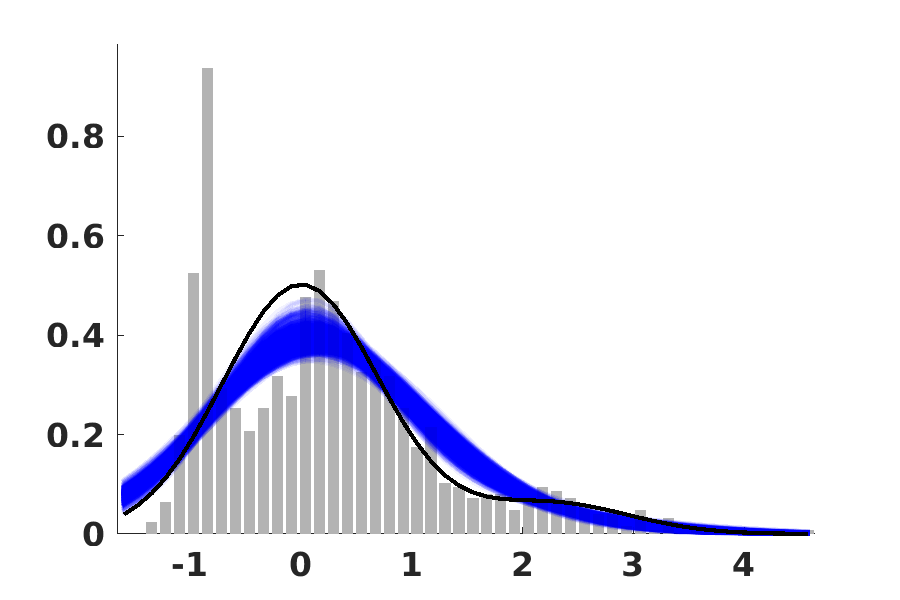}       &
                \includegraphics[width=2.6in]{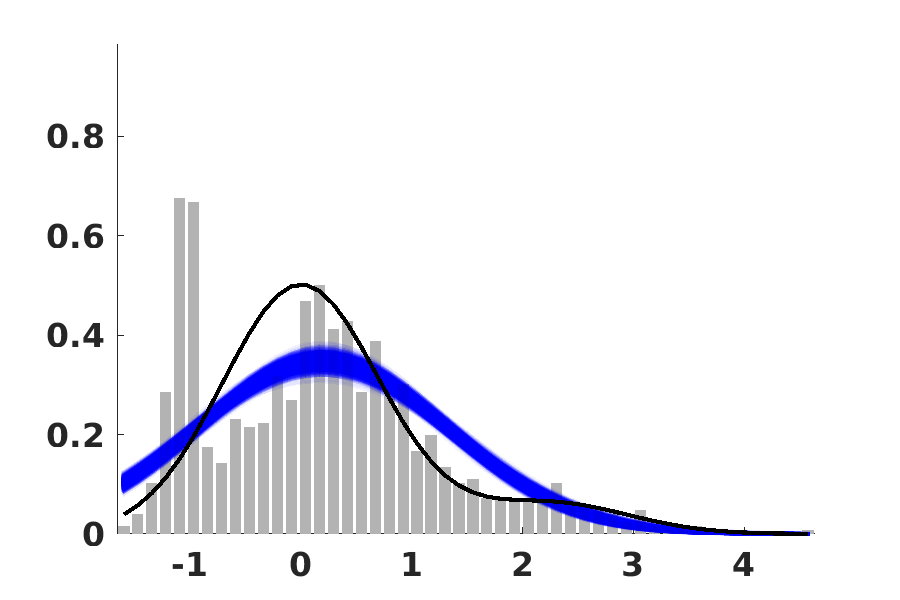}     
        \end{tabular}
        \end{center}
        {\footnotesize {\em Notes:} The histograms depict $\mathbb{E}[\lambda_i|Y_{1:N,0:T}]$, $i=1,\ldots,N$, for two different model specifications. The shaded areas are hairlines obtained by generating draws from the posterior distribution of $\xi$ and plotting the corresponding  random effects densities $p(\lambda|\xi)$.  The black lines represent the true $p(\lambda)$. }\setlength{\baselineskip}{4mm}                                     
\end{figure}

The panels of Figure~\ref{f_MC1results_phatlambda} show the true RE density $p(\lambda)$, hairlines that represent $p(\lambda|\xi)$ generated from posterior draws of $\xi$, and histograms of the point estimates $\mathbb{E}[\lambda_i|Y_{1:N,0:T}]$. The left panel corresponds to the flexible specification, whereas the panel on the right displays results for the Normal specification. In both cases we allow for heteroskedasticity. The posterior distribution of $p(\lambda|\xi)$ under the flexible specification concentrates near the true density, whereas, not surprisingly, the parametric specification yields larger discrepancies between the true RE density and the draws from the posterior distribution. Because of the shrinkage effect of the prior distribution, we generally expect the cross-sectional distribution of $\mathbb{E}[\lambda_i|Y_{1:N,0:T}]$ (histograms) to be less dispersed than the distribution of $\lambda_i$ (density plots). Moreover, if we observe sequences of all zeros for multiple units $i$, posterior inference of the corresponding $\lambda_i$s should be the same. This will create a spike in the left tail of the $\mathbb{E}[\lambda_i|Y_{1:N,0:T}]$ distribution. Both features are present in the figure.\footnote{\color{black} We provide illustrative analytical examples of these effects in the Online Appendix.} 

\section{Empirical Analysis}
\label{sec:empirics}

We now use different versions of the dynamic panel Tobit model to forecast loan charge-off rates (charge-offs divided by the stock of loans in the previous period, multiplied by 400). As mentioned in the introduction, a charge-off occurs if a loan is deemed unlikely to be collected because the borrower has become substantially delinquent after a period of time. The prediction of charge-off rates is interesting from the perspectives of banks, regulators, and investors, because charge-offs generate losses on loan portfolios and are, in fact, a large contributor to bank losses. If these charge-off rates are large, the bank may be entering a period of distress and require additional capital.\footnote{The accounting details are more complicated: bank balance sheets contain a contra asset account called ``Allowance for Loan and Lease Losses'' (ALLL). Provisions for LLL are created based on estimated credit losses and reduce the income of the bank. Charge-offs reduce the ALLL and the gross loans on the balance sheet, leaving the net amount unchanged. At this stage, the charge-offs do not lead to a further reduction of income. Whether or not a bank takes a loss provision or a charge-off is to some extent a managerial/accounting decision, although regulators require loans they classify as losses to be charged off. We abstract from strategic accounting aspects; see \cite{Moyer1990} for a seminal paper.}

We consider a panel of ``small'' banks, which we define to be banks with total assets of less than one billion dollars.\footnote{Monitoring potential loan losses in small banks is useful by itself. Moreover, the delinquency rates of small banks could foreshadow those rates of large banks since the small banks tend to have more subprime borrowers who are more vulnerable to minor deterioration in economic condition.} For these banks it is reasonable to assume that they operate in local markets. The forecasts are generated from model (\ref{eq:paneltobit})  where $y_{it}$ are charge-off rates. As potential explanatory variables we consider the quarter-on-quarter inflation in the house price index $\Delta \ln \mbox{HPI}_{it-1}$, the change in the unemployment rate $\Delta \mbox{UR}_{it-1}$, and the growth rate in personal income $\Delta \ln \mbox{INC}_{it-1}$.
Here $\Delta$ is the temporal difference operator. The term $\beta'x_{it-1}$ therefore captures variation in regional economic conditions which we measure at the state level. Banks located in regions with poor economic conditions may be more likely to encounter loan losses because of a higher fraction of borrowers that are unable to repay their loans. Our baseline model is based on $x_{it} = [\Delta \ln \mbox{HPI}_{it}, \Delta \mbox{UR}_{it}]'$, but we also consider a specification that includes personal income as a third explanatory variable and a specification without any explanatory variables.

The heterogeneous intercept $\lambda_i$ can be interpreted as a bank-specific measure of the quality of the loan portfolio: the smaller $\lambda_i$, the higher the quality of the loan portfolio and the less likely a charge-off is to occur. The autoregressive component in the model captures the persistence of the composition of the loan portfolio over time, and the covariates shift the density of repayment probabilities. We consider various choices of $p \big(\lambda_i,y_{i0}^*,\sigma_i| x_{i,-1},\xi \big)$; see Section~\ref{subsec:details.specifications}.  The data set is described in Section~\ref{subsec:empirics.data}. Section~\ref{subsec:empirics.density} presents density forecast comparisons for various model specifications. Estimates of the heterogeneous and homogeneous parameters are reported in Section~\ref{subsec:empirics.estimates}. Posterior predictive checks are conducted in Section~\ref{subsec:predchecks}. Finally, Section~\ref{subsec:empirics.intervalforecasts} contains the set forecast results.

\subsection{Data}
\label{subsec:empirics.data}

The raw data are obtained from ``call reports'' (FFIEC 031 and 041) that the banks have to file with their regulator and are available through the website of the Federal Reserve Bank of Chicago. Due to missing observations and outliers we restrict our attention to four loan categories: credit card (CC) loans, other consumer credit (CON), construction and land development (CLD), and residential real estate (RRE). We construct rolling panel data sets for each loan category that have a time dimension of twelve quarterly observations: one observation $y_0$ to initialize the estimation, $T=10$ observations for estimation, and one observation to evaluate the one-step-ahead forecast. The number of banks $N$ in the cross section varies depending on market size and date availability. The earliest sample considered in the estimation starts ($t=0$) in 2001Q2 and the most recent sample starts in 2016Q1. A detailed description of the construction of the data set is provided in the Online Appendix.

In the remainder of this section, we present two types of results: (i) forecast evaluation statistics and parameter estimates for RRE and CC charge-off rates based on samples that cover the Great Recession and range from 2007Q2 ($t=0$) to 2009Q4 ($t=T$);\footnote{There are, in general, large uncertainties during the Great Recession. Thus, accurate density and set forecasts are important.} (ii) scatter plots summarizing forecast evaluation statistics for the 111 rolling samples that we constructed (based on data availability) for the above-mentioned four loan categories. 

\begin{table}[t!]
        \caption{Summary Statistics for Baseline Samples}
        \label{tab:baseline.summarystats}
        \begin{center}
                \begin{tabular}{lrccccc} \hline \hline
                                          & $N$   & Zeros [\%] & All Zeros [\%] & Mean & 75th & Max  \\ \hline                  
                        RRE & 2,576 & 76 & 61 & 0.25 & 0.00 & 33.1 \\ 
                        CC              & 561   & 43 & 22 & 3.27 & 4.07 & 260  \\ \hline
                \end{tabular}
        \end{center}
        {\footnotesize {\em Notes:} The estimation sample ranges from 2007Q2 ($t=0$) to 2009Q4 ($t=T=10$). We forecast 2010Q1 observations. ``Zeros'' refers to the fraction of zeros in the overall sample of observations (all $i$ and all $t$), ``All Zeros'' is the fraction of banks for which charge-off rates are zero in all periods. Mean, 75th percentile, and maximum are computed based on the overall sample.}\setlength{\baselineskip}{4mm}
\end{table}

Table~\ref{tab:baseline.summarystats} contains some summary statistics for the two baseline samples. 
For the small banks in our sample, RRE loans are an important part of their loan portfolio. For approximately 45\% (25\%) of the banks RREs account for 20\% to 50\% (more than 50\%) of their loan portfolio. CC loans, on the other hand, make up less than 2\% of the loans held by the banks in our sample. Both baseline samples contain a substantial fraction of zero charge-off observations: 76\% for RREs and 43\% for CC, which makes it challenging to estimate the coefficients of our panel data models. Moreover, 61\% of the banks in the RRE sample never write off any loans between 2007 and 2009. The distribution of charge-off rates, across banks and time, is severely skewed. For RREs the 75th percentile is 0 and the maximum is 33.1\% annualized. For CCs the corresponding figures are 4.07\% and 260\%, respectively. A table with summary statistics for the remaining samples is provided in the Online Appendix.

\subsection{Density Forecasts and Model Selection}
\label{subsec:empirics.density}

\noindent {\bf Selected Samples.} We begin the empirical analysis by comparing the density forecast performance of several variants of (\ref{eq:paneltobit}) for the two baseline samples using $x_{it} = [\Delta \ln \mbox{HPI}_{it}, \Delta \mbox{UR}_{it}]'$. This comparison includes forecasts from a Tobit model and a linear model with homogeneous intercepts and homoskedastic innovation variances. Table~\ref{tab:dfcst} reports LPS (the larger the better) and CRPS (the smaller the better). Several observations stand out. First, allowing for heteroskedasticity improves the density forecasts unambiguously. Second, in both RRE and CC samples, all four heteroskedastic specifications lead to very similar density forecasting performance.

\begin{table}[t!]
        \caption{Density Forecast Performance}
        \label{tab:dfcst}
        \begin{center}
                \begin{tabular}{llcccc} \\  \hline \hline
                        & \multicolumn{2}{c}{RRE} &   \multicolumn{2}{c}{CC}  \\ 
                        Specification       &     LPS      &       CRPS  &     LPS         &       CRPS \\ \hline
                        \multicolumn{5}{l}{Heteroskedastic Models} \\ \hline
                        Flexible CRE            &       -0.523 &       0.240  &       -1.921 &       1.957  \\   
                        Normal CRE      &       -0.521 &       0.240  &       -1.901 &       1.895  \\
                        Flexible RE             &       -0.525 &       0.238  &       -1.925 &       1.970  \\ 
                        Normal RE       &       -0.524 &       0.237  &       -1.912 &       1.936  \\ \hline

                        \multicolumn{5}{l}{Homoskedastic Models} \\ \hline
                        Flexible CRE            &       -0.751 &       0.272   &       -2.512 &       2.495  \\
                        Normal CRE      &       -0.751 &       0.272  &       -2.463 &       2.343  \\ 
                        Flexible RE             &       -0.751 &       0.270  &       -2.630 &       2.613  \\ 
                        Normal RE       &       -0.752 &       0.270  &       -2.535 &       2.391   \\ 
                        Pooled Tobit            &       -0.831 &       0.310  &       -2.642  &       2.620  \\ 
                        Pooled Linear           &       -1.594 &       0.374  &       -3.010 &       2.789  \\ 

                        \hline
                \end{tabular}
        \end{center}
        {\footnotesize {\em Notes:} The estimation sample ranges from 2007Q2 ($t=0$) to 2009Q4 ($t=T=10$). We use $x_{it} = [\Delta \ln \mbox{HPI}_{it}, \Delta \mbox{UR}_{it}]'$ and forecast 2010Q1 observations.}\setlength{\baselineskip}{4mm}
\end{table}  

\begin{figure}[t!]
        \caption{Log Predictive Density Scores -- All Samples}
        \label{fig:lps.scatters}
        \begin{center}
                        \includegraphics[width=2in]{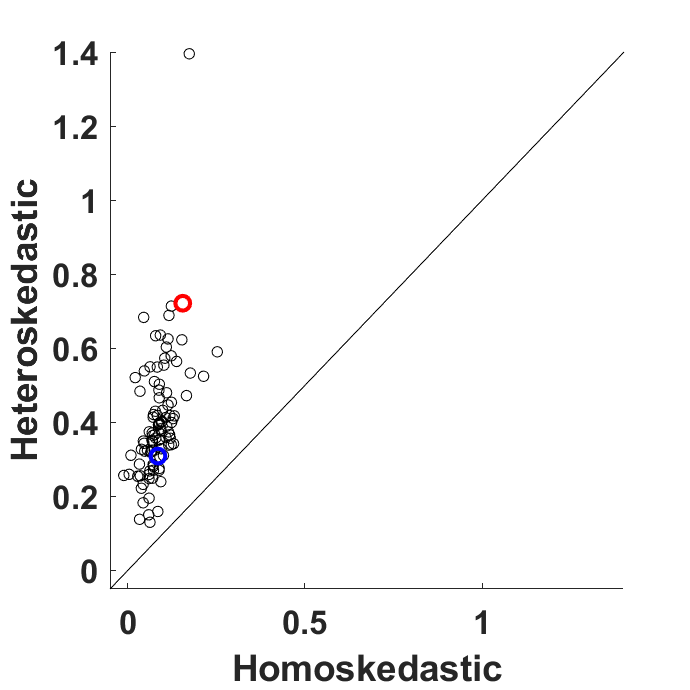}     
        \end{center}
        {\footnotesize {\em Notes:} Flexible CRE specification. The figure illustrates pairwise comparisons of log predictive scores. We also show the 45-degree line. Log probability scores are depicted as differentials relative to pooled Tobit. The blue (red) circle corresponds to RRE (CC). We use $x_{it} = [\Delta \ln \mbox{HPI}_{it}, \Delta \mbox{UR}_{it}]'$.}\setlength{\baselineskip}{4mm}                                     
\end{figure}

\noindent {\bf All Samples.} Figure~\ref{fig:lps.scatters} summarizes the LPS comparisons for all 111 samples. We focus on the comparison of predictive scores from the heteroskedastic specifications versus homoskedastic specifications using flexibly modeled correlated random effects. The solid line is the 45-degree line and the blue and red circles correspond to the scores associated with the baseline RRE and CC samples reported in Table~\ref{tab:dfcst}. The figure shows that the results for the baseline samples are qualitatively representative: incorporating heteroskedasticity is important for density forecasting. 
We provide a figure in the Online Appendix that illustrates that LPS differentials between Normal versus flexible CREs  and CREs versus REs are small.
In view of these results, we subsequently focus on the flexible CRE specification with heteroskedasticity. 

\begin{figure}[t!]
        \caption{RRE Charge-Off Rate Predictive Tail Probabilities, Spatial Dimension}
        \label{fig:tailprob.spatial}
        \begin{center}
                \begin{tabular}{cc}
                        2010Q1 & 2018Q1 \\
                        \includegraphics[width=3in,trim=0in 0in 0in 0in, clip]{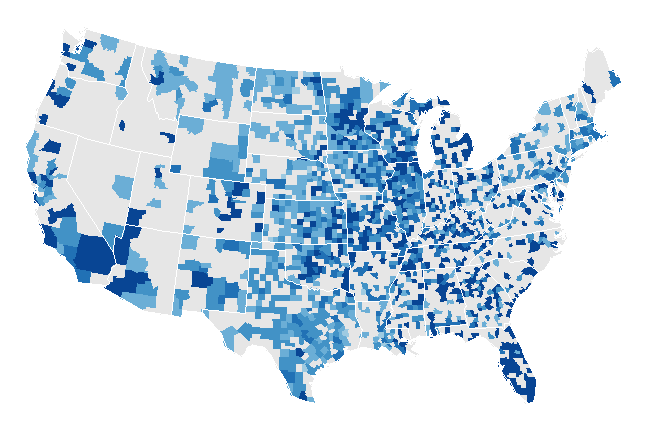} &
                        \includegraphics[width=3in,trim=0in 0in 0in 0in, clip]{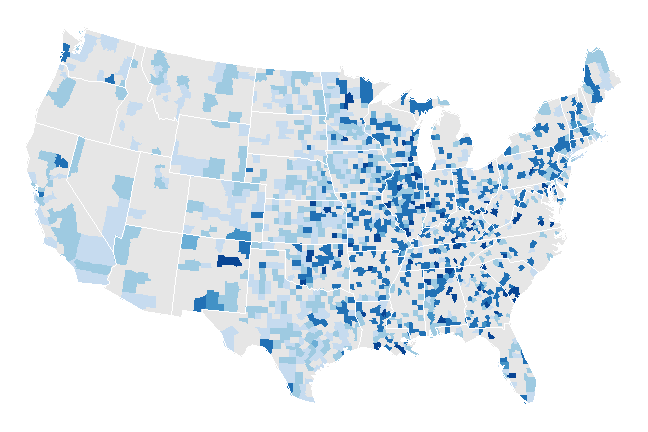} \\
                        \multicolumn{2}{c}{\includegraphics[width=6.4in,trim=0in 0in 0in 0in, clip]{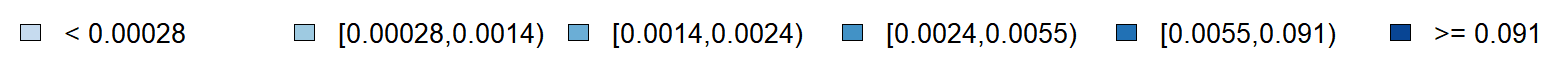}}
                \end{tabular}
        \end{center}
        {\footnotesize {\em Notes:} Predictive tail probabilities are defined as $\mathbb{P}\{y_{iT+1} \ge c |Y_{1:N,0:T},X_{1:N,-1:T}\}$, where $c=1$\%. Flexible CRE specification with heteroskedasticity. The estimation samples range from 2007Q2 ($t=0$) to 2009Q4 ($t=T=10$) and 2015Q2 ($t=0$) to 2017Q4 ($t=T=10$).  }\setlength{\baselineskip}{4mm}                                     
\end{figure}

\noindent {\bf Tail Probabilities for Selected Samples.} From the density forecasts we can compute probability forecasts for particular events. We consider the tail event $\mathbb{I}\{ y_{iT+1} \ge c\}$ for $c=1$\% for now. Figure~\ref{fig:tailprob.spatial} visualizes the probabilities of the tail event for RRE charge-off rates for 2010Q1 and 2018Q1, emphasizing the spatial dimension.\footnote{Similar maps for CC charge-off rates are available in the Online Appendix.} We associate each bank $i$ with a particular county. If there are multiple banks in one county, we average the predicted probabilities. 2010Q1 is the immediate aftermath of the Great Recession and the counties that are covered by our sample appear predominantly in dark blue, indicating that predicted probabilities of the event exceed 9.1\%. Banks in California, Florida, 
and the Midwest from Minnesota, Wisconsin, and Michigan down to Arkansas, Mississippi, and Alabama are predicted to write off a considerable fraction of their RRE loans. Eight years later, the situation has improved considerably, as the map now appears in light blue instead of dark blue, in particular in hard hit states such as California and Florida.

While this paper focuses on forecasting problems, the predictive densities derived from our empirical model can be embedded into more complex decision problems that more closely capture the objectives of policy makers or regulators. In this case, the predictive density is used to compute posterior expected losses associated with policy decisions. The accuracy of the loss calculation is tied to the empirical adequacy of the predictive density, which is what we are evaluating in this section.

\subsection{Parameter Estimates for Selected Samples}
\label{subsec:empirics.estimates}

\begin{figure}[t!]
        \caption{Heterogeneous Coefficient Estimates, RRE Charge-Off Rates}
        \label{fig:lambda.sigma}
        \begin{center}
                \begin{tabular}{ccc}
                        $\mathbb{E}[\lambda_i/(1-\rho)|\cdot]$ & $\mathbb{E}\big[\ln (\sigma_i/\sqrt{1-\rho^2})|\cdot\big]$ & Scatter \\[1ex] 
                        \includegraphics[width=2in]{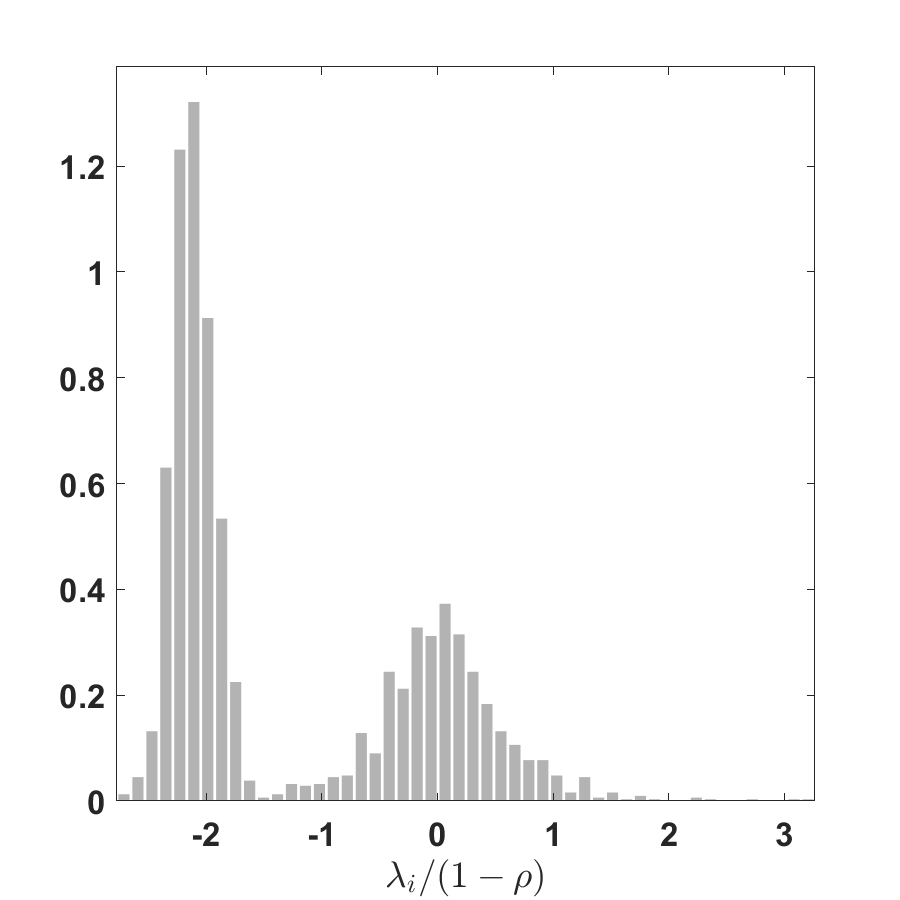}      & \includegraphics[width=2in]{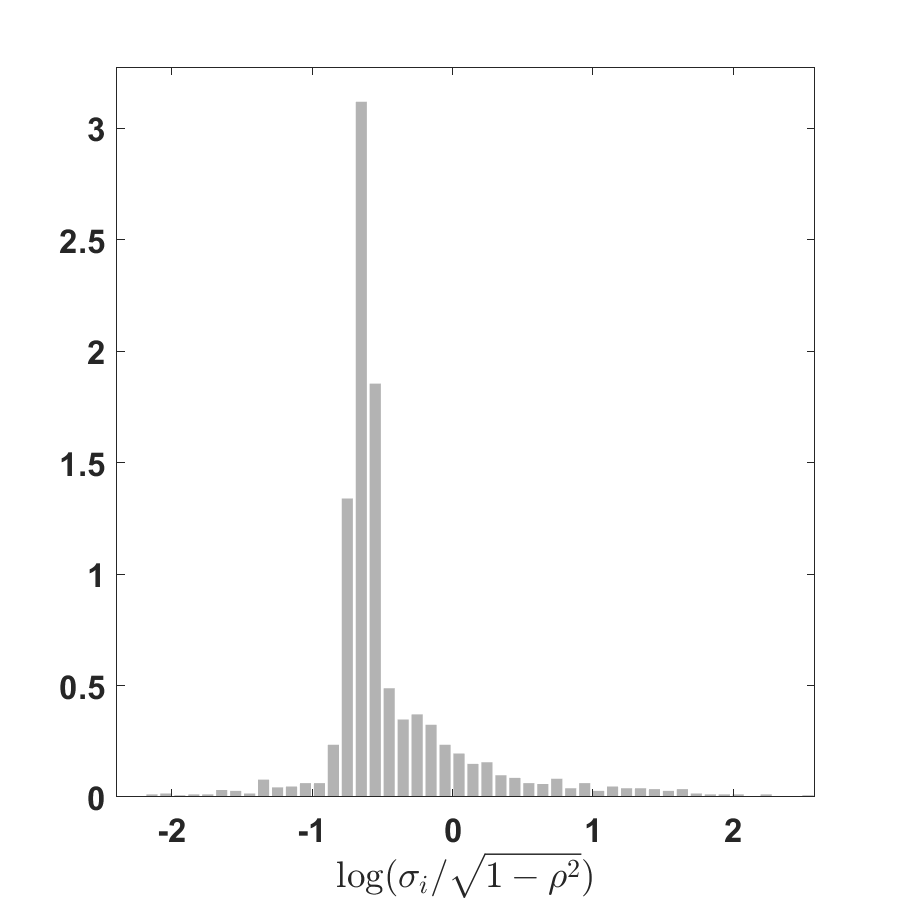} & 
                        \includegraphics[width=2in]{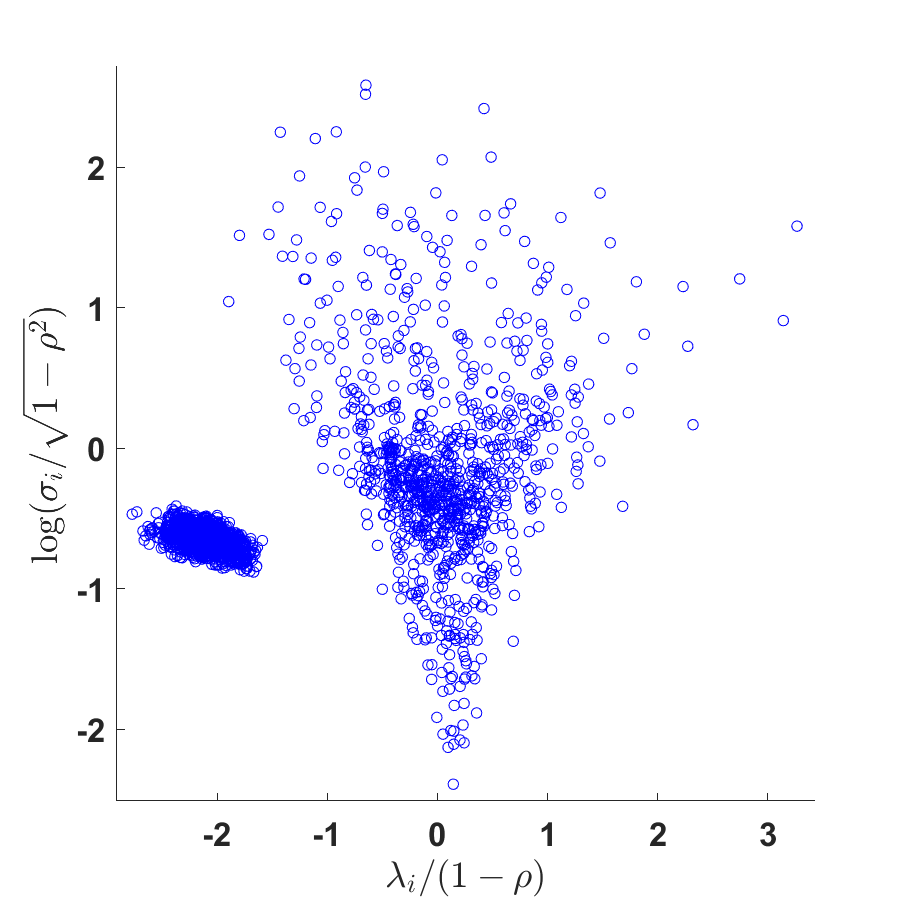}                                   
                \end{tabular}
        \end{center}
        {\footnotesize {\em Notes:} Heteroskedastic flexible CRE specification. The estimation sample ranges from 2007Q2 ($t=0$) to 2009Q4 ($t=T=10$). A few extreme observations are not visible in the plots. The conditioning set is $(Y_{1:N,0:T},X_{1:N,-1:T})$.}\setlength{\baselineskip}{4mm}                                     
\end{figure}

\noindent {\bf Heterogeneous Parameters.} The distributions of posterior mean estimates of the heterogeneous coefficients for the 2007Q2 sample of RRE charge-off rates are depicted in Figure~\ref{fig:lambda.sigma}.\footnote{Similar plots for CC charge-off rates are available in the Online Appendix.} We use the AR coefficient $\rho$ to rescale $\lambda_i$ and $\sigma_i$. 
The panels on the left and in the center of the figure show histograms for the posterior means of $\lambda_i$ and $\sigma_i$, respectively, whereas the right panel contains a scatter plot that illustrates the correlation between the posterior means of intercepts and shock standard deviations. 

A notable feature of the histogram for the posterior means of $\lambda_i/(1-\rho)$ is the spike in the left tail of the distribution. Such spikes were also present in the Monte Carlo simulation; see Figure~\ref{f_MC1results_phatlambda}. The spike corresponds to banks with predominantly zero charge-off rates. For these banks, the sample contains very little information about $\lambda_i$ other than that it has to be sufficiently small to explain the zero charge-off rates. In turn, the posterior mean estimate is predominantly driven by the prior. Similar spikes are visible in the histogram for the posterior means of the re-scaled log standard deviations and the right panel shows that the $\sigma_i$ spike and the $\lambda_i$ spike are associated with the same banks. Small estimates of $\sigma_i$ are associated with near zero estimates of $\lambda_i$, whereas large estimates of $\sigma_i$ are associated with a broad range of $\lambda_i$ estimates. The large dispersion of $\sigma_i$ estimates is consistent with the substantially better density forecast performance of the heteroskedastic models. 

\begin{figure}[t!]
        \caption{$\widehat{\lambda_i/\sigma_i}$ and $\mathbb{P}\big\{i \in \mbox{High} \big\}$ versus Log Assets (Bank Size)}
        \label{fig:lambdasigma.vs.logassets}
        \begin{center}
                \begin{tabular}{cccc}
                        &RRE & CC &\\
                        \rotatebox{90}{\hspace*{5cm} $\widehat{\lambda_i/\sigma_i}$} &
                        \includegraphics[width=2.7in,trim=0.43in 0in 0in 0in,clip]{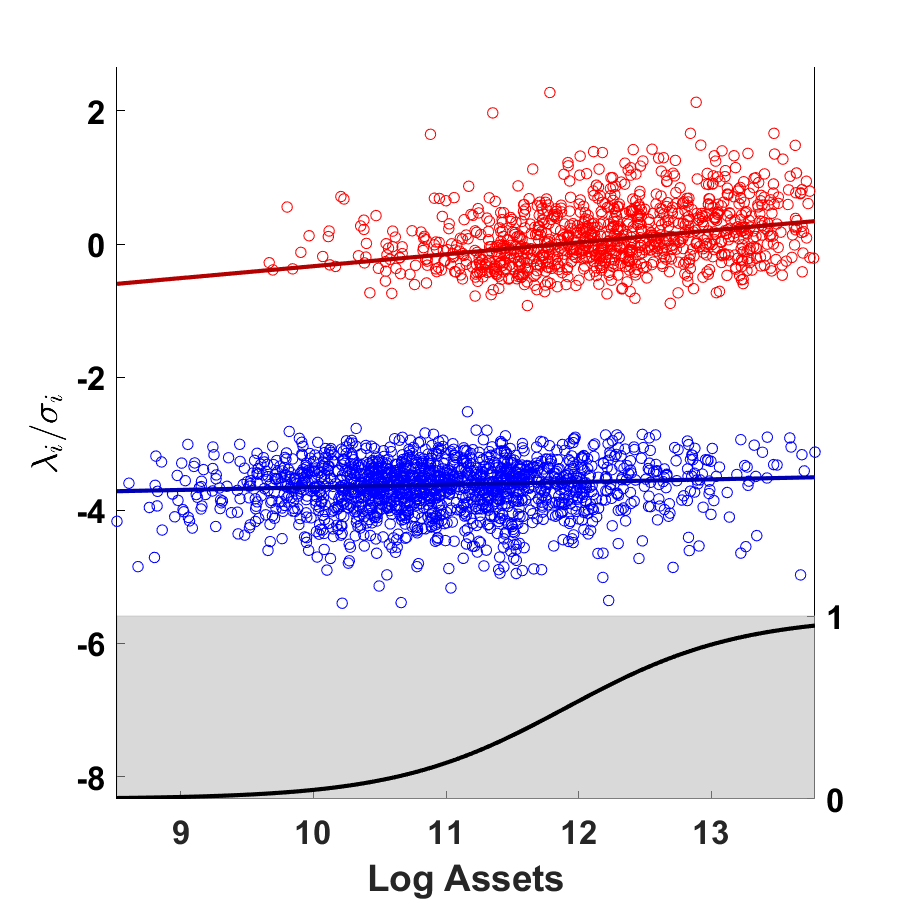} &
                        \includegraphics[width=2.7in,trim=0.43in 0in 0in 0in,clip]{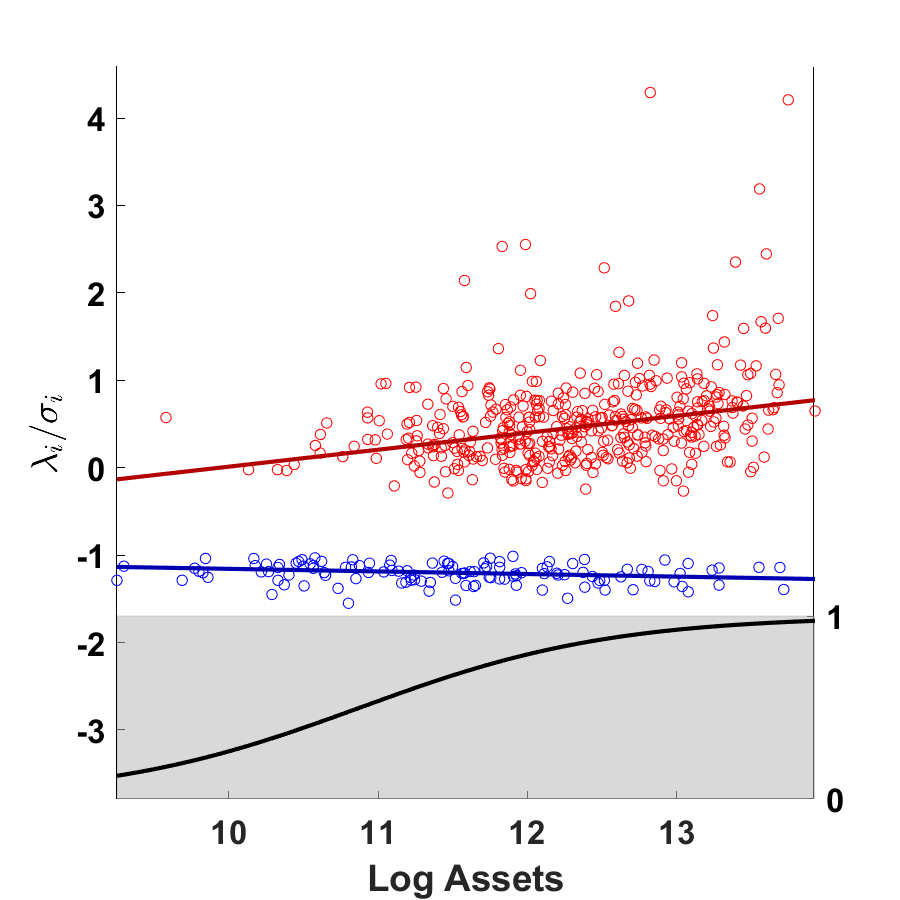} 
                        \rotatebox{270}{\hspace*{-3.0cm} $\mathbb{P}\big\{i \in \mbox{High} \big\}$}
                \end{tabular}
        \end{center}
        {\footnotesize {\em Notes:} Heteroskedastic flexible CRE specification. The estimation sample ranges from 2007Q2 ($t=0$) to 2009Q4 ($t=T=10$). Bank assets are measured at \(t=0\). We form a low-$\lambda$ (blue) and high-$\lambda$ (red) group (left scale). The lines in the top segment of the plot are LAD regression lines. The black lines in the grey shaded areas are predicted probabilities from the logit models (right scale). }\setlength{\baselineskip}{4mm}                                     
\end{figure}



Because regional economic conditions have already been controlled for by $\beta'x_{it-1}$, the estimates of $\lambda_i$ are more likely to be related to bank characteristics. Popular explanations for the heterogeneity in loan losses across banks, here captured by the heterogeneity of $\hat{\lambda}_i$, are attitude toward risk, i.e., some banks might have a greater propensity to take risk or have better opportunities to diversify returns on their loan portfolio, and quality of credit management; see \cite{KeetonMorris1987} for an early contribution and \cite{Ghosh2015,Ghosh2017} more recently.
        
In Figure~\ref{fig:lambdasigma.vs.logassets} we illustrate the relationship between the posterior mean estimate of $\widehat{\lambda_i/\sigma_i}$, which for $\rho=0$ and $\beta=0$ determines the probability of non-zero charge-offs, and bank size measured by the log of total assets. The top segments of the two panels contain scatter plots with group-wise least-absolute-deviations (LAD) regression lines (left scale). As we have seen previously in Figure~\ref{fig:lambda.sigma} there are two groups of $\hat{\lambda}_i$ estimates. For simplicity, we refer to these groups as low-$\lambda$ and high-$\lambda$ groups respectively. For both the RRE and CC samples the positive relationship between bank size and riskiness of the loan portfolio $\widehat{\lambda_i/\sigma_i}$ is more pronounced for banks in the high-$\lambda$ group. The slope coefficients are 0.18 and 0.19, respectively. The shaded areas at the bottom of the panels contain fitted probabilities (right scale) from a logit model that uses log assets as right-hand-side variable. The larger the assets, the higher the probability that it belongs to the high-$\lambda$ group. These results suggest that larger banks in our sample tend to hold riskier loan portfolios.

\begin{table}[t!]
        \caption{Regressions of $\widehat{\lambda_i/\sigma_i}$ on Bank Characteristics}
        \label{tab:lambda.regressions}
        \begin{center}
        \scalebox{0.76}{
        \begin{tabular}{lccccccccccccc} \\ \hline \hline
                & \multicolumn{6}{c}{RRE} & & \multicolumn{6}{c}{CC} \\
                & \multicolumn{2}{c}{Low} & \multicolumn{2}{c}{High} & \multicolumn{2}{c}{Logit} & & \multicolumn{2}{c}{Low} & \multicolumn{2}{c}{High} & \multicolumn{2}{c}{Logit} \\ \cline{2-7} \cline{9-14}
            Log Assets & 0.05$^*$ & (0.02) & 0.20$^*$ & (0.03) & 1.55$^*$ & (0.11) && 0.00 & (0.02) & 0.21$^*$ & (0.03) & 1.27$^*$ & (0.32) \\
            Loan Fraction & 0.07 & (0.12) & 0.17 & (0.15) & 5.07$^*$ & (0.53) && 10.4 & (6.63) & 0.61 & (0.50) & 1254$^*$ & (179) \\ 
            Capital-Asset & 0.30 & (0.51) & -1.36 & (0.99) & -12.0$^*$ & (2.83) && 0.53 & (0.41) & -1.42 & (1.02) & -7.75 & (7.53) \\        
            Loan-Asset & 0.09 & (0.14) & 0.56$^*$ & (0.23) & 6.43$^*$ & (0.66) && -0.20 & (0.14) & -0.07 & (0.21) & 6.42$^*$ & (2.23) \\ 
            ALLL-Loan & 5.82 & (3.97) & 12.0$^*$ & (5.10) & 88.1$^*$ & (18.2) && -0.01 & (2.23) & -1.45 & (4.02) & 85.8$^*$ & (38.2) \\
            Diversification & 0.34 & (0.35) & -0.20 & (0.13) & -0.10 & (0.63) && -0.05 & (0.31) & 1.19$^*$ & (0.42) & 0.82 & (4.69) \\
            Ret.\ on Assets & -26.1$^*$ & (7.55) & -1.08 & (10.06) & -122$^*$ & (36.1) && 28.7$^*$ & (7.73) & 11.1 & (12.4) & -225 & (130)\\ 
            OCA & -19.4$^*$ & (9.63) & 8.25 & (10.05) & 47.9 & (31.01) && 6.38 & (7.23) & 16.92 & (11.8) & -125 & (124) \\
            Intercept & -4.16$^*$ & (0.32) & -2.86$^*$ & (0.47) & -23.8$^*$ & (1.60) && -1.27$^*$ & (0.28) & -2.18$^*$ & (0.45) & -20.4$^*$ & (4.79) \\[1ex]
            Pseudo $R^2$ & \multicolumn{2}{c}{0.03} & \multicolumn{2}{c}{0.06} & \multicolumn{2}{c}{0.32} & & \multicolumn{2}{c}{0.18} & \multicolumn{2}{c}{0.11} &  \multicolumn{2}{c}{0.47} \\                                                  
            \hline      
        \end{tabular}
    }
    \end{center}
        {\footnotesize {\em Notes:} Heteroskedastic flexible CRE specification. The estimation sample ranges from 2007Q2 ($t=0$) to 2009Q4 ($t=T=10$). Bank characteristics are measured at \(t=0\). Low (High) refers to small (large) $\widehat{\lambda_i/\sigma_i}$ group of banks (cutoff is approx -2 for RRE and -1 for CC); see red and blue dots in Figure~\ref{fig:lambdasigma.vs.logassets}. For banks in Low and High groups we regress $\widehat{\lambda_i/\sigma_i}$ on the variables listed in the first column using a least absolute deviations estimator. Logit refers to estimates of a logit model for $\mathbb{I}\{ i \in \mbox{High} \}$. Standard errors are in parenthesis. $*$ indicates significance at 5\% level. Pseudo $R^2$ are computed as follows: $\mbox{LAD} = 1 - \sum |\hat{u}_i(\mbox{all})|/\sum |\hat{u}_i(\mbox{intcpt})|$ (Koenker and Machado, 1999), $\mbox{Logit} = 1-\mbox{loglh(all)}/\mbox{loglh(intcpt)}$ (McFadden, 1973). }\setlength{\baselineskip}{4mm}
\end{table}           
                
In Table~\ref{tab:lambda.regressions} we present estimates from LAD regressions of $\widehat{\lambda_i/\sigma_i}$ on multiple bank characteristics (measured in period $t=0$), separately for the low-$\lambda$ and the high-$\lambda$ group of banks.\footnote{Data definitions and summary statistics for the bank characteristics are provided in the Online Appendix.} We also report estimates for a logit model for $\mathbb{I}\{ i \in \mbox{High} \}$. According to the logit estimates bank size (log assets), the ratio of RRE or CC loans to all loans, lending specialization (ratio of total loans to total assets), and lack of credit quality (ratio of ALLL to total loans) increase the probability that a bank belongs to the high-$\lambda$ group. Capitalization (capital-to-asset ratio) and profitability (return on assets) lower the probability that a bank belongs to the high-$\lambda$ group. For the group-specific regressions only a few bank variables appear to be significant. Foremost, it is bank size measured by log assets. For the RRE high-$\lambda$ group it also includes lending specialization, and for the CC high-$\lambda$ group it includes diversification (share of non-interest income to total income). Operational efficiency, measured by the ratio of overhead costs to assets (OCA) is predominantly insignificant.

\cite{Ghosh2017} studies macroeconomic and bank-level determinants of non-performing loans, i.e., loans past due 90 days or more, for the 100 largest commercial banks over the period 1992Q4 to 2016Q1. With the exception of log assets and loan fractions, we followed his study in constructing our bank-level regressors. Although our sample differs from his in several dimensions (selection of banks, measure of loan performance, and time period), we provide a brief comparison of the results for real estate loans as follows.

\cite{Ghosh2017} finds the following significant relationships for real estate loans: log capital-to-assets (positive), log loans-to-assets (negative), log inverse credit quality (positive), log return on assets (negative). In our logit regression the same bank characteristics have significant coefficients, but the signs of the estimates for the capital-to-asset and the loan-to-asset ratio differ. As \cite{Ghosh2017} points out, the effect of bank capitalization on loan quality is theoretically ambiguous. On the one hand, managers in banks with low capital bases have a moral hazard incentive to engage in risky lending practices (negative relationship). On the other hand, managers in highly capitalized banks may feel confident to engage in risky lending (positive relationship). With respect to the loan-to-asset ratio, our positive estimate for RRE contradicts the notion that banks that are specialized in lending do a better job in selecting high-quality loans, and the positive relationship may reflect that these banks could have more liberal lending policies. 

We also report goodness-of-fit ($R^2$) measures in Table~\ref{tab:lambda.regressions}. For the LAD regressions we report \cite{KoenkerMachado1999}'s quantile regression $R^2$. For the logit regressions we compute \cite{McFadden1973}'s pseudo $R^2$. For the RRE loans the variation in loan quality ($\widehat{\lambda_i/\sigma_i}$) explained by bank characteristics is low. The $R^2$s for the group-specific LAD regressions are only 0.03 and 0.06, respectively. For the CC sample, bank characteristics are more successful in explaining variations in loan quality. The $R^2$ values are 0.18 and 0.11, respectively. The logit regressions attain pseudo $R^2$ values of 0.32 and 0.47 which indicate that the bank characteristics considered here are partly successful in determining whether a bank belongs to the low-$\lambda$ or high-$\lambda$ group.

%
%
%
%
%

\noindent {\bf Common Parameters.} Parameter estimates of the common coefficients for the flexible CRE specification with heteroskedasticity are reported in Table~\ref{tab:para_estimates} for the 2007Q2 samples. We report posterior means and 90\% credible intervals. For each sample we consider three specifications: (i) the baseline specification with $\Delta \ln \mbox{HPI}_{it-1}$ and $\Delta \mbox{UR}_{it-1}$; (ii) an extended version that also includes $\Delta \ln \mbox{INC}_{it-1}$; (iii) and a version without regressors. 

\begin{table}[t!]
        \caption{Estimates of Common Parameters}
        \label{tab:para_estimates}
        \begin{center}
                \scalebox{0.86}{
                        \begin{tabular}{lccccccccc} \\  \hline \hline
                                & \multicolumn{2}{c}{$y_{it-1}^*$} & \multicolumn{2}{c}{$\Delta \ln \mbox{HPI}_{it-1}$} & \multicolumn{2}{c}{$\Delta \mbox{UR}_{it-1}$} & \multicolumn{2}{c}{$\Delta \ln \mbox{INC}_{it-1}$} & LPS \\ 
                                &   Mean & CI &   Mean & CI &   Mean & CI &   Mean & CI \\ \hline
                                RRE  &   0.21 & [  0.18,  0.25]&   -0.03 & [ -0.04, -0.02]&   0.15 & [ 0.13, 0.17]  & & & -0.5232\\
                                &   0.22 & [  0.18,  0.26] &
                                -0.03 & [ -0.04, -0.02]&  0.15 & [ 0.12, 0.17]&   .001 & [ -.005,  .007] & -0.5214 \\            
                                &   0.29 & [  0.27,  0.31]  & & & & & & & -0.5214 \\[2ex]  
                                CC   &   0.41 & [  0.36,  0.46]& -0.09 & [-0.15, -0.04]&   0.46 & [0.30, 0.62]  & & & -1.9214 \\
                                &   0.41 & [  0.36,  0.45]& -0.10 & [-0.16, -0.04]& 0.46 & [0.30, 0.63]&   .010 & [ -.030,  .051] & -1.9216 \\                                       
                                &   0.48 & [  0.43,  0.52] & & & & & & & -1.9268\\ 
                                \hline
                        \end{tabular}
                }
        \end{center}
        {\footnotesize {\em Notes:} Heteroskedastic flexible CRE specification. The estimation sample ranges from 2007Q2 ($t=0$) to 2009Q4 ($t=T=10$). The table contains posterior means and 90\% credible intervals in brackets.}\setlength{\baselineskip}{4mm}
\end{table}  

Both samples exhibit mild autocorrelation. The point estimate of $\rho$ is 0.21 for RRE and 0.41 for CC. To report the estimates of $\beta$ we undo the standardization of the regressors. The numerical values can be interpreted as follows. For the RRE sample, under the extended specification that includes personal income growth a 1\% quarter-on-quarter fall of house prices leads to an increase in charge-off rates by 0.03 percentage points. A 1\% increase in the unemployment rate raises the charge-off rates by 0.15 percentage points. Finally, a 1\% growth of personal income increases the charge-off rates by 0.001 percentage points. For both samples, the coefficients on persistence, house-price inflation, and unemployment rate changes are ``significant," whereas the coefficient on the income growth regressor is ``insignificant'' in that it is small and its sign is ambiguous. Adding income growth hardly alters the coefficient estimates for house-price inflation and unemployment rate changes. The estimates for the CC sample are qualitatively similar to RRE but about three times larger in magnitude. 

In the last column of Table~\ref{tab:para_estimates} we report the LPS, now up to four decimal places, that were previously used for the comparison of density forecasts in Table~\ref{tab:dfcst}. The values for the three configurations of $x_{it}$ are very close. For the CC sample the LPS criterion favors our baseline specification with $x_{it} = [\Delta \ln \mbox{HPI}_{it}, \Delta \mbox{UR}_{it}]'$, whereas for the RRE sample strictly speaking the model without regressors is preferred. In the Online Appendix we show scatter plots of $\hat{\lambda}_i + \beta'x_{it-1}$ versus $\hat{\lambda}_i$ which indicate that only a very small fraction is explained by local economic conditions. Despite the quantitatively small effect of local economic conditions on charge offs we proceed with $x_{it} = [\Delta \ln \mbox{HPI}_{it}, \Delta \mbox{UR}_{it}]'$, whose coefficients are ``significant," and examine the effects of changes in house prices and unemployment more carefully.

Because the Tobit model is nonlinear, the average effect of a change in the regressors (``treatment effect'') depends on $\lambda_i$. 
We consider a change of the regressor from its sample value $x_{iT}$ to $\tilde{x}_{iT} = x_{iT} + \iota' \Delta x$, where the unit-length vector $\iota$ determines the direction of the perturbation of $x_{iT}$ and $\Delta x>0$ the magnitude. Accounting for censoring, we decompose the treatment effect on $y_{iT+1}$ as follows:
\begin{eqnarray}
        \frac{\tilde{y}_{iT+1} - y_{iT+1}}{\Delta x} 
        &=& \beta'\iota \mathbb{I}\{\lambda_i + \rho y_{iT}^* + \beta'x_{iT} + u_{iT+1} > 0  \}  \\
        &&+ \frac{\lambda_i + \rho y_{iT}^* + \beta'\tilde{x}_{iT} + u_{iT+1}}{\Delta x} 
         \bigg(\mathbb{I}\{\lambda_i + \rho y_{iT}^* + \beta'\tilde{x}_{iT} + u_{iT+1} > 0  \} \nonumber \\
        && \hspace*{1cm} - \mathbb{I}\{\lambda_i + \rho y_{iT}^* + \beta'x_{iT} + u_{iT+1} > 0  \} \bigg) \nonumber \\
        &=& I_i + II_i. \nonumber 
\end{eqnarray}
Term $I_i$ captures the intensive margin, i.e., a bank that has non-zero charge-offs conditional on $x_{iT}$ and $\tilde{x}_{iT}$. In this region the Tobit model is linear and the effect is $\beta'\iota$. The second term, $II_i$, captures the extensive margin of banks switching between zero and positive charge-offs. 

\begin{figure}[t!]
        \caption{Effects (Terms I and II) of HPI and UR on CC Charge-Off Rates}
        \label{fig:xit.marginal.effects.CC}
        \begin{center}
                \begin{tabular}{cc}
                        HPI Fall & Unemployment Increase \\
                        \includegraphics[width=3in]{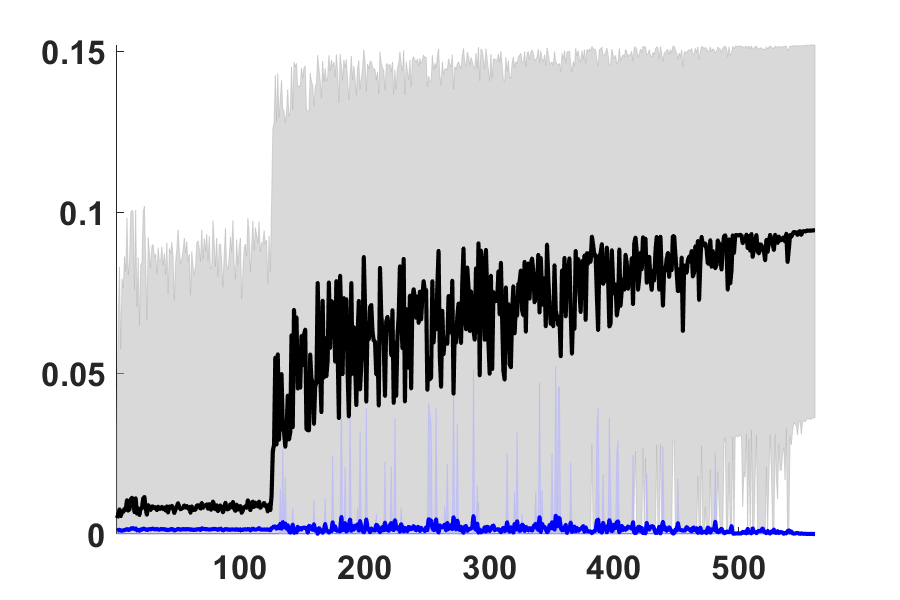} & 
                        \includegraphics[width=3in]{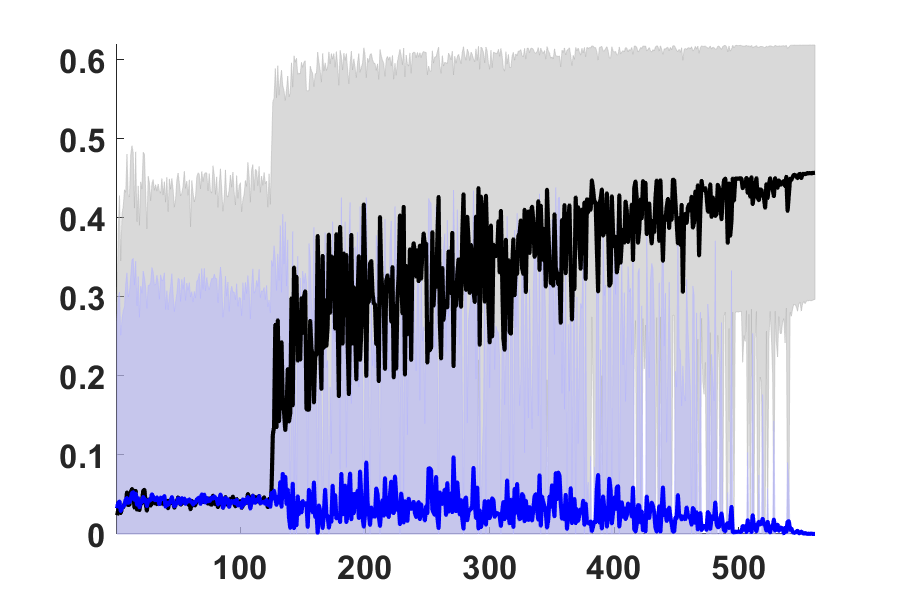} 
                \end{tabular}
        \end{center}
        {\footnotesize {\em Notes:} Heteroskedastic flexible CRE specification. The estimation sample ranges from 2007Q2 ($t=0$) to 2009Q4 ($t=T=10$). The banks $i=1,\ldots,N$ along the $x$-axis are sorted based on the posterior means  $\widehat{\lambda_i/\sigma_i}$. Terms $I_i$ are shown in black/grey and terms $II_i$ in dark/light blue. The units on the $y$-axis are in percent. The solid lines indicate the posterior means of the treatment effect components and the shaded areas delimit 90\% credible bands. }\setlength{\baselineskip}{4mm}                                     
\end{figure}


Figure~\ref{fig:xit.marginal.effects.CC} depicts the posterior mean and the 90\% credible band of the two components of the treatment effect for the banks in the 2007Q2 CC sample.\footnote{A similar figure for the RRE sample is available in the Online Appendix.} We sort the banks based on the posterior means  $\widehat{\lambda_i/\sigma_i}$, which for $\rho=0$ and $\beta=0$ would determine the probability of a positive charge-off. We consider two choices for $\iota'\Delta x$: a 5\% drop in house prices (left panels) and a 5\% rise in the unemployment rate within one quarter (right panels). These are severe shocks to the local economies. 
For the first approximately 120 banks the posterior mean of $I_i$ (black/grey) is close to zero. These are the banks with low values of $\hat{\lambda}_i$ that appear as a mass in the left tail of the density plot in the left panel of Figure~\ref{fig:lambda.sigma}. Under the baseline conditions $x_{iT}$ they are unlikely to have non-zero charge-offs.
For the remaining banks the posterior mean of the term $I$ treatment effect rises under the HPI fall scenario from 0.03\% to 0.1\%, where the latter value is the coefficient estimate reported in Table~\ref{tab:para_estimates}. The credible intervals are fairly wide, ranging from 0\% to 0.15\%. 

The posterior mean for component $II$ (dark/light blue) of the treatment effect is qualitatively similar under the two economic scenarios. For the first 120 banks term $II$ is small because much of $\beta'(\tilde{x}_{iT}-x_{iT})$ has to compensate for the low estimate of $\lambda_i$ before the latent variable $y_{iT+1}^*$ becomes positive. For the remaining banks the term is also small, but for a different reason: with high probability these banks already have positive charge-offs under the baseline economic conditions. Quantitatively, the effects are larger under the very severe unemployment scenario. The switch of low $\lambda_i$ banks from zero to positive charge-offs leads to a posterior mean of the average treatment effect of 0.04\%. As $\widehat{\lambda_i/\sigma_i}$ increases, the expected value of term $II$ decreases because it becomes more likely that the bank has positive charge-offs even under the baseline scenario.

\subsection{Posterior Predictive Checks for Selected Samples}
\label{subsec:predchecks}

\begin{figure}[t!]
        \caption{Posterior Predictive Checks: Cross-sectional Distribution of Sample Statistics}
        \label{fig:predictive.checks}
        \begin{center}
                \begin{tabular}{cccc}
                        & Density $y_{iT+1}|(y_{iT+1} > 0)$   & Distr. of Frequency   & Correlation of $(y_{it},y_{it-1})$ \\ 
                        &                                     & of Zero Charge-Offs   & if Both Are Positive\\[1ex]  
                        \rotatebox{90}{\phantom{abcde} RRE} &   
                        \includegraphics[width=2in]{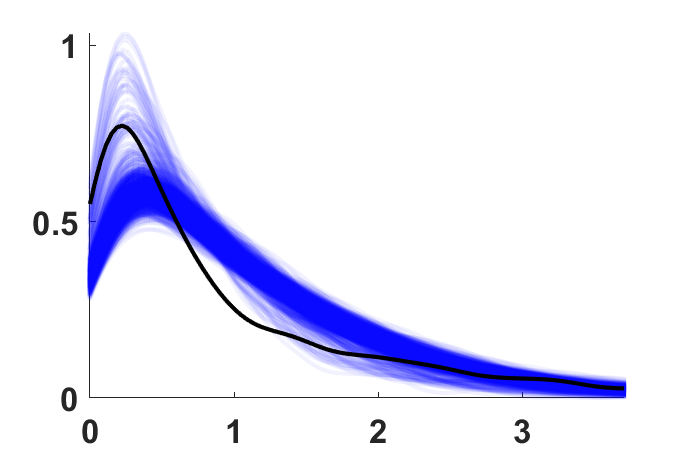}      & \includegraphics[width=2in]{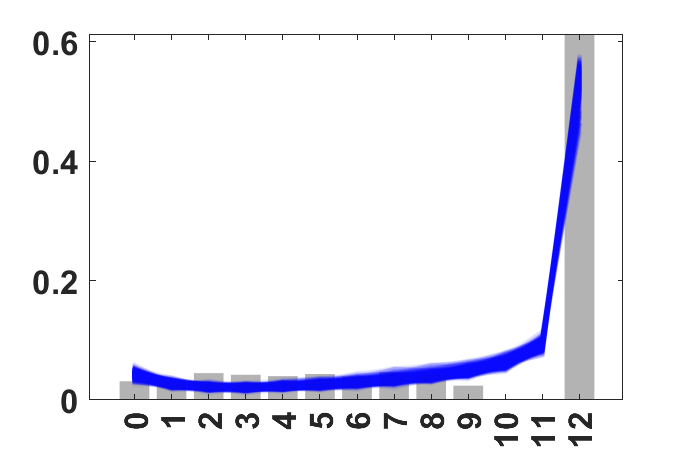} & 
                        \includegraphics[width=2in]{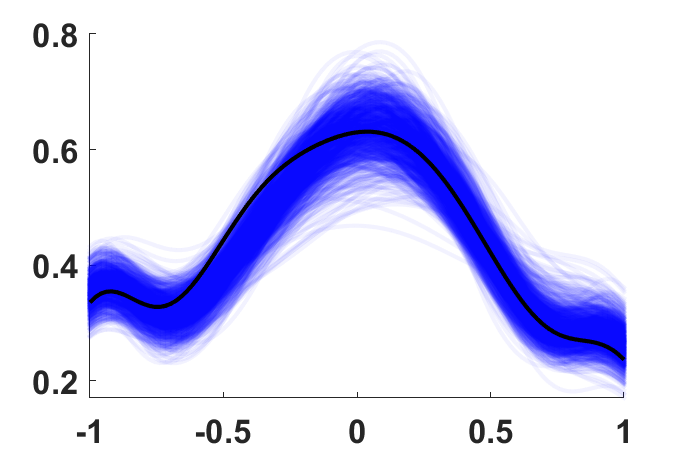} \\[1ex]                                 
                        \rotatebox{90}{\phantom{abcdef} CC} &                                                     
                        \includegraphics[width=2in]{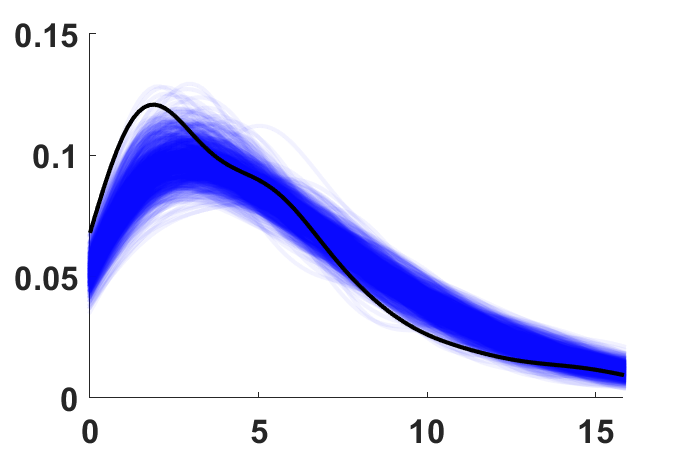}      & \includegraphics[width=2in]{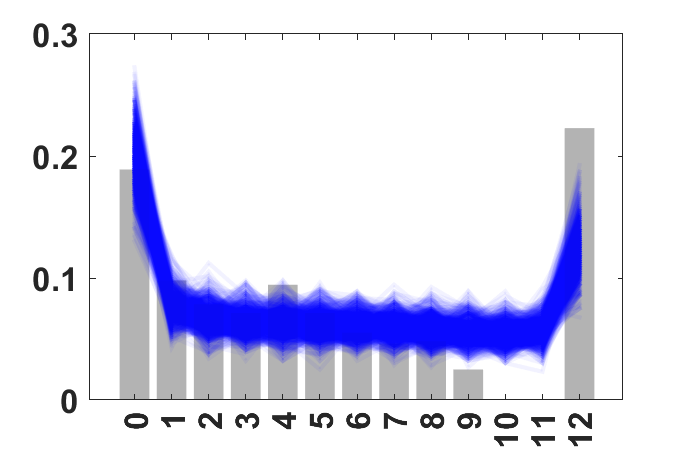} & 
                        \includegraphics[width=2in]{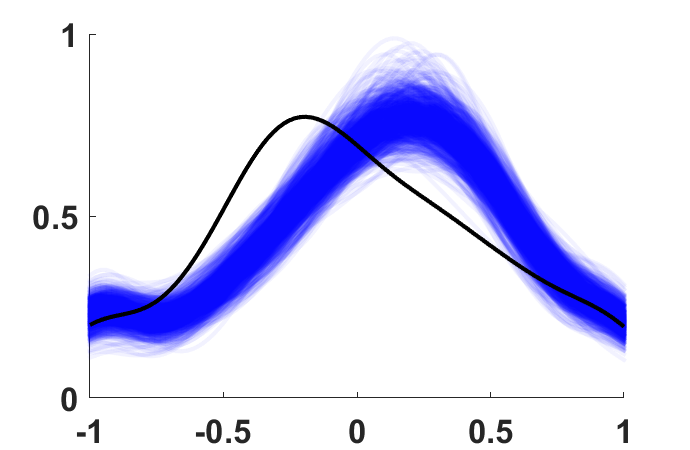} \\
                \end{tabular}
        \end{center}
        {\footnotesize {\em Notes:} Heteroskedastic flexible CRE specification. The estimation sample ranges from 2007Q2 ($t=0$) to 2009Q4 ($t=T=10$). The black lines (left and right panels) and the histogram (center panels) are computed from the actual data. Each hairline corresponds to a simulation of a sample $\tilde{Y}_{1:N,0:T+1}$ of the panel Tobit model based on a parameter draw from the posterior distribution.}  \setlength{\baselineskip}{4mm}                                     
\end{figure}

In order to assess the fit of the estimated panel Tobit model, we report posterior predictive checks in Figure~\ref{fig:predictive.checks}. A posterior predictive check examines the extent to which the estimated model can generate artificial data with sample characteristics that are similar to the characteristics of the actual data that have been used for estimation.\footnote{Textbook treatments of posterior predictive checks can be found, for instance, in \cite{lancaster2004} and \cite{Geweke2005}.} Consider the top left panel of the figure. Here, the particular characteristic, or sample statistic, under consideration is the cross-sectional density of $y_{iT+1}$ conditional on $y_{iT+1}>0$. The black line is computed from the actual RRE loan sample. Each blue hairline is generated as follows: (i) take a draw of $(\rho,\beta,\xi)$ from the posterior distribution; (ii) conditional on these draws generate $\lambda_{1:N}$, $Y_{1:N,0}^*$, and $\sigma^2_{1:N}$; (iii) simulate a panel of observations $\tilde{Y}_{1:N,0:T+1}$; (iv) compute a kernel density estimate based on $\tilde{Y}_{1:N,T+1}$. The swarm of hairlines visualizes the posterior predictive distribution. A model passes a posterior predictive check if the observed value of the sample statistic does not fall too far into the tails of the posterior predictive distribution. Rather than formally computing $p$-values, we focus on a qualitative assessment of the model fit. 

By and large, the estimated models for RRE and CC charge-off rates do a fairly good job in reproducing the cross-sectional densities of $y_{iT+1}$ in that some of the hairlines generated from the posterior cover the observed densities. The only discrepancies arise for charge-off values close to zero. With high probability, the densities computed from simulated data have less mass than the observed RRE and CC densities. Moreover, the modes of the simulated densities are slightly to the right and lower than the modes in the two actual densities. The hairlines depict the densities conditional on $y_{iT+1}>0$. In the observed RRE sample the fraction of $y_{iT+1}=0$ is 0.71. The corresponding 90\% interval obtained from the estimated model is $[0.73, \, 0.79]$. For CC charge-off rates, the fraction in the data is 0.43 and the corresponding 90\%  interval obtained from the estimated model is $[0.37, \, 0.47]$.

The center panels of Figure~\ref{fig:predictive.checks} focus on the estimated models' ability to reproduce the number of zero charge-off observations. For each unit $i$ we compute the number of periods in which $y_{it}=0$. Because $T=10$ the maximum number of zeros between $t=0$ and $t=T+1$ is 12. The histogram is generated from the actual data, whereas the hairlines are computed from the simulated data. For instance, 61\% of the banks do not write off any RRE loans in the twelve quarters of the sample and roughly 5\% of the banks write off RRE loans in every period. Overall, the estimated models do remarkably well in reproducing the patterns in the data. For RRE loans, the model captures the large number of all-zero samples and the fairly uniform distribution of the number of samples with zero to nine instances of $y_{it}=0$. The only deficiency is that the model cannot explain the absence of samples with ten or eleven instances of zero charge-off rates. In the case of CC loans, the estimated model underpredicts the number of all-zero samples but generally is able to match the rest of the distribution.

The last column of Figure~\ref{fig:predictive.checks} provides information about the models' ability to capture some of the dynamics of the charge-off data. Here the test statistic is the first-order sample autocorrelation of the $y_{i,0:T+1}$ sequence, conditional on both $y_{it}$ and $y_{it-1}$ being greater than zero. The panels in the figure depict the cross-sectional density of these sample autocorrelations. For the RRE loans the density computed from the actual data is covered by the hairlines generated from the posterior predictive distribution. For the CC loans the estimated model generates somewhat higher sample autocorrelations than what is present in the data. 

In the Online Appendix (see Figure~\ref{appfig:predictive.checks}) we consider three additional predictive checks based on (i) the time series mean of $y_{it}$ after observing a zero (and, if applicable, before observing the next zero), (ii)  the time series mean of $y_{it}$ before observing a zero (and, if applicable, after observing the previous zero), (iii) a robust estimate of the first-order autocorrelation of $y_{i,0:T+1}$ provided there are sufficiently many non-zero observations. With the exception of the autocorrelations in the CC sample, the two estimated models are able to reproduce the cross-sectional densities of the sample statistics.  

\subsection{Set Forecasts}
\label{subsec:empirics.intervalforecasts}

\begin{figure}[t!]
        \caption{Set Forecasts, Banks Sorted by $\mathbb{E}[y_{iT+1}|Y_{1:N,0:T},X_{1:N,-1:T}]$}
        \label{fig:interval.forecasts.sorted}
        \begin{center}
                \begin{tabular}{cc}
                            RRE & CC \\
                        \includegraphics[width=3in]{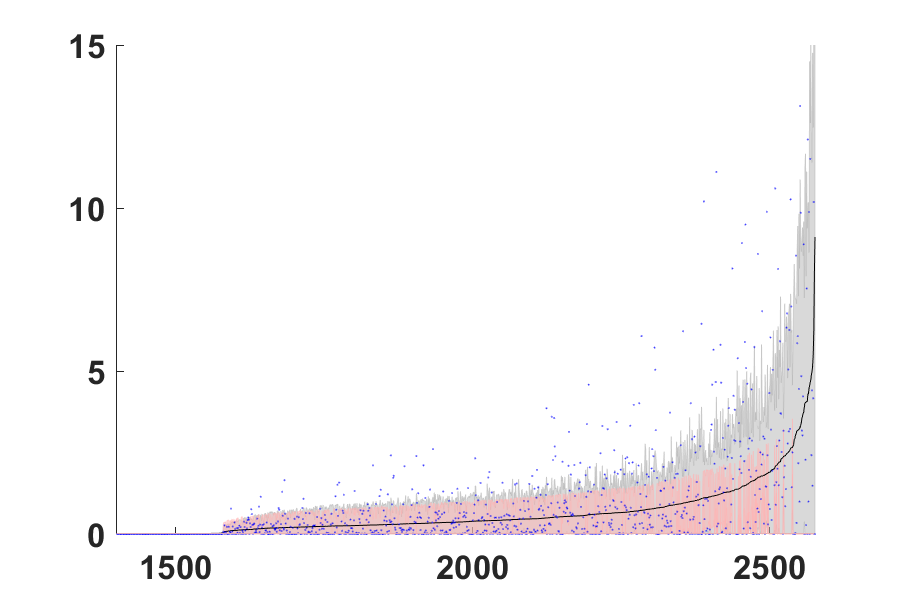} &
                        \includegraphics[width=3in]{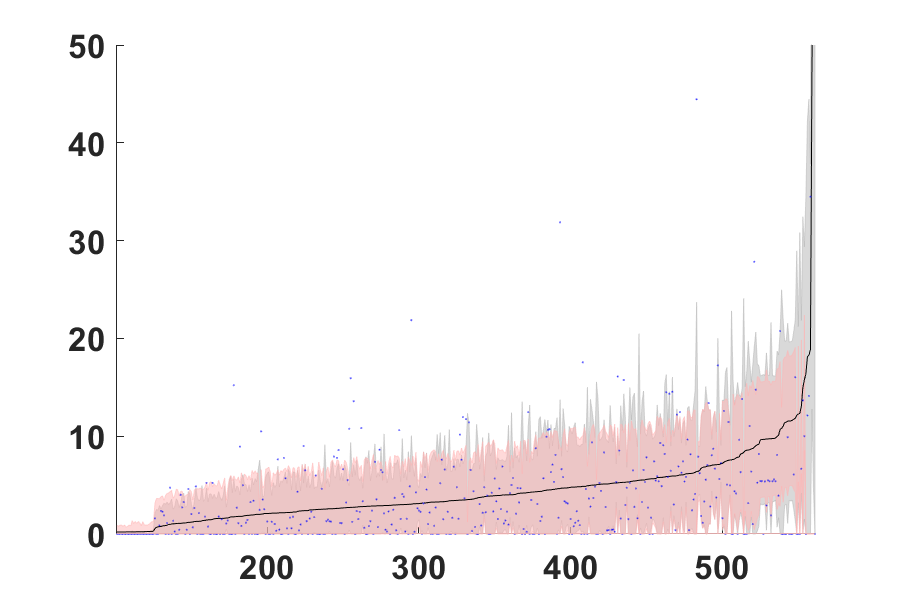} 
                \end{tabular}
        \end{center}
        {\footnotesize {\em Notes:} Flexible CRE specification with heteroskedasticity. The estimation sample ranges from 2007Q2 ($t=0$) to 2009Q4 ($t=T=10$). We forecast 2010Q1 observations. The nominal coverage probability is 90\%. Posterior mean forecasts (solid black line), actuals (blue dots), and set forecasts targeting pointwise (grey) and average (pink) coverage probability. }\setlength{\baselineskip}{4mm}                                     
\end{figure}

\noindent {\bf Selected Samples.} Set forecasts for 2010Q1, constructed as HPD sets from the posterior predictive distribution, are visualized in Figure~\ref{fig:interval.forecasts.sorted}. The nominal credible level is 90\%. We distinguish forecasts targeting pointwise coverage probability (grey) from forecasts targeting average coverage probability (pink). For each bank $i$ we plot the set forecast, the posterior mean forecast and the actual realization of the charge-off rate. The banks are sorted according to $\mathbb{E}[y_{iT+1}|Y_{1:N,0:T},X_{1:N,-1:T}]$. We don't show forecasts for the first 1,400 (100) banks for the RRE (CC) sample because they are essentially zero. 

A comparison of the grey and the pink sets in  Figure~\ref{fig:interval.forecasts.sorted} shows the effect of targeting average versus pointwise coverage. The upper bound as a function of $i$ increases less under targeting average coverage probability, because the criterion allows us to shorten very wide predictive sets and lengthen narrow sets, while reducing the average length. For the RRE sample set forecasts for banks with large expected charge-off rates $\mathbb{E}[y_{iT+1}|Y_{1:N,0:T},X_{1:N,-1:T}]$ become considerably shorter. In fact for $i>2,500$ many of them become $\{0\}$.
Although we plot the actual values of the charge-off rates in Figure~\ref{fig:interval.forecasts.sorted}, it is not possible to glean how close the empirical coverage frequency is to the nominal coverage probability. 
Thus, in Table~\ref{tab:interval.forecasts} we report both the average length of the sets and the empirical coverage frequency. 
For both samples the set forecasts that are constructed by targeting the average coverage probability have a cross-sectional coverage frequency that is close to the nominal coverage probability of 90\% and they tend to be shorter than the ones obtained by targeting pointwise coverage probability.\footnote{We also computed evaluation statistics for the homoskedastic specification. It turns out that the set forecasts generated by the homoskedastic specifications are substantially larger than the sets obtained from the models with heteroskedasticity, without improving the coverage probability. This finding is consistent with the density forecast results in Table~\ref{tab:dfcst}.} 

\begin{table}[t!]
        \caption{Set Forecast Performance}
        \label{tab:interval.forecasts}
        \begin{center}
                \begin{tabular}{llccccc} \\  \hline \hline
                                 &&           &             & \multicolumn{3}{c}{Fraction of Sets of the Form} \\
                             && Coverage & Ave. Len. & $\quad \{0\} \quad$ & $\; \; \; [0,\, b] \; \; \;$ & $\{0\} \cup [a,\,b] $ \\ \hline
RRE &  Target Ave Coverage   &   0.88 &   0.31 &   0.68 &   0.28 &   0.04  \\         
&Target Ptwise Coverage   &   0.94 &   0.75 &   0.61 &   0.36 &   0.03  \\[1ex] 
                 
                 CC &  Target Ave Coverage   &   0.91 &   6.48 &   0.02 &   0.81 &   0.17  \\         
                     &Target Ptwise Coverage &   0.91 &   7.74 &   0.19 &   0.56 &   0.25  \\ \hline
                \end{tabular}
        \end{center}
        {\footnotesize {\em Notes:} Flexible CRE specification with heteroskedasticity. The estimation sample ranges from 2007Q2 ($t=0$) to 2009Q4 ($t=T=10$). We forecast 2010Q1 observations. The nominal coverage probability is 90\%.}\setlength{\baselineskip}{4mm}
\end{table}

We also report the frequency of the three types of set forecasts. Due to the large number of zero observations in the RRE sample, there is a large fraction of banks, between 60\% and 68\%, for which the posterior predictive probability of observing $y_{iT+1}=0$ exceeds 90\%. This leads to a forecast of $\{0\}$. For the CC sample the fraction of $\{0\}$ forecasts is considerably smaller. 

As one switches from targeting pointwise coverage probability to average coverage probability the composition of the set types changes. 
Roughly speaking, the forecaster should widen the ``narrow'' sets (small $\sigma_i$) by lowering their HPD threshold, and tighten the wide sets (large $\sigma_i$) by raising their HPD threshold. For the RRE sample with a relatively high fraction of zeros, when targeting pointwise coverage, the average coverage probability is largely above 90\%, so this mechanism manifests itself as reducing wider pointwise sets to $\{0\}$, which decreases the average coverage probability and average length at the same time. Thus, there is an increase in the fraction of $\{0\}$ forecasts; also see the right tail in the left panel of Figure~\ref{fig:interval.forecasts.sorted}. 

For the CC sample with a relatively low fraction of zeros, when targeting pointwise coverage, the average coverage probability is already close to 90\%. Switching from targeting pointwise to targeting average coverage, the majority of $\{0\}$ forecasts  are converted into $[0,b]$ forecasts by adding a small continuous portion and thereby increasing the pointwise coverage of these units to more than 90\%; see  the left tail in the right panel of Figure~\ref{fig:interval.forecasts.sorted}. Moreover, about one third of the disconnected forecasts are converted into connected forecasts, which is due to a lengthening of the sets for  small $\sigma_i$ units. In the end, the fraction of $[0,b]$ forecasts increases substantially in this case.
        
\begin{figure}[t!]
        \caption{Set Forecasts: Targeting Pointwise vs. Average Coverage -- All Samples}
        \label{fig:ifcst.scatters}
        \begin{center}
                \begin{tabular}{cc}
                        Improvement in Length and Coverage & Reduction in Length Only \\
                \includegraphics[width=3in]{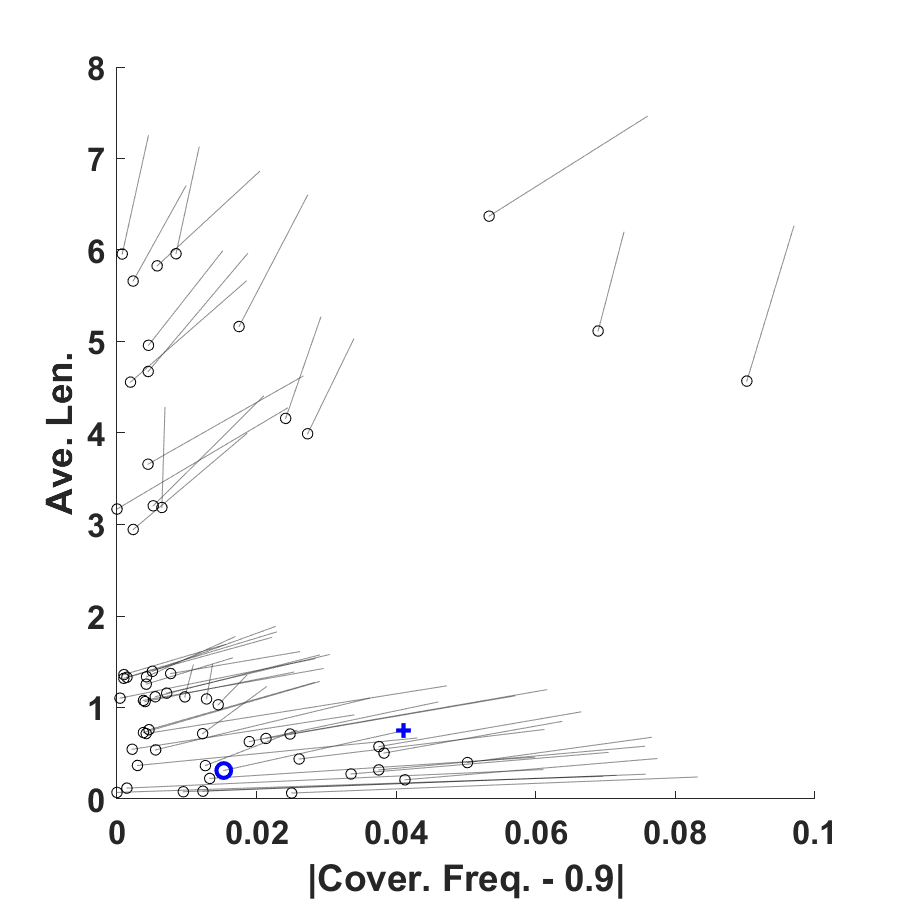} &
                \includegraphics[width=3in]{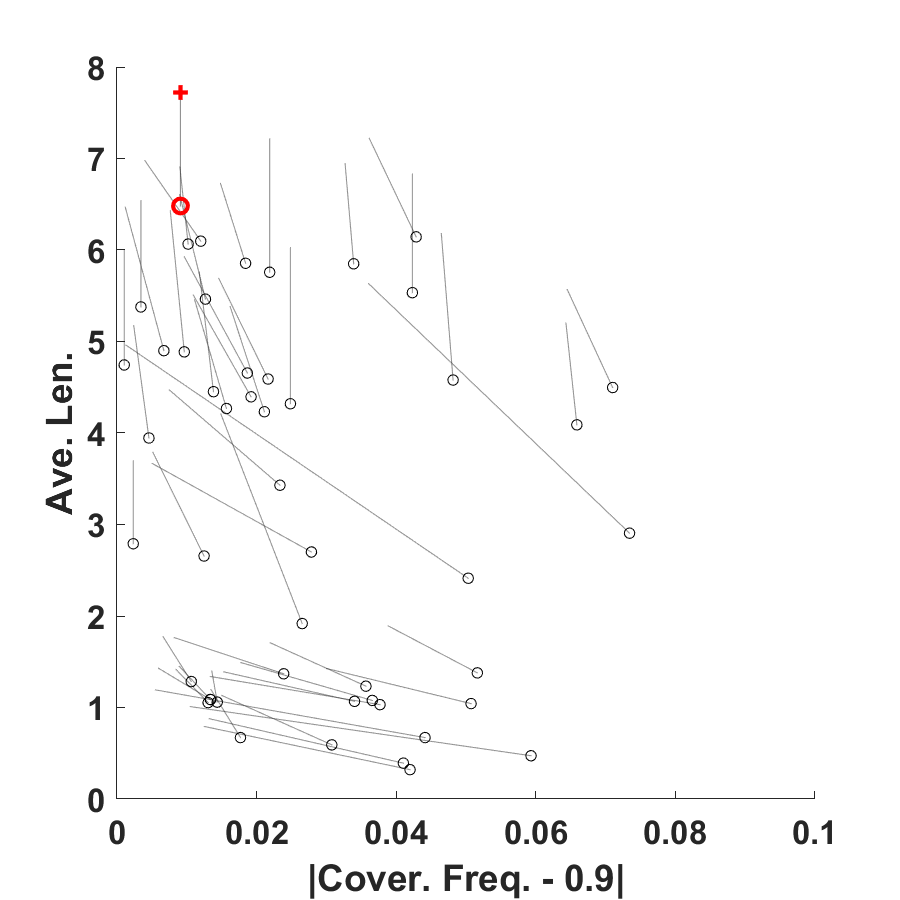} 
                \end{tabular}
        \end{center}
        {\footnotesize {\em Notes:} Flexible CRE specification with heteroskedasticity. The blue (red) symbols correspond to the RRE (CC) baseline sample. The two endpoints of each hairline indicate the coverage probability and length for a particular  estimation sample. Circled endpoints correspond to targeting average coverage probability, and unmarked endpoints (or crosses for the baseline samples) represent pointwise coverage targeting. Hairlines in the left panel represent samples for which the coverage frequency gets closer to the nominal coverage probability of 90\% and the length becomes shorter. The remaining samples are represented by the hairlines in the right panel. }\setlength{\baselineskip}{4mm}                       
\end{figure}

\noindent {\bf All Samples.} In Figure~\ref{fig:ifcst.scatters} we provide information about the coverage frequency and average length size of the set forecast for all samples. We focus on a comparison between targeting pointwise versus average coverage probability in the flexible CRE specification with heteroskedasticity. Each hairline corresponds to one of the 111 different samples and the two endpoints of the hairlines indicate average length and deviation of the empirical coverage frequency from the 90\% nominal credible level. The circled endpoints correspond to targeting average coverage, and unmarked endpoints (or crosses for the baseline samples) represent pointwise coverage targeting. The left panel comprises samples for which targeting average coverage brings the empirical coverage frequency closer to 90\% {\em and} reduces the average length. Here the hairlines point into the lower left corner of the graph. The remaining samples are represented by the hairlines in the right panel. 
Targeting the average coverage unambiguously reduces the average length. For 52\% of the samples it also improves the empirical coverage frequency (left panel). For the  
 remaining 48\% of the samples the deterioration of the coverage frequency is relatively small. The median improvement in coverage probability in the left panel is 0.022, whereas the median deterioration in the right panel is only 0.007. We conclude that, by and large, directly targeting the average posterior coverage probability improves the empirical coverage frequency in the cross section and produces shorter set forecasts.

\section{Conclusion}
\label{sec:conclusion}

The limited dependent variable panel with unobserved individual effects is a common data structure but not extensively studied in the forecasting literature. This paper constructs forecasts based on a flexible dynamic panel Tobit model to forecast individual future outcomes based on a panel of censored data with large $N$ and small $T$ dimensions. Our empirical application to loan charge-off rates of small banks shows that the estimation of heterogeneous intercepts and conditional variances improves density and set forecasting performance in the more than 100 samples considered. Posterior predictive checks conducted for two particular samples indicate that the Tobit model is able to capture salient features of the charge-off panel data sets. Our framework can be extended {\color{black}to allow for stronger forms of simultaneity between the dependent variable and regressors} and to account for dynamic panel versions of more general multivariate censored regression models. 
We can also allow for missing observations in our panel data set. Finally, even though we focused on the analysis of charge-off data, there are many other potential applications for our methods.


\setstretch{1}
\bibliography{ref}

\clearpage
\setstretch{1.3}
\appendix
\setcounter{saveeqn}{\value{section}}\renewcommand{\theequation}{\mbox{%
                \Alph{saveeqn}.\arabic{equation}}} \setcounter{saveeqn}{1} %
\setcounter{equation}{0}
\renewcommand*\thepage{A-\arabic{page}}
\setcounter{page}{1}
\renewcommand*\thetable{A-\arabic{table}}
\setcounter{table}{0}
\renewcommand*\thefigure{A-\arabic{figure}}
\setcounter{figure}{0}

\bc

{\Large {\bf Supplemental Online Appendix to \\ ``Forecasting with a Panel Tobit Model'' }}

{\bf Laura Liu, Hyungsik Roger Moon, and Frank Schorfheide}

\ec

\noindent This Online Appendix consists of the following sections:

\begin{itemize}
            \item[A]  Proof of Theorem~\ref{prop:post.emp.cov} and a Simple Example
        \item[B]  Computational Details 
        \item[C]  Supplemental Information on the Monte Carlo
        \item[D]  Data Set
        \item[E]  Additional Empirical Results
\end{itemize} 


\section{Theorem~\ref{prop:post.emp.cov}}
\label{appsec:theory}

\subsection{Proof of Theorem~\ref{prop:post.emp.cov}.} 

Let $\vartheta = (\theta,\xi)$. The Bayes
model specifies a joint distribution for the observations $(Y_{1:N,0:T},Y_{1:N,T+h})$ and the parameters $(\vartheta,\lambda_{1:N},\sigma^2_{1:N})$. This joint distribution can be factored into conditional distributions as follows
\begin{eqnarray}
\lefteqn{p(Y_{1:N,0:T},Y_{1:N,T+1},\vartheta,\lambda_{1:N})} \label{appeq:factorize.joint}\\
&=&  p(Y_{1:N,0:T})p(\vartheta|Y_{1:N,0:T}) \left( \prod_{i=1}^N p(\lambda_{i},\sigma^2_{i}|\vartheta,Y_{i,0:T})  p(y_{iT+h}|\lambda_{i},\sigma^2_{i},\vartheta,Y_{i,0:T}) \right). \nonumber
\end{eqnarray}
Sampling in a Bayesian framework involves drawing parameters from the appropriate distribution and generating data conditional on these parameters. {\color{black}According to Assumption (i), the future observations are sampled from the predictive density. This sampling can be implemented as follows: let $\tilde{\vartheta}_N$ be a draw from the posterior $p(\vartheta|Y_{1:N,0:T})$ and sample the future observations from $p(Y_{1:N,T+h}|Y_{1:N,0:T}, \tilde{\vartheta}_N)$.}

We start with the bound
\begin{eqnarray}
 \lefteqn{\left| \frac{1}{N} \sum_{i=1}^N \mathbb{I} \big\{ y_{iT+h} \in C_{iT+h|T}(Y_{1:N,0:T}) \big\} - (1-\alpha) \right| } \\
        & \le & \left| \frac{1}{N} \sum_{i=1}^N \left(\mathbb{I} \big\{ y_{iT+h} \in C_{iT+h|T}(Y_{1:N,0:T}) \big\} - \mathbb{P}_{Y_{1:N,0:T},\tilde \vartheta_N}^{y_{iT+h}} \{ y_{iT+h} \in C_{iT+h|T}(Y_{1:N,0:T}) \} \right)\right| \nonumber \\
        && + \left| \frac{1}{N} \sum_{i=1}^N \mathbb{P}_{Y_{1:N,0:T},\tilde \vartheta_N}^{y_{iT+h}} \{ y_{iT+h} \in C_{iT+h|T}(Y_{1:N,0:T}) \} - (1-\alpha)  \right| \nonumber \\
        &=& B_1\big(Y_{1:N,0:T},Y_{1:N,T+h}, \tilde{\vartheta}_N \big) + B_2\big(Y_{1:N,0:T},Y_{1:N,T+h}, \tilde{\vartheta}_N \big). \nonumber 
\end{eqnarray}  
The desired result follows if we can show that for any $\epsilon > 0$
\be
  \lim_{N \longrightarrow \infty} \; \mathbb{P}^{Y_{1:N,0:T},Y_{1:N,T+h}, \tilde{\vartheta}_N} \big\{  B_j\big(Y_{1:N,0:T},Y_{1:N,T+h}, \tilde{\vartheta}_N \big) > \epsilon \big\}=0, \quad j=1,2.
\ee
        
\noindent {\bf Analysis of Term $B_1(\cdot)$.}  Note that $0\le B_1(\cdot)<1$. We write
\begin{eqnarray*}
  \lefteqn{\lim_{N \longrightarrow \infty} \; \mathbb{P}^{Y_{1:N,0:T},Y_{1:N,T+h}, \tilde{\vartheta}_N} \big\{  B_1\big(Y_{1:N,0:T},Y_{1:N,T+h}, \tilde{\vartheta}_N \big) > \epsilon \big\} } \\
  &=& \lim_{N \longrightarrow \infty} \;  \int 
        \mathbb{P}_{Y_{1:N,0:T}, \tilde{\vartheta}_N}^{Y_{1:N,T+h}} \big\{  B_1\big(Y_{1:N,0:T},Y_{1:N,T+h}, \tilde{\vartheta}_N \big) > \epsilon \big\}
        p(Y_{1:N,0:T}, \tilde{\vartheta}_N)d(Y_{1:N,0:T}, \tilde{\vartheta}_N) \\
  &=&  \int 
  \bigg[ \lim_{N \longrightarrow \infty} \; \mathbb{P}_{Y_{1:N,0:T}, \tilde{\vartheta}_N}^{Y_{1:N,T+h}} \big\{  B_1\big(Y_{1:N,0:T},Y_{1:N,T+h}, \tilde{\vartheta}_N \big) > \epsilon \big\} \bigg]
  p(Y_{1:N,0:T}, \tilde{\vartheta}_N)d(Y_{1:N,0:T}, \tilde{\vartheta}_N) \\
  &=&  \int 
  0 \cdot 
  p(Y_{1:N,0:T}, \tilde{\vartheta}_N)d(Y_{1:N,0:T}, \tilde{\vartheta}_N) \\
  &=& 0,        
\end{eqnarray*}
as required.
The second equality follows from the Dominated Convergence Theorem and the third equality follows from a Weak Law of Large Numbers for independently distributed random variables. Conditional on $(Y_{1:N,0:T},\tilde{\vartheta}_N)$,
$y_{iT+h}$ is sampled independently from $p(y_{iT+h}|\tilde{\vartheta}_N,Y_{1:N,0:T})$; see (\ref{appeq:factorize.joint}).
        
\noindent {\bf Analysis of Term $B_2(\cdot)$.}   To capture the probability mass at zero, define $a_{i0,N}=-\infty$ and $b_{i0,N} = 0$. 
Let $\tilde{\vartheta}_N$ be a draw from the posterior $p(\vartheta|Y_{1:N,0:T})$. Recall that by construction of the set forecast
\[
   \frac{1}{N} \sum_{i=1}^N \sum_{k=0}^{K_i}  
   \int_{a_{ik,N}}^{b_{ik,N}} \int p(y_{iT+h}^*|Y_{i,0:T},\vartheta)  p(\vartheta|Y_{1:N,0:T}) d\vartheta d y_{iT+h}^* = 1-\alpha.
\]
Then,
\begin{eqnarray}
\lefteqn{ \left| \frac{1}{N} \sum_{i=1}^N \sum_{k=0}^{K_i} \int_{a_{ik,N}}^{b_{ik,N}} p(y_{iT+h}^*|Y_{i,0:T},\tilde{\vartheta}_N) dy_{iT+h}^* - (1-\alpha) \right| } \label{eq:theory.deltaN} \\
&=& \left| \frac{1}{N} \sum_{i=1}^N \sum_{k=0}^{K_i} \int_{a_{ik,N}}^{b_{ik,N}} p(y_{iT+h}^*|Y_{i,0:T},\tilde{\vartheta}_N)dy_{iT+1}^* - \int \left[ \int_{a_{ik,N}}^{b_{ik,N}} p(y_{iT+h}^*|Y_{i,0:T},\vartheta) d y_{iT+h}^* \right] p(\vartheta|Y_{1:N,0:T}) d\vartheta  \right| \nonumber \\
&=& \left| \frac{1}{N} \sum_{i=1}^N \sum_{k=0}^{K_i} \int \left[\int_{a_{ik,N}}^{b_{ik,N}} p(y_{iT+1}^*|Y_{i,0:T},\tilde{\vartheta}_N)dy_{iT+1}^* -  \int_{a_{ik,N}}^{b_{ik,N}} p(y_{iT+1}^*|Y_{i,0:T},\vartheta) d y_{iT+1}^* \right] p(\vartheta|Y_{1:N,0:T}) d\vartheta  \right|, \nonumber 
\end{eqnarray}
where we exchanged the order of integration in the second term on the right-hand side of the first equality.
Combining the definition of $F_{ik,N}(\vartheta)$ in (\ref{eq:theory.def.FiN}) with (\ref{eq:theory.deltaN}) and noting that $0 \le F_{ik,N}(\vartheta) \le 1$, we obtain
\begin{eqnarray}
\lefteqn{\left| \frac{1}{N} \sum_{i=1}^N \sum_{k=0}^{K_i} \int_{a_{ik,N}}^{b_{ik,N}} p(y_{iT+1}^*|Y_{i,0:T},\tilde{\vartheta}_N) dy_{iT+1}^* - (1-\alpha) \right|} \label{eq:theory.DeltaN.2} \\
&=& \left| \frac{1}{N} \sum_{i=1}^N \sum_{k=0}^{K_i} \int \left[F_{ik,N}(\tilde{\vartheta}_N) -  F_{ik,N}(\vartheta) \right] p(\vartheta|Y_{1:N,0:T}) d\vartheta  \right| \nonumber \\
&\le&  \frac{1}{N} \sum_{i=1}^N \sum_{k=0}^{K_i} \int \left| F_{ik,N}(\tilde{\vartheta}_N) -  F_{ik,N}(\vartheta) \right| p(\vartheta|Y_{1:N,0:T}) d\vartheta \nonumber \\
&\le& \frac{1}{N} \sum_{i=1}^N \sum_{k=0}^{K_i} \int_{{\cal N}_N(\bar{\vartheta}_N)} \left| F_{ik,N}(\tilde{\vartheta}_N) -  F_{ik,N}(\vartheta) \right| p(\vartheta|Y_{1:N,0:T}) d\vartheta + \int_{{\cal N}^c_N(\bar{\vartheta}_N)} p(\vartheta|Y_{1:N,0:T}) d\vartheta \nonumber \\
&=& I + II, \nonumber
\end{eqnarray}
say. The last inequality uses the bound $\left| F_{ik,N}(\tilde{\vartheta}_N) -  F_{ik,N}(\vartheta) \right| \le 1$ for the second term.

According to Assumption~(ii), we can choose a stochastic sequence of shrinking neighborhoods ${\cal N}_N(\bar{\vartheta}_N)$ such that 
\[
II \stackrel{p}{\longrightarrow} 0
\]
as $N \longrightarrow \infty$.
Now consider term $I$. Write 
\begin{eqnarray*}
        I &=& \frac{1}{N} \sum_{i=1}^N \sum_{k=0}^{K_i} \mathbb{I}\{ \tilde{\vartheta}_N \in {\cal N}_N(\bar{\vartheta}_N) \}\int_{{\cal N}_N(\bar{\vartheta}_N)} \left| F_{ik,N}(\tilde{\vartheta}_N) -  F_{ik,N}(\vartheta) \right| p(\vartheta|Y_{1:N,0:T}) d\vartheta \\
        && + \frac{1}{N} \sum_{i=1}^N \sum_{k=0}^{K_i} \mathbb{I}\{ \tilde{\vartheta}_N \in {\cal N}^c_N(\bar{\vartheta}_N) \} \int_{{\cal N}_N(\bar{\vartheta}_N)} \left| F_{ik,N}(\tilde{\vartheta}_N) -  F_{ik,N}(\vartheta) \right| p(\vartheta|Y_{1:N,0:T}) d\vartheta \\
        &=& Ia + Ib,
\end{eqnarray*}
say. It is straightforward to establish that term $Ib$ converges to zero. Recall that the posterior mode is a function of $Y_{1:N,0:T}$. For any $\epsilon > 0$
\begin{eqnarray*}
\mathbb{P}^{Y_{1:N,0:T},\tilde{\vartheta}_N} \{ Ib > \epsilon \}  
 &\le& \mathbb{P}^{Y_{1:N,0:T},\tilde{\vartheta}_N} \left\{ \mathbb{I}\big\{ \tilde{\vartheta}_N \in {\cal N}^c_N(\bar{\vartheta}_N) \big\} \left( \frac{1}{N} \sum_{i=1}^N K_i \right) > \epsilon \right\} \\
 &=& \mathbb{P}^{Y_{1:N,0:T},\tilde{\vartheta}_N} \big\{  \tilde{\vartheta}_N \in {\cal N}^c_N(\bar{\vartheta}_N) \big\} \\
 &=& \int \mathbb{P}_{Y_{1:N,0:T}}^{\tilde{\vartheta}_N} \big\{  \tilde{\vartheta}_N \in {\cal N}^c_N(\bar{\vartheta}_N) \big\} 
 p(Y_{1:N,0:T}) dY_{1:N,0:T}
 \\
& \longrightarrow & 0.
\end{eqnarray*}
The convergence statement in the last line follows from Assumption~(ii) and the Dominated Convergence Theorem:
\begin{eqnarray*}
  \lefteqn{ \lim_{N \longrightarrow \infty} \; 
  \int \mathbb{P}_{Y_{1:N,0:T}}^{\tilde{\vartheta}_N} \big\{  \tilde{\vartheta}_N \in {\cal N}^c_N(\bar{\vartheta}_N) \big\} 
  p(Y_{1:N,0:T}) dY_{1:N,0:T} } \\
  &=&
  \int \bigg[ \lim_{N \longrightarrow \infty} \;\mathbb{P}_{Y_{1:N,0:T}}^{\tilde{\vartheta}_N} \big\{  \tilde{\vartheta}_N \in {\cal N}^c_N(\bar{\vartheta}_N) \big\}  \bigg] p(Y_{1:N,0:T}) dY_{1:N,0:T} \\
  &=&
  \int 0 \cdot  p(Y_{1:N,0:T}) dY_{1:N,0:T}.
\end{eqnarray*}

To bound term $Ia$ we use the Lipschitz condition in Assumption~(iii):
\begin{eqnarray}
Ia & \le &  \frac{1}{N}  \sum_{i=1}^N \sum_{k=1}^{K_i} M_{ik,N} ({\cal N}_N(\bar{\vartheta}_N)) \mathbb{I}\{ \tilde{\vartheta}_N \in {\cal N}_N(\bar{\vartheta}_N) \}\int_{{\cal N}_N(\bar{\vartheta}_N)} \| \tilde{\vartheta}_N -  \vartheta \| p(\vartheta|Y_{1:N,0:T}) d\vartheta \nonumber \\
&\le&  \frac{1}{N}  \sum_{i=1}^N \sum_{k=1}^{K_i} M_{ik,N}({\cal N}_N(\bar{\vartheta}_N)) \mathbb{I}\{ \tilde{\vartheta}_N \in {\cal N}_N(\bar{\vartheta}_N) \}  \\
&& \times \int_{{\cal N}_N(\bar{\vartheta}_N)} \big(\| \tilde{\vartheta}_N -  \bar{\vartheta}_N \| + \| \bar{\vartheta}_N -  \vartheta \|\big) p(\vartheta|Y_{1:N,0:T}) d\vartheta \nonumber \\
&\le& \left(\frac{1}{N} \sum_{i=1}^N \sum_{k=1}^{K_i} M_{ik,N} ({\cal N}_N(\bar{\vartheta}_N))\right)  \mathbb{I}\{ \tilde{\vartheta}_N \in {\cal N}_N(\bar{\vartheta}_N) \}\| \tilde{\vartheta}_N -  \bar{\vartheta}_N \| \nonumber \\
&& + \left(\frac{1}{N} \sum_{i=1}^N \sum_{k=1}^{K_i} M_{ik,N}({\cal N}_N(\bar{\vartheta}_N)) \right)  \int_{{\cal N}_N(\bar{\vartheta}_N)}  \| \bar{\vartheta}_N -  \vartheta \| p(\vartheta|Y_{1:N,0:T}) d\vartheta \nonumber \\ 
&\le& \left(\frac{1}{N} \sum_{i=1}^N \sum_{k=1}^{K_i} M_{ik,N} ({\cal N}_N(\bar{\vartheta}_N))\right) 2 \delta_N. \nonumber 
\end{eqnarray}
The last inequality follows from the definition of the neighborhood ${\cal N}_N(\bar{\vartheta}_N)$. Using Assumptions (ii) and  (iv), we can deduce that 
\be
Ia \stackrel{p}{\longrightarrow} 0,
\ee
in $\mathbb{P}^{Y_{1:N,0:T}}$ probability, which completes the proof. $\blacksquare$

\subsection{A Simple Example}

Consider a simple  model without censoring:
\be
y_{it} = \lambda_i + \theta y_{it-1} + u_{it}, \quad y_{i0} \sim N(0,1), \quad \lambda_i \sim N(\xi,1),  \quad u_{it} \sim N(0,1), \quad T=1.
\ee
Define the vector of homogeneous parameters as $\vartheta = [\theta,\xi]'$. We use a 
prior of the form
\[
   p(\vartheta) \sim N(0,I).
\]
In this example the predictive distribution is unimodal, which means that the HPD set constructed from the continuous part of the predictive density is a single interval. In turn,
the summation of predictive interval segments over $k$ is unnecessary. Let $z_{it} = [1,y_{it-1}]'$. The distribution of $y_{i1}|y_{i0},\vartheta$ after integrating out $\lambda_i$ is
\[
y_{i1}|(y_{i0},\vartheta) \sim iid N \big( z_{i1}'\vartheta, 2 \big), \quad i=1,\ldots,N.
\]
Convergence in probability statements in Theorem~\ref{prop:post.emp.cov} refer to the marginal distribution of the data characterized by the density
\begin{eqnarray*}
   p(Y_{1:N,0:1}) 
   &=& (2\pi)^{-N/2-1} \left( \int 
   \exp \left\{ -\frac{1}{2 \cdot 2} \left(\sum_{i=1}^N (y_{i1} - z_{i1}' \vartheta)^2 \right) - \frac{1}{2}\vartheta'\vartheta \right\} d\vartheta \right) \\
   && \times (2\pi)^{-N/2} \exp\left\{ - \frac{1}{2} \sum_{i=1}^N y_{i0}^2 \right\}.
\end{eqnarray*}

\noindent {\bf Assumption (ii)} 
This leads to the likelihood function
\begin{eqnarray*}
        p(Y_{1:N,0:1}|\vartheta)
        &\propto& \exp \left\{ - \frac{1}{2 \cdot 2} \left( \vartheta' \left(\sum_{i=1}^N z_{i1} z_{i1}'\right) \vartheta - 2 \vartheta' \left( \sum_{i=1}^N z_{i1} y_{i1} \right) \right) \right\}.
\end{eqnarray*}
Under the Normal prior for $\vartheta$ we obtain the following posterior mean and (scaled) variance:
\be
\bar{\vartheta}_N =  \left(\frac{1}{2} \sum_{i=1}^N z_{i1}z_{i1}' + I \right)^{-1} \left( \frac{1}{2} \sum_{i=1}^N z_{i1} y_{i1} \right) , \quad
\bar{V}_N = \left( \frac{1}{2N} \sum_{i=1}^N z_{i1} z_{i1}' + \frac{1}{N} I \right)^{-1}.
\ee
The overall posterior distribution is given by
\be
\vartheta | Y_{1:N,0:1} \sim N \big( \bar{\vartheta}_N, \bar{V}_N/N \big).
\ee
We can define the shrinking neighborhood as the set
\be
{\cal N}_N(\bar{\vartheta}_N) = \big\{ \vartheta \; \big| \; (\vartheta - \bar{\vartheta}_N)' \bar{V}_N^{-1} (\vartheta - \bar{\vartheta}_N)' \le 2 N^{-\eta}  \}, \quad 0 < \eta < 1.
\ee
Thus, for $\vartheta \in {\cal N}_N(\bar{\vartheta}_N)$ we have 
\[
\lambda_{min}(\bar{V}_N^{-1}) \| \vartheta - \bar{\vartheta}_N \|^2 \le 2 N^{-\eta}
\]
or
\[
\| \vartheta - \bar{\vartheta}_N \| \le \sqrt{\frac{2}{\lambda_{min}(\bar{V}_N^{-1})}} N^{-\eta/2} \equiv \delta_N.
\]
The argument can be completed by showing that 
\[
    \lambda_{min}(\bar{V}_N^{-1}) \stackrel{p}{\longrightarrow} \epsilon_*, \quad \epsilon_* > 0
\]
under $\mathbb{P}^{Y_{1:N,0}}$.

\noindent {\bf Assumption (iii)} We now construct the Lipschitz constant. 
Consider 
\begin{eqnarray*}
        F_{i,N}(\theta,\xi) &=& \int_{a_{i,N}}^{b_{i,N}} \int_{\lambda_i}  p_N(y_{i2}|\lambda_i + \theta y_{i1},1)
        p(\lambda_i|y_{i,0:1},\theta,\xi)  d\lambda_i d y_{i2} \\
        &=&  \int_{\lambda_i}  \left[\int_{a_{i,N}}^{b_{i,N}} p_N(y_{i2}|\lambda_i+\theta y_{i1},1)dy_{i2} \right]
        p(\lambda_i|y_{i,0:1},\theta,\xi)  d\lambda_i  \\
        &=& \int_{\lambda_i}  \Phi_N\big(g(\lambda_i + \theta y_{i1};u_{i,N})\big)  
        p(\lambda_i|y_{i,0:1},\theta,\xi)   d\lambda_i\\
        &&   - \int_{\lambda_i} \Phi_N\big(g(\lambda_i+\theta y_{i1};l_{i,N})\big)  
        p(\lambda_i|y_{i,0:1},\theta,\xi)  d\lambda_i,  
\end{eqnarray*}
where
\[
g(\lambda_i + \theta y_{i1};\zeta) = \zeta - \lambda_i -\theta y_{i1}, \quad \zeta \in \{a_{i,N},b_{i,N}\}.
\]
To find a Lipschitz constant, we construct a bound for
\[
\left\| \frac{\partial}{\partial (\theta,\xi)} F_{i,N}(\theta,\xi)  \right\|.
\]
Define
\[
F_{i,N,\zeta}(\theta,\xi) = \int_{\lambda_i}  \Phi_N\big(g(\lambda_i + \theta y_{i1};\zeta)\big)  
p(\lambda_i|y_{i,0:1},\theta,\xi) d\lambda_i, \quad  \zeta \in \{a_{i,N},b_{i,N}\}.
\]
Exchanging the order of differentiation and integration, write
\begin{eqnarray*}
        \frac{\partial}{\partial \theta} F_{i,N,\zeta}(\theta,\xi)      
        &=& \int_{\lambda_i} \phi_N\big(g(\lambda_i +\theta y_{i1}; \zeta)\big) \left( \frac{\partial}{\partial \theta} g(\lambda_i+\theta y_{i1}; \zeta)  \right) 
        p(\lambda_i|y_{i,0:1},\theta,\xi)   d\lambda_i \\
        &&+\int_{\lambda_i}  \Phi_N\big(g(\lambda_i+\theta y_{i1}; \zeta)\big)  
        \left( \frac{\partial}{\partial \theta}p(\lambda_i|y_{i,0:1},\theta,\xi) \right) d\lambda_i      \\
        \frac{\partial}{\partial \xi} F_{i,N,\zeta}(\theta,\xi)         
        &=& \int_{\lambda_i} \phi_N\big(g(\lambda_i +\theta y_{i1}; \zeta)\big) \left( \frac{\partial}{\partial \xi} g(\lambda_i+\theta y_{i1}; \zeta)  \right) 
        p(\lambda_i|y_{i,0:1},\theta,\xi)   d\lambda_i \\
        &&+\int_{\lambda_i}  \Phi_N\big(g(\lambda_i+\theta y_{i1}; \zeta)\big)  
        \left( \frac{\partial}{\partial \xi}p(\lambda_i|y_{i,0:1},\theta,\xi) \right) d\lambda_i.
\end{eqnarray*}
Now note that 
\[
0 \le \phi_N(\cdot) \le (2 \pi)^{-1/2}, \quad 0 \le \Phi_N(\cdot) \le 1,
\]
and
\[
\frac{\partial}{\partial \theta} g(\lambda_i+\theta y_{i1};\zeta) = y_{i1}, \quad
\frac{\partial}{\partial \xi} g(\lambda_i+\theta y_{i1}; \zeta) = 0.
\]
Finally,
\[
\int_{\lambda_i} \left( \frac{\partial}{\partial \theta}p(\lambda_i|y_{i,0:1},\theta,\xi) \right) d\lambda_i =
\frac{\partial}{\partial \theta} \int_{\lambda_i} p(\lambda_i|y_{i,0:1},\theta,\xi)  d\lambda_i = 0.
\]
The same result holds for differentiation with respect to $\xi$.
In turn, we obtain
\be
\left| \frac{\partial}{\partial \theta} F_{i,N,\zeta}(\theta,\xi) \right|
\le \left| \frac{y_{i1}}{\sqrt{2 \pi}} \right|, \quad
\left| \frac{\partial}{\partial \xi} F_{i,N,\zeta}(\theta,\xi) \right|
= 0.
\ee
Noting that
\[
F_{i,N}(\theta,\xi) = F_{i,N,u_{i,N}}(\theta,\xi) - F_{i,N,l_{i,N}}(\theta,\xi),
\]
we can now define the Lipschitz constant
\[
M_{i,N} = 2 \left| \frac{y_{i1}}{\sqrt{2 \pi}} \right| = \sqrt{\frac{2}{\pi}} |y_{i1}|,
\]
which does not depend on ${\cal N}_N(\bar{\vartheta}_N)$. Thus, Assumption (iii) is satisfied. 

\noindent {\bf Assumption (iv)}  Notice that in our model $\mathbb{E}[h(y_{i1})] = \mathbb{E}[h(y_{11})]$ for any $i$ because the cross-sectional units are exchangeable. Moreover, $\mathbb{E}[h(y_{i1})|\vartheta] = \mathbb{E}[h(y_{11})|\vartheta]$ for any $i$. 
Choose $M$ such that $M > \mathbb{E}[|y_{11}|]$.
Now consider the bound
\begin{eqnarray*}
        \lefteqn{\mathbb{I} \left\{  \frac{1}{N} \sum_{i=1}^N M_{i,N}  > \sqrt{2/\pi} M \right\} }\\
        &=& \mathbb{I} \left\{  \frac{1}{N} \sum_{i=1}^N \sqrt{2/\pi} |y_{i1}|  > \sqrt{2/\pi} M \right\}  \\  
        &=& \mathbb{I} \left\{  \frac{1}{N} \sum_{i=1}^N  \big( |y_{i1}| - \mathbb{E}\big[|y_{11}| \, \big| \, \vartheta \big] + \mathbb{E}\big[|y_{11}| \, \big| \, \vartheta \big]-\mathbb{E}[|y_{11}|]  \big)  >  M - \mathbb{E}[|y_{11}|] \right\}  
\end{eqnarray*}
Let $\tilde{M} = (M - \mathbb{E}[|y_{11}|])/2$ and write 
\begin{eqnarray*}
        \lefteqn{\mathbb{I} \left\{  \frac{1}{N} \sum_{i=1}^N \sum_{k=0}^{1} M_{i,N}  > \sqrt{8/\pi} M \right\} }\\
        &\le& \mathbb{I} \left\{  \frac{1}{N} \sum_{i=1}^N  \big( |y_{i1}| - \mathbb{E}\big[|y_{11}| \, \big| \, \vartheta \big] \big) > \tilde{M} \right\}
        + \mathbb{I} \left\{   \big( \mathbb{E}\big[|y_{11}| \, \big| \, \vartheta \big]-\mathbb{E}[|y_{11}|]  \big)  >  \tilde{M} \right\}  
\end{eqnarray*}

We now analyze the two indicator functions separately. First,
\begin{eqnarray*}
        \lefteqn{ \lim_{N \longrightarrow \infty} \mathbb{P}^{Y_{1:N,0:1},\vartheta} \left\{  \frac{1}{N} \sum_{i=1}^N  \big( |y_{i1}| - \mathbb{E}\big[|y_{11}| \, \big| \, \vartheta \big] \big) > \tilde{M} \right\} } \\
        &=& \lim_{N \longrightarrow \infty} \mathbb{E}^\vartheta \left[ \mathbb{P}^{Y_{1:N,0:1}}_\vartheta \left\{  \frac{1}{N} \sum_{i=1}^N  \big( |y_{i1}| - \mathbb{E}\big[|y_{11}| \, \big| \, \vartheta \big] \big) > \tilde{M} \right\} \right] \\
        &=& \mathbb{E}^\vartheta \left[ \lim_{N \longrightarrow \infty} \mathbb{P}^{Y_{1:N,0:1}}_\vartheta \left\{  \frac{1}{N} \sum_{i=1}^N  \big( |y_{i1}| - \mathbb{E}\big[|y_{11}| \, \big| \, \vartheta \big] \big) > \tilde{M} \right\} \right] \\
        &=& 0.
\end{eqnarray*}
The exchange of the limit and expectation is justified by the Dominated Convergence Theorem. Conditional on $\vartheta$ the random variables $|y_{i1}|$ are independently and identically distributed and using a weak law of large numbers for $\frac{1}{N} \sum_{i=1}^N |y_{i1}|$ delivers the desired result. 

Second, we need to control 
\[
  \mathbb{P}^\vartheta \left\{   \big( \mathbb{E}\big[|y_{11}| \, \big| \, \vartheta \big]-\mathbb{E}[|y_{11}|]  \big)  >  \tilde{M} \right\}.  
\]
Under our prior distribution, the random variable $\mathbb{E}\big[|y_{11}| \, \big| \, \vartheta \big]$ is stochastically bounded, which means that for any $\epsilon > 0$ we can choose a $\tilde{M}$ such that
\[
    \mathbb{P}^\vartheta \left\{   \big( \mathbb{E}\big[|y_{11}| \, \big| \, \vartheta \big]-\mathbb{E}[|y_{11}|]  \big)  >  \tilde{M} \right\}  < \epsilon.
\]
This delivers the desired result.

\section{Computational Details}
\label{appsec:computations}

\subsection{Gibbs Sampling}
\label{subsec:computations.gibbs}

The Gibbs sampler for the flexible RE/CRE specification with heteroskedasticity is initialized as follows: 
\begin{itemize}
\item $Y_{1:N,0:T}^*$ with $Y_{1:N,0:T}$;
\item $\rho$ with a generalized method of moments (GMM) estimator $\hat{\rho}$, such as the orthogonal differencing in \cite{ArellanoBover1995} (implementation details can be found in the working paper version of \cite{LiuMoonSchorfheide2018b});
\item $\lambda_i$ with $\hat{\lambda}_i=\frac{1}{T} \sum_{t=1}^T(y_{it}^*-\hat\rho y_{it-1}^*)$;
\item $\sigma_i^2$ with the variance of the GMM orthogonal differencing residues for each individual $i$, i.e., let $y_{it}^\perp, t=1,\cdots,T-1,$ denote the data after orthogonal differencing transformation, then $\hat{\sigma}^2_i= \widehat{\mathbb V}_i(y_{it}^\perp-\hat\rho y_{it-1}^\perp)$,  the time-series variances of $y_{it}^\perp-\hat\rho y_{it-1}^\perp$;
\item for $z=\lambda,\sigma$, $\alpha_z$ with its prior mean; $\gamma_{z,i}$ with $k$-means clustering where $k=10$;  $\{\Phi_{k},\Sigma_{k},\pi_{\lambda,k}\}_{k=1}^K$ and $\{\psi_{k},\omega_{k},\pi_{\sigma,k}\}_{k=1}^K$ are drawn from the conditional posteriors described in Section~\ref{subsec:details.posterior}. 
\end{itemize}
The Gibbs samplers for the other dynamic panel Tobit specifications are special cases in which some of the parameter blocks drop out. The Gibbs sampler for the pooled Tobit and linear specifications are initialized via pooled OLS, which ignores the censoring.
We generate a total of $M_0+M = 10,000$ draws using the Gibbs sampler and discard the first $M_0=1,000$ draws.

\subsection{Set Forecasts}
\label{subsec:computations.intervalforecasts}

To simplify the notation, we drop $X_{1:N,-1:T}$ from the conditioning set in the remainder of this section. The HPD sets generated by the algorithms presented in this subsection always include zero and be of the form
\[
C_i =   \{0\} \cup \left( \bigcup_{k=1}^{K_i} [a_{ik}, \, b_{ik}] \right)
\]
with the understanding that (i) $C_i = \{0\}$ if $K_i=0$, (ii) $a_{i1}$ may be equal to zero, and (iii) 
\[
   a_{i1} < b_{i1} < a_{i2} < b_{i2} < \ldots < a_{iK_i} < b_{iK_i}.
\]

Based on posterior draws $(\lambda_i^{(j)},\sigma_i^{2(j)},y_{iT}^{*(j)},\theta^{(j)})$, we can compute 
the conditional mean and variances $\mu_{iT+h|T}^{(j)}$, and $\sigma_{iT+h|T}^{2(j)}$, which are the primitives for the subsequent algorithms.
The conditional predictive distribution of $y_{iT+h}$ is given by a truncated Normal of the form
\begin{eqnarray}
\lefteqn{p(y_{iT+h} |\mu_{iT+h|T}^{(j)},\sigma_{iT+h|T}^{2(j)})}\label{eq:pred.distr.y}\\
&=& \Phi_N \big(- \mu_{iT+h|T}^{(j)} / \sigma_{iT+h|T}^{(j)} \big)\delta_0(y_{iT+h})+p_N(y_{iT+h} |  \mu_{iT+h|T}^{(j)},\sigma_{iT+h|T}^{2(j)}) \mathbb{I}\{ y_{iT+h} > 0\}, \nonumber
\end{eqnarray}
where $\delta_0(y)$ is the Dirac function that is $0$ for $y \not = 0$, and has the properties that $\delta_0(y) \ge 0$ and $\int \delta_0(y) dy = 1$. Using a sampler for a truncated Normal distribution, it is straightforward to generate draws from the conditional predictive density. 

To construct highest posterior density (HPD) sets, we need to evaluate the posterior predictive density, integrating out  $(\mu_{iT+h|T},\sigma_{iT+h|T}^{2})$ under the posterior distribution. We do so using the Monte Carlo averages
\begin{eqnarray} 
  \pi_{i0} &=&
  \frac 1 {M}\sum_{j=1}^{M}\Phi_N \big(- \mu_{iT+h|T}^{(j)} / \sigma_{iT+h|T}^{(j)} \big)  \label{eq:pi0.mcmc} \\
  \pi_i(y) &=& \frac{1}{M} \sum_{j=1}^{M}p_N\left(y\left|\mu_{iT+h|T}^{(j)},\;\sigma_{iT+h|T}^{2(j)}\right.\right) \label{eq:piy.mcmc}
\end{eqnarray}
such that
\be
\pi_{i0} \delta_0(y)  + 
\pi_i(y) \mathbb{I}\{ y> 0\}  \approx p(y|Y_{1:N,0:T}). 
\label{eq:py.mcmc}
\ee
We also define the weights
\be
W_i^{(j)} = 1 - \Phi_N \big( -\mu_{iT+h|T}^{(j)} / \sigma_{iT+h|T}^{(j)} \big), \label{eq:Wij}
\ee
which have the property that $\frac{1}{M} \sum_{j=1}^M W_i^{(j)} = 1-\pi_{i0}$.

\noindent {\bf Algorithm for $1-\alpha$ Set Forecasts Targeting Pointwise Coverage Probability:}

\medskip

\noindent For $i=1,\ldots,N$:
\begin{enumerate}
        
        \item For $j=1,\ldots,M$: compute $(\mu_{iT+h|T}^{(j)},\sigma_{iT+h|T}^{2(j)})$
        based on a draw $(\lambda_i^{(j)},\sigma_i^{2(j)},y_{iT}^{*(j)},\theta^{(j)})$ from the posterior distribution.
        \item Evaluate the weights $\{W_{i}^{(j)}\}_{j=1}^M$ in (\ref{eq:Wij}) and compute    $\pi_{i0}$ in (\ref{eq:pi0.mcmc}).
        \item If $\pi_{i0}\ge1-\alpha$, then $C_i =   \{0\}$.
        \item If $\pi_{i0}<1-\alpha$, then
        \begin{enumerate}        
                \item Draw  $\{y_{iT+h}^{(j)}\}_{j=1}^{M}$ from the normalized continuous part of the predictive distribution $\pi_i(y)\mathbb{I}\{ y > 0\}/\int \pi_i(y)\mathbb{I}\{ y > 0\}dy$  and form the pairs $\{(y_{iT+h}^{(j)},W_i^{(j)} ) \}_{j=1}^M$. 
                \item Sort $\{(y_{iT+h}^{(j)},W_i^{(j)})\}_{j=1}^{M}$ in  ascending order based on $y_{iT+h}^{(j)}$.
                \item For $j=1,\ldots,{M}$: compute $\pi_{i}^{(j)} = \pi_i(y_{iT+h}^{(j)}) \approx 
                p(y_{iT+h}^{(j)}|Y_{1:N,0:T})$ based on (\ref{eq:piy.mcmc}).
                
                \item Let $\Pi_i = \{ (\pi_i^{(j)},y_{iT+h}^{(j)}, W_i^{(j)})\}_{j=1}^{M}$. Sort the elements in $\Pi_i$ based on $\pi_i^{(j)}$ in descending order. Denote the sorted elements in $\Pi_i$ by $(\pi_i^{(s)},y_{iT+h}^{(s)},W_i^{(s)})$.
                \item Note that by construction $\sum_{s=1}^M W_i^{(s)} = 1-\pi_{i0}$. Let $\bar{\Pi}_i$ be the set of largest density values:
                \[
                \bar{\Pi}_i = \bigg\{ (\pi_i^{(s)},y_{iT+h}^{(s)},W_i^{(s)}) \; \big| \; s=1,\ldots,\bar{s}, \; \sum_{s=1}^{\bar{s}} W_i^{(s)} \approx (1-\alpha - \pi_{i0})M \bigg\} .
                \]
                \item Recall that the $(j)$ superscript refers to draws sorted according to $y_{iT+h}^{(j)}$. For $j=1,\ldots,{M}$:
                \begin{enumerate}
                        \item If (A) $j=1$ and $(\pi_i^{(j)},y_{iT+h}^{(j)}, W_i^{(j)})\in\bar{\Pi}_i$, OR (B) $j>1$, $(\pi_i^{(j-1)},y_{iT+h}^{(j-1)}, W_i^{(j-1)})\notin\bar{\Pi}_i$, and $(\pi_i^{(j)},y_{iT+h}^{(j)}, W_i^{(j)})\in\bar{\Pi}_i$,
then $y_{iT+h}^{(j)}$ is the start of an interval, denoted by $a_{ik}$, where $k$ is an index for the intervals. 
                        \item If (A) $j={M}$ and $(\pi_i^{(j)},y_{iT+h}^{(j)}, W_i^{(j)})\in\bar{\Pi}_i$, OR (B) $j<{M}$, $(\pi_i^{(j)},y_{iT+h}^{(j)}, W_i^{(j)})\in\bar{\Pi}_i$, and $(\pi_i^{(j+1)},y_{iT+h}^{(j+1)}, W_i^{(j+1)})\notin\bar{\Pi}_i$, then $y_{iT+h}^{(j)}$ is the end of an interval, denoted by $b_{ik}$. 
                \end{enumerate} 
                This leads to $K_i$ intervals of the form $[a_{ik},b_{ik}]$, $k=1,\ldots,K_i$. If $a_{i1} = y_{iT+h}^{(1)}$,  then let $a_{i1} = 0$. 
                \item Delete intervals that are singletons and adjust $K_i$ accordingly.
                Note that $K_i$ may be zero for some $i$'s.
                \item In the end, unit $i$'s set forecast takes form\[
                C_{it+h|T} =   \{0\} \cup \left( \bigcup_{k=1}^{K_i} [a_{ik}, \, b_{ik}] \right).\] 
        \end{enumerate}
\end{enumerate}

\noindent {\bf Algorithm for $1-\alpha$ Set Forecasts Targeting Average Coverage Probability:}
\begin{enumerate}
        \item For $i=1,\ldots,N$: 
        \begin{enumerate}
                \item For $j=1,\ldots,M$: compute $(\mu_{iT+h|T}^{(j)},\sigma_{iT+h|T}^{2(j)})$
                based on a draw $(\lambda_i^{(j)},\sigma_i^{2(j)},y_{iT}^{*(j)},\theta^{(j)})$ from the posterior distribution.
                \item Evaluate the weights $\{W_{i}^{(j)}\}_{j=1}^M$ in (\ref{eq:Wij}) and compute    $\pi_{i0}$ in (\ref{eq:pi0.mcmc}).
        \end{enumerate}
        \item Define $\pi_0 = \frac{1}{N} \sum_{i=1}^N \pi_{i0}$ (average probability of zero). Note that $\frac{1}{N M} \sum_{i=1}^N \sum_{j=1}^M W_{i}^{(j)} = 1 -\frac{1}{N}\sum_{i=1}^N \pi_{i0} = 1 -\pi_0$.        
        \item If $\pi_0\ge1-\alpha$ then:
        \begin{enumerate}
                \item Sort the units $i$ in descending order based $\pi_{i0}$. 
                \item Assign the set $\{0\}$ to the units with the largest $\pi_{i0}$ values until the desired coverage is reached. All other units $i$ are assigned $\emptyset$.
        \end{enumerate}
        \item Elseif $\pi_0< 1-\alpha$, then:
        \begin{enumerate}
                \item For $i=1,\ldots,N$:
                \begin{enumerate}
                        \item Draw  $\{y_{iT+h}^{(j)}\}_{j=1}^{M}$ from the normalized continuous part of the predictive distribution  normalized continuous part of the predictive distribution $\pi_i(y)\mathbb{I}\{ y > 0\}/\int \pi_i(y)\mathbb{I}\{ y > 0\}dy$ and form the pairs $\{(y_{iT+h}^{(j)}, W_i^{(j)} ) \}_{j=1}^M$. 
                        \item Sort $\{(y_{iT+h}^{(j)}, W_i^{(j)})\}_{j=1}^{M}$ in  ascending order based on $y_{iT+h}^{(j)}$.
                        \item For $j=1,\ldots,{M}$: compute $\pi_{i}^{(j)} = \pi_i(y_{iT+h}^{(j)}) \approx 
                        p(y_{iT+h}^{(j)}|Y_{1:N,0:T})$ based on (\ref{eq:piy.mcmc}).
                \end{enumerate}
                \item Let $\Pi = \big\{ (\pi_i^{(j)},y^{(j)}_{iT+h}, W_i^{(j)}) \; | \; i=1,\ldots,N \; \mbox{and} \; j=1,\ldots,{M}  \big\}$. Sort the elements in $\Pi$ based on $\pi_i^{(j)}$ in descending order. Denote the sorted elements in $\Pi$ by $(\pi^{(s)},y^{(s)}_{T+h},  W^{(s)})$. We dropped the $i$ subscript from the triplet, because we are pooling across $i$.
                \item Let $\bar{\Pi}$ be the set of largest density values:
                \[
                \bar{\Pi}=\left\{\big(\pi^{(s)},y_{T+h}^{(s)}, W^{(s)}\big) \, \bigg| \, s=1,\cdots,\bar{s}, \, \sum_{s=1}^{\bar{s}}  W^{(s)} \approx (1-\alpha-\pi_0) NM \right\}.
                \] 
                \item For $i=1,\ldots,N$: 
                
                \begin{enumerate}
                        \item For $j=1,\ldots,{M}$:
                        \begin{enumerate}
                                \item If (A) $j=1$ and $(\pi_i^{(j)},y_{iT+h}^{(j)},  W_i^{(j)})\in\bar{\Pi}$, OR (B) $j>1$, $(\pi_i^{(j-1)},y_{iT+h}^{(j-1)},  W_i^{(j-1)})\notin\bar{\Pi}$, and $(\pi_i^{(j)},y_{iT+h}^{(j)},  W_i^{(j)})\in\bar{\Pi}$, then $y_{iT+h}^{(j)}$ is the start of an interval, denoted by $a_{ik}$, where $k$ is an index for the intervals. 
                                \item If (A) $j={M}$ and $(\pi_i^{(j)},y_{iT+h}^{(j)},  W_i^{(j)})\in\bar{\Pi}$ OR (B) $j<{M}$, $(\pi_i^{(j)},y_{iT+h}^{(j)},  W_i^{(j)})\in\bar{\Pi}$, and $(\pi_i^{(j+1)},y_{iT+h}^{(j+1)},  W_i^{(j+1)})\notin\bar{\Pi}$, then $y_{iT+h}^{(j)}$ is the end of an interval, denoted by $b_{ik}$. 
                        \end{enumerate} 
                        This leads to $K_i$ intervals of the form $[a_{ik},b_{ik}]$, $k=1,\ldots,K_i$. If $a_{i1} = y_{iT+h}^{(1)}$, then let $a_{i1} = 0$. 
                        \item Delete intervals that are singletons and adjust $K_i$ accordingly. Note that $K_i$ may be zero for some $i$'s. 
\item In the end, unit $i$'s set forecast takes form \[
                        C_i =   \{0\} \cup \left( \bigcup_{k=1}^{K_i} [a_{ik}, \, b_{ik}] \right).
                        \]
                \end{enumerate}
        \end{enumerate}
\end{enumerate}

\subsection{Density Forecasts}
\label{subsec:computations.densityforecasts}
The log-predictive density can be approximated by 
\be
  \ln p \big(y_{iT+h}| Y_{1:N,0:T} \big)
  \approx \left\{ \begin{array}{ll}
                    \ln \mathbb{P}\big[y_{iT+h}=0| Y_{1:N,0:T} \big], & \mbox{if} \; y_{iT+h} = 0, \\
                    \ln \left( \frac{1}{M} \sum_{j=1}^M p_N\big(y_{iT+h} | \mu_{iT+h|T}^{(j)},\sigma_{iT+h|T}^{2(j)}\big) \right),
         & \mbox{otherwise}. \end{array}
         \right. 
\ee

Define the empirical distribution function based on the draws from the posterior predictive distribution as
\be
   \hat{F}(y_{iT+h}) = \frac{1}{M} \sum_{j=1}^M \mathbb{I} \{ y_{iT+h}^{(j)} \le y_{iT+h} \}.
\ee
Then the probability integral transform associated with the density forecast of $y_{iT+h}$ can be approximated as 
\be
  PIT(y_{iT+h}) \approx \hat{F}(y_{iT+h}).
\ee
The continuous ranked probability score associated with the density can be approximated as
\be
CRPS(\hat{F},y_{iT+h}) = \int_{0}^\infty
\big( \hat{F}(x) - \mathbb{I}\{y_{iT+h} \le x \}  \big)^2 dx.
\ee

Because the density $\hat{F}(y_{iT+h})$ is a step function, we can express the integral as a Riemann sum. To simplify
the notation we drop the $iT+h$ subscripts and add an $o$ superscript for the observed value at which the score is evaluated.
Drawing a figure helps with the subsequent formulas. 
Define
\[
    M_* = \sum_{j=1}^M \mathbb{I}\{ y^{(j)} \le y^o\}. 
\]

\noindent {\em Case 1:} $M_*=M$. Then,
\begin{eqnarray}
CRPS(\hat{F},y^o) = \sum_{j=2}^M \big[ \hat{F}(y^{(j-1)}) - 0 \big]^2(y^{(j)}-y^{(j-1)})
 + \big[1-0 \big]^2 (y^{o}-y^{(M)}). \label{eq:crps-case1}
\end{eqnarray}

\noindent {\em Case 2:} $M_*=0$. Then,
\begin{eqnarray}
CRPS(\hat{F},y^o) = \big[0-1 \big]^2 (y^{(1)}-y^{o}) + \sum_{j=2}^M \big[ \hat{F}(y^{(j-1)}) -1\big]^2(y^{(j)}-y^{(j-1)}).\label{eq:crps-case2}
\end{eqnarray}

\noindent {\em Case 3:} $1 \le M_* \le M-1$. Then,
\begin{eqnarray}
\lefteqn{ CRPS(\hat{F},y^o) }  \label{eq:crps-case3} \\
&=& \sum_{j=2}^{M_*} \big[ \hat{F}(y^{(j-1)}) - 0 \big]^2(y^{(j)}-y^{(j-1)}) + \big[ \hat{F}(y^{(M_*)}) - 0 \big]^2 (y^{o}-y^{(M_*)}) \nonumber \\
                   && \quad + \big[ \hat{F}(y^{(M_*)}) - 1 \big]^2 (y^{(M_*+1)}-y^{o}) 
                   + \sum_{j=M_*+2}^{M} \big[ \hat{F}(y^{(j-1)}) - 1 \big]^2(y^{(j)}-y^{(j-1)}). \nonumber
\end{eqnarray}

Equivalently, based on \cite{GneitingRaftery} Equation (21), we have
\begin{eqnarray}
 CRPS(\hat{F},y^o) &=& \frac{1}{M} \sum_{j=1}^{M} |y^{(j)}-y^{o}| - \frac{1}{M^2} \sum_{1\le i< j \le M} (y^{(j)}-y^{(i)}). \label{eq:crps-GR}
\end{eqnarray}
To see their equivalence, note that (\ref{eq:crps-GR}) can be re-written as follows:
\begin{eqnarray}
\lefteqn{\frac{1}{M}\sum_{j=1}^{M} |y^{(j)}-y^{o}| - \frac{1}{M^2}\sum_{1\le i< j \le M} (y^{(j)}-y^{(i)})} \label{eq:crps-GR2} \\
&=& \frac{1}{M}\left[\sum_{j > M_*} y^{(j)} - \sum_{j \le M_*} y^{(j)} + \big(M_* -  (M-M_*)\big)y^o \right] 
- \frac{1}{M^2} \sum_{j=1}^M (2j-M-1) y^{(j)}.\nonumber \\
&=& \frac{1}{M^2}\left[-\sum_{j=1}^{M_*}(2j-1)y^{(j)}
   +\sum_{j=M_*+1}^{M}(2M-2j+1)y^{(j)}\right] + \frac{2M_*-M}{M}y^o. \nonumber
\end{eqnarray}
Considering that $\hat{F}(y^{(j)})$ is the empirical distribution, we have
\begin{eqnarray*}
 \hat{F}(y^{(j)}) &=& \frac{j}{M}.
\end{eqnarray*}
First, let us look at the more general Case 3. After replacing $\hat{F}(y^{(j)})$, the RHS of  (\ref{eq:crps-case3}) becomes
\begin{eqnarray*}
\lefteqn{\sum_{j=2}^{M_*} \big[ \hat{F}(y^{(j-1)}) - 0 \big]^2(y^{(j)}-y^{(j-1)}) + \big[ \hat{F}(y^{(M_*)}) - 0 \big]^2 (y^{o}-y^{(M_*)})} \nonumber \\
\lefteqn{+ \big[ \hat{F}(y^{(M_*)}) - 1 \big]^2 (y^{(M_*+1)}-y^{o}) 
+ \sum_{j=M_*+2}^{M} \big[ \hat{F}(y^{(j-1)}) - 1 \big]^2(y^{(j)}-y^{(j-1)})} \\      
&=& \sum_{j=2}^{M_*} \frac{(j-1)^2}{M^2} (y^{(j)}-y^{(j-1)}) + \frac{M_*^2}{M^2} (y^{o}-y^{(M_*)}) \\
&&  + \frac{(M-M_*)^2}{M^2} (y^{(M_*+1)}-y^{o}) 
    + \sum_{j=M_*+2}^{M} \frac{(M-(j-1))^2}{M^2}(y^{(j)}-y^{(j-1)})\\
&=&\frac{1}{M^2}\bigg[-y^{(1)}+\sum_{j=2}^{M_*}\big((j-1)^2-j^2 \big)y^{(j)}
   +\sum_{j=M_*+1}^{M-1}\big((M-(j-1))^2-(M-j)^2 \big)y^{(j)}\\
&&   +y^{(M)}+\big(M_*^2 -(M-M_*)^2\big)y^o\bigg]\\
&=& \frac{1}{M^2}\bigg[-\sum_{j=1}^{M_*}(2j-1)y^{(j)}
   +\sum_{j=M_*+1}^{M}(2M-2j+1)y^{(j)}\bigg] + \frac{2M_*-M}{M}y^o, 
\end{eqnarray*}
which is the same as (\ref{eq:crps-GR2}).
Similarly, for Case 1, after substituting $\hat{F}$, the RHS of  (\ref{eq:crps-case1}) becomes
\begin{eqnarray*}
\lefteqn{\sum_{j=2}^M \big[ \hat{F}(y^{(j-1)}) - 0 \big]^2(y^{(j)}-y^{(j-1)})
        + \big[1-0 \big]^2 (y^{o}-y^{(M)})} \\
&=& \sum_{j=2}^{M} \frac{(j-1)^2}{M^2} (y^{(j)}-y^{(j-1)}) + (y^{o}-y^{(M)})\\
&=&\frac{1}{M^2}\big[-y^{(1)}+\sum_{j=2}^{M}\big((j-1)^2-j^2 \big)y^{(j)}\big]+y^o\\
&=& -\frac{1}{M^2}\sum_{j=1}^{M_*}(2j-1)y^{(j)} + y^o, 
\end{eqnarray*}
which is equal to (\ref{eq:crps-GR2}) when $M_*=M$. And for Case 2, after substituting $\hat{F}$, the RHS of  (\ref{eq:crps-case2}) becomes
\begin{eqnarray*}
        \lefteqn{\big[0-1 \big]^2 (y^{(1)}-y^{o}) + \sum_{j=2}^M \big[ \hat{F}(y^{(j-1)}) -1\big]^2(y^{(j)}-y^{(j-1)})}\\
&=&  (y^{(1)}-y^{o}) + \sum_{j=2}^{M} \frac{(M-(j-1))^2}{M^2}(y^{(j)}-y^{(j-1)})\\
&=&\frac{1}{M^2}\big[\sum_{j=1}^{M-1}\big((M-(j-1))^2-(M-j)^2 \big)y^{(j)}+y^{(M)}\big]- y^o\\
&=& \frac{1}{M^2}\sum_{j=1}^{M}(2M-2j+1)y^{(j)} -y^o, 
\end{eqnarray*}
which is equal to (\ref{eq:crps-GR2}) when $M_*=0$.

\clearpage

\section{Supplemental Information on the Monte Carlo}

{\color{black} \noindent {\bf Implementation of Forecasts.} To generate forecasts, we first sample draws from the posterior distribution of the model parameters and the latent variable $y_{iT}^*$, and then, conditional on each of these draws, simulate a trajectory $\{ y_{iT+s}^*, y_{iT+s} \}_{s=1}^h$ from the  predictive distribution. While we ignore the censoring in the estimation of the pooled linear specification, we do account for it when we generate forecasts from the linear model. In a final step, the simulated trajectories are converted into density or set forecasts that reflect parameter uncertainty, potential uncertainty about $y_{iT}^*$, and uncertainty about future shocks.

\noindent {\bf Distribution of $\lambda_i$ versus $\mathbb{E}[\lambda_i|Y_{1:N,0:T}]$.} The following two examples help to interpret the comparison of the $p(\lambda)$s and the histograms of $\mathbb{E}[\lambda_i|Y_{1:N,0:T}]$ in Figure~\ref{f_MC1results_phatlambda} in the main paper. First, suppose that the model is static, linear, and homoskedastic, i.e., $y_{it} = \lambda_i + u_{it}$, $u_{it} \sim N(0,\sigma^2)$ and $\lambda_i \sim N(\phi_\lambda,1)$, and $\phi_\lambda$ is known (which implies $p(\lambda)$ is known).  Therefore, the maximum likelihood estimator (MLE) $\hat{\lambda}_i = \lambda_i + \frac{1}{T} \sum_{t=1}^T u_{it}$ has the cross-sectional distribution $\hat{\lambda}_i \sim N \big( \phi_\lambda, 1 + \sigma^2/T \big)$ and the posterior means have the distribution
\[
\mathbb{E}[\lambda_i|Y_{1:N,1:T}] 
= \frac{T/\sigma^2}{T/\sigma^2 + 1} \hat{\lambda}_i + \frac{1}{T/\sigma^2 + 1} \phi_\lambda
\sim N \left( \phi_\lambda, \frac{1}{1 + \sigma^2/T}   \right).
\]
In this example, the distribution of the posterior mean estimates is less dispersed than the distribution of the $\lambda_i$'s, but centered at the same mean, which is qualitatively consistent with Figure~\ref{f_MC1results_phatlambda}. 

Second, to understand the effect of censoring, suppose that $y^*_{it} = \lambda_i + u_{it}$
and we observe a sequence of zeros. The likelihood associated with this sequence of zeros is given by
$\Phi_N^T(-\lambda_i/\sigma)$. The posterior mean for a sequence of zeros is then given by
\[
\mathbb{E}[\lambda_i|Y_{1:N,1:T} = 0] = \frac{ \int \lambda \Phi_N^T(-\lambda/\sigma) p(\lambda) d \lambda }{\int \Phi_N^T(-\lambda/\sigma) p(\lambda) d\lambda}
\]
and provides a lower bound for the estimator $\hat{\lambda}_i$. If the $\lambda_i$'s are sampled from the prior, we should observe this posterior mean with probability
$\int \Phi_N^T\big(-\lambda/\sigma\big) p(\lambda) d \lambda$. Thus, according to this example, there should be a spike in the left tail of the distributions of $\mathbb{E}[\lambda_i|Y_{1:N,1:T}]$. This spike is clearly visible in the two panels of 
Figure~\ref{f_MC1results_phatlambda}.
}

\noindent {\bf Sensitivity to Fraction of Zeros in Sample.} To examine the sensitivity of the MCMC algorithm to the fraction of zeros in the sample, we changed the design of the Monte Carlo experiment to raise the fraction of zeros. Recall from Table~\ref{t_MC1design} in the main text that 
\[
\mbox{Fraction of zeros = 45\%} \quad : \quad p(\lambda_i|y_{i0}^*) = \frac{1}{9}p_N\left(\lambda_i|2.25,0.5\right)+\frac{8}{9}p_N\left(\lambda_i|0,0.5\right).
\]
To increase the number of zeros to 60\% and 75\%, respectively, we consider
\begin{eqnarray*}
        \mbox{Fraction of zeros = 60\%} &:& p(\lambda_i|y_{i0}^*) = \frac{1}{9}p_N\left(\lambda_i|1.85,0.5\right)+\frac{8}{9}p_N\left(\lambda_i|-0.4, 0.5\right) \\
        \mbox{Fraction of zeros = 75\%} &:& p(\lambda_i|y_{i0}^*) = \frac{1}{9}p_N\left(\lambda_i|1.3,0.5\right)+\frac{8}{9}p_N\left(\lambda_i|-0.95,0.5\right).
\end{eqnarray*}
Under the baseline configuration, the number fraction of trajectories with all zeros was 15\%. Under the alternative scenarios, this fraction increases to 23\% and 34\%, respectively. 

\begin{figure}[t!]
        \caption{Convergence Diagnostics Based on $\rho^{(j)}$ Sequence}
        \label{appf_MC123_Convergence}
        \begin{center}
                \begin{tabular}{cccc}
                        &45\% Zeros & 60\% Zeros & 75\% Zeros \\
                        \rotatebox{90}{\hspace*{0.5cm} Trace Plot $\rho^{(j)}$} &
                        \includegraphics[width=2in]{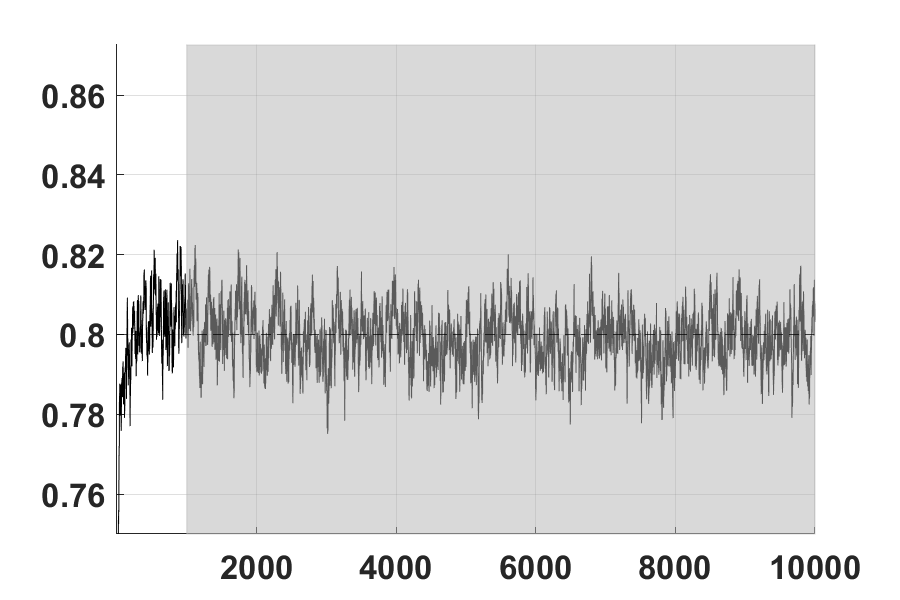}       &
                        \includegraphics[width=2in]{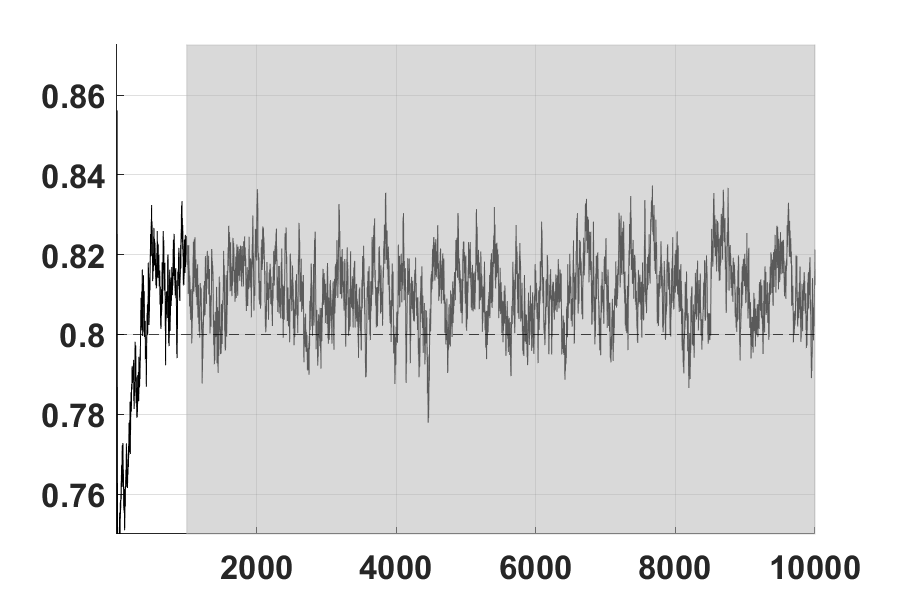}       &
                        \includegraphics[width=2in]{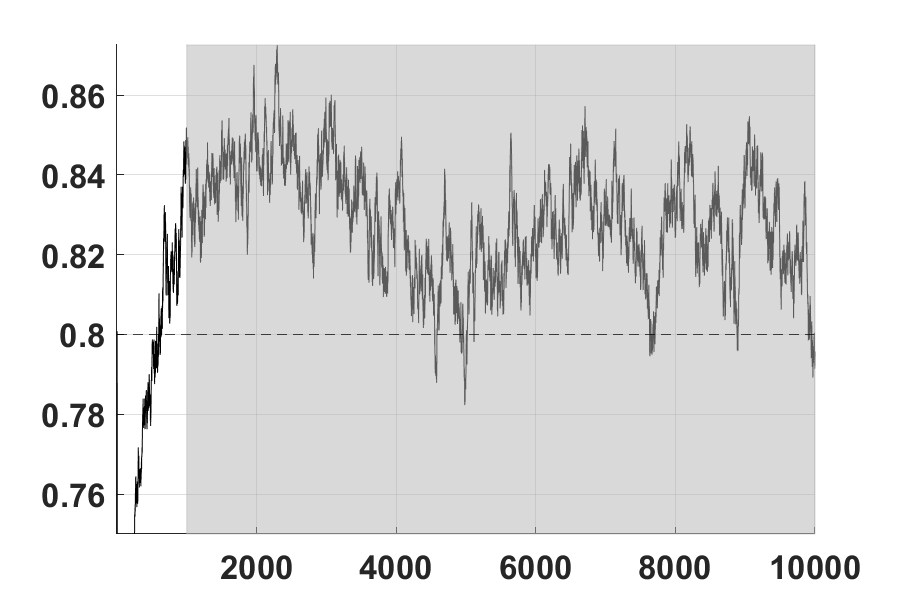}\\
                        \rotatebox{90}{\hspace*{0.5cm} Autocorr. $\rho^{(j)}$} &
                        \includegraphics[width=2in]{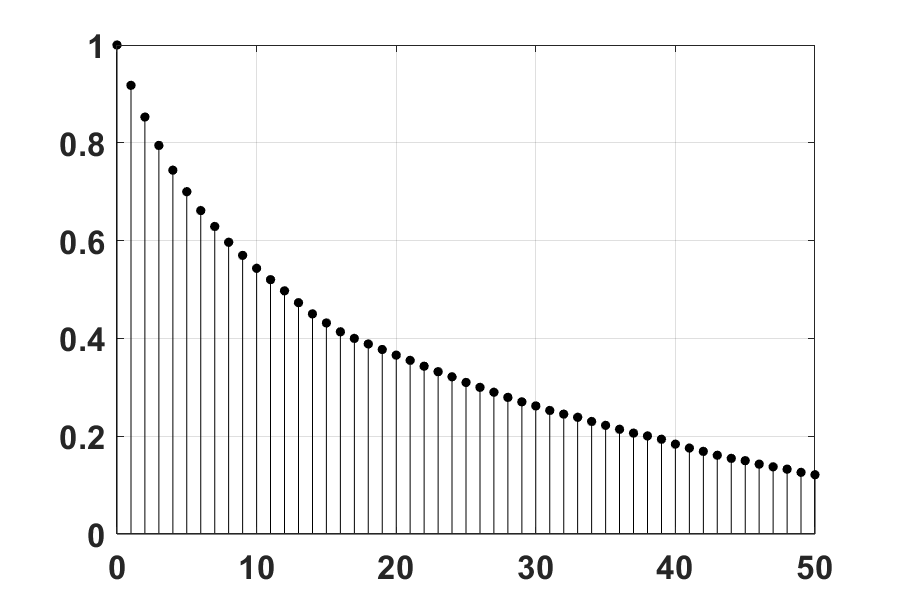}       &
                        \includegraphics[width=2in]{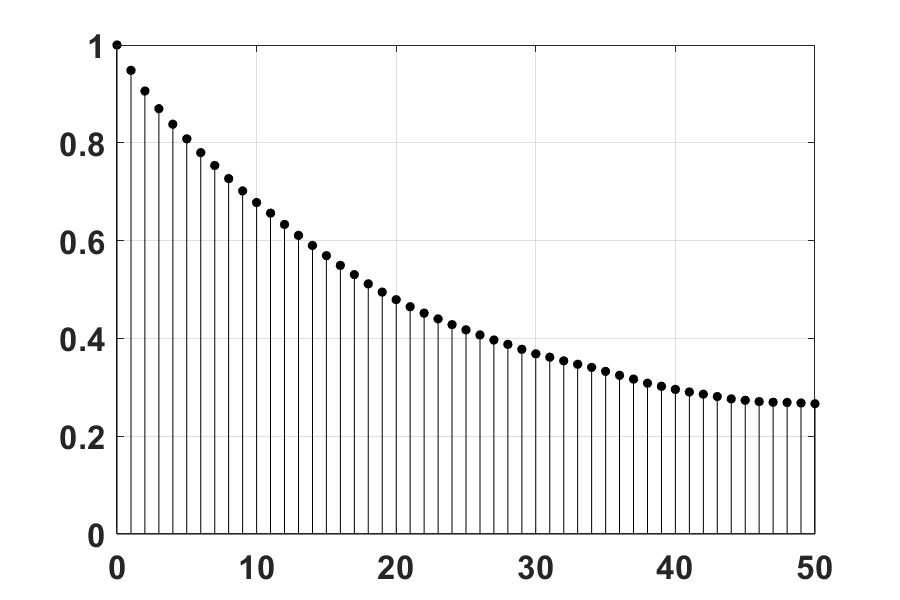}       &
                        \includegraphics[width=2in]{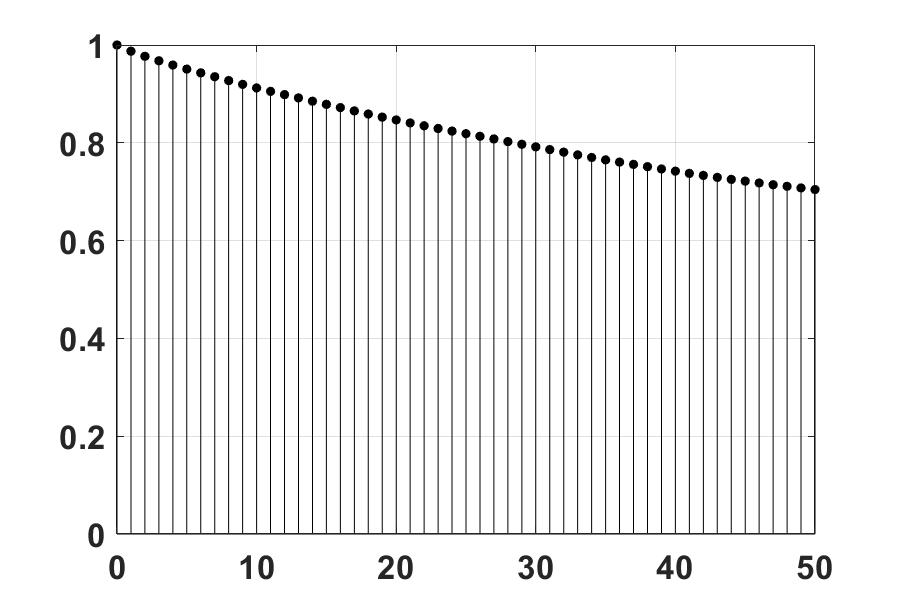} \\
                        \rotatebox{90}{\hspace*{0.3cm} Posterior of $\rho$} &
                        \includegraphics[width=2in]{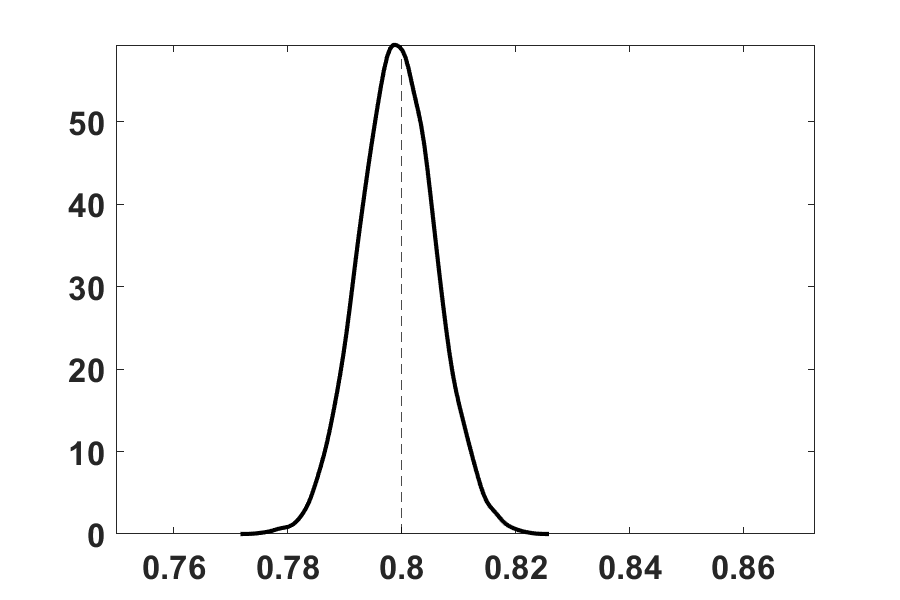}       &
                        \includegraphics[width=2in]{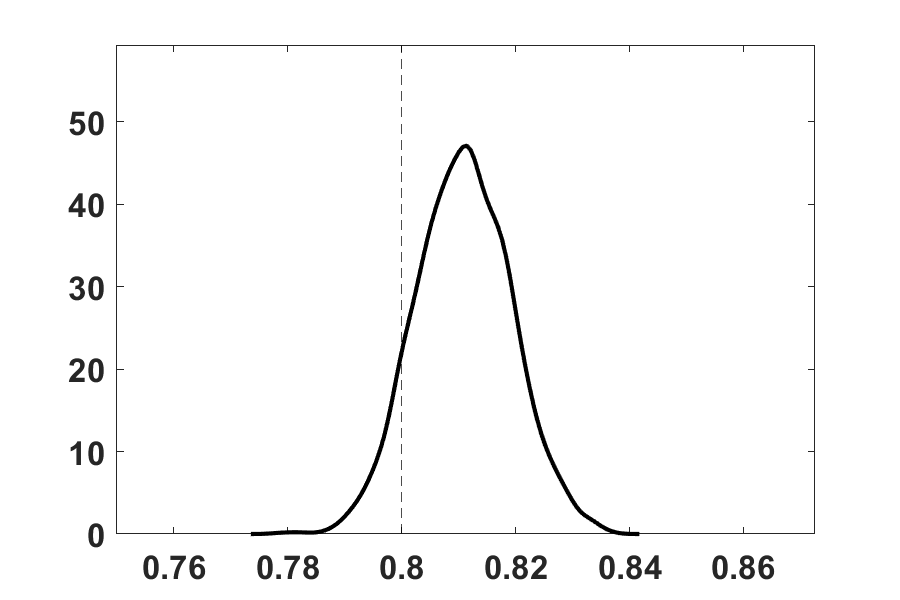}       &
                        \includegraphics[width=2in]{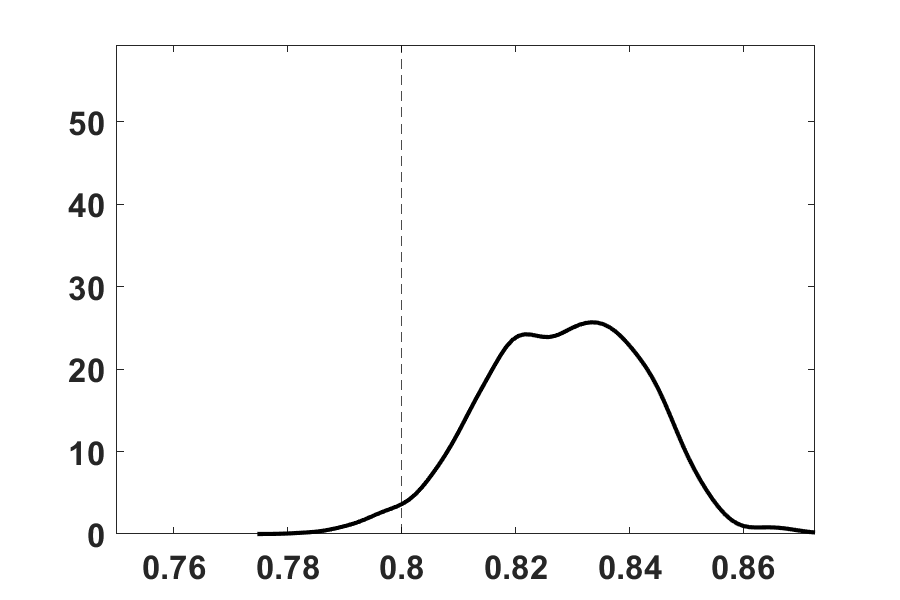}                  
                \end{tabular}
        \end{center}
        {\footnotesize {\em Notes:} The dashed horizontal lines in the first row and the dashed vertical lines in the last row indicate the ``true'' value of $\rho=0.8$. $(j)$ in the superscript indicates the MCMC draws. The first 1,000 draws are discarded as burn-in, so the shaded regions in the first row indicate the MCMC draws kept for posterior analyses. }\setlength{\baselineskip}{4mm}                                     
\end{figure}

Under the alternative DGPs, the MCMC remains stable, despite the larger number of zeros in the samples. In Figure~\ref{appf_MC123_Convergence} we show some convergence diagnostics based on the sequence of draws $\rho^{(j)}$. The first row contains trace plots, the second row autocorrelation functions, and the third row posterior density estimates. As the number of zeros increases, the chain becomes more persistent and the spread of the posterior increases because $\rho$ is effectively estimated from fewer observations. Nonetheless, the algorithm remains well behaved. While 75\% appears to be a large fraction, notice that the sample size is $T\cdot N = 10,000$. Thus, we still have 2,500 non-zero observations.\footnote{We also tried a design with 95\% zeros. Not surprisingly, we experienced convergence problems for this rather extreme design.}

Table~\ref{appt_MC123results_ifcst_dfcst} reproduces and extends the results reported in Table~\ref{t_MC1results_ifcst_dfcst} of the main text. The overall message from the baseline MC design is preserved under the alternative specifications of the DGP. 
The forecasts get more precise as we increase the fraction of zeros. The more zeros in the sample and the longer the zero spells, the stronger the evidence that the next observation will also be a zero. In fact, under all three designs, 100\% of the units with all-zero observations assign a probability of no less than 95\% to $y_{iT+1} = 0$.\footnote{For units with all zeros, the chance of predicting zeros is large in practice, though in principle, these units still convey a slight amount of information about the common parameters and the left tail of the underlying distribution of cross-sectional heterogeneity.}
This improves the density forecasts (lower LPS and CRPS) and shortens the predictive sets. The downside of more zeros is that the estimation of the homogeneous parameter $\rho$ becomes more difficult. Both bias and standard deviation of $\hat{\rho}$ across Monte Carlo repetitions increase which is mirrored in the shape of the posterior depicted in the last row of Figure~\ref{appf_MC123_Convergence}. As mentioned before, this is plausible: the fewer non-zero observations, the less information about $\rho$ is in the sample.

\begin{table}[h!]
        \caption{Monte Carlo Experiment: Forecast Performance and Parameter Estimates}
        \label{appt_MC123results_ifcst_dfcst}
        \begin{center}
                \scalebox{0.92}{
                \begin{tabular}{lcccccccc} \\  \hline \hline
                        & \multicolumn{2}{c}{Density Forecast} &   \multicolumn{2}{c}{Set Forecast} &   \multicolumn{2}{c}{Set Forecast} 
                        &   \multicolumn{2}{c}{Estimates}  \\ 
                        & & &   \multicolumn{2}{c}{``Average"} &   \multicolumn{2}{c}{``Pointwise"} \\  
                        &       LPS         &       CRPS &      Cov.    &      Length  &      Cov.    &      Length  &     Bias$(\hat{\rho})$        &       StdD$(\hat{\rho})$  \\ \hline
                        \multicolumn{9}{c}{Fraction of Zeros in Panel is 45\% (From Paper)} \\ \hline
                        Flexible \& Heterosk. &  -0.757 &   0.277 &   0.910 &   1.260 &   0.933 &   1.503 &  -0.002 &   0.005  \\ 
Normal \& Heterosk.   &  -0.758 &   0.277 &   0.908 &   1.248 &   0.932 &   1.498 &  -0.006 &   0.005  \\ 
Flexible \& Homosk.   &  -0.902 &   0.294 &   0.929 &   1.506 &   0.942 &   1.698  &   0.007 &   0.008 \\ 
Normal \& Homosk.     &  -0.903 &   0.294 &   0.929 &   1.501 &   0.942 &   1.699 &   0.001 &   0.007   \\ 
                        \hline
                        \multicolumn{9}{c}{Fraction of Zeros in Panel is 60\%} \\ \hline
                       Flexible \& Heterosk. &  -0.552 &   0.194 &   0.909 &   0.706 &   0.948 &   1.023  &   0.005 &   0.006 \\ 
Normal \& Heterosk.   &  -0.553 &   0.194 &   0.908 &   0.702 &   0.948 &   1.024  &   0.001 &   0.006 \\ 
Flexible \& Homosk.   &  -0.655 &   0.206 &   0.931 &   0.878 &   0.955 &   1.162  &   0.012 &   0.009  \\ 
Normal \& Homosk.     &  -0.656 &   0.207 &   0.931 &   0.880 &   0.956 &   1.169 &   0.009 &   0.009  \\ 
                        \hline
                        \multicolumn{9}{c}{Fraction of Zeros in Panel is 75\%} \\ \hline
                        Flexible \& Heterosk. &  -0.316 &   0.109 &   0.909 &   0.219 &   0.970 &   0.567  &   0.015 &   0.009 \\ 
Normal \& Heterosk.   &  -0.316 &   0.109 &   0.909 &   0.220 &   0.971 &   0.571  &   0.013 &   0.009 \\ 
Flexible \& Homosk.   &  -0.375 &   0.117 &   0.931 &   0.310 &   0.974 &   0.660 &   0.020 &   0.012    \\ 
Normal \& Homosk.     &  -0.376 &   0.117 &   0.932 &   0.315 &   0.975 &   0.668 &   0.022 &   0.013  \\ 
                        \hline
                \end{tabular}       
            }
        \end{center}
        {\footnotesize {\em Notes:} ``Cov.'' is coverage frequency and ``Length'' is an average across $i$.}\setlength{\baselineskip}{4mm}
\end{table}

\begin{figure}[t!]
        \caption{Posterior Means and Estimated RE Distributions for $\lambda_i$, Flexible \& Heterosk.\ Specification}
        \label{f_MC123results_phatlambda}
        \begin{center}
                \begin{tabular}{ccc}
                        45\% Zeros & 60\% Zeros & 75\% Zeros \\
                        \includegraphics[width=2in]{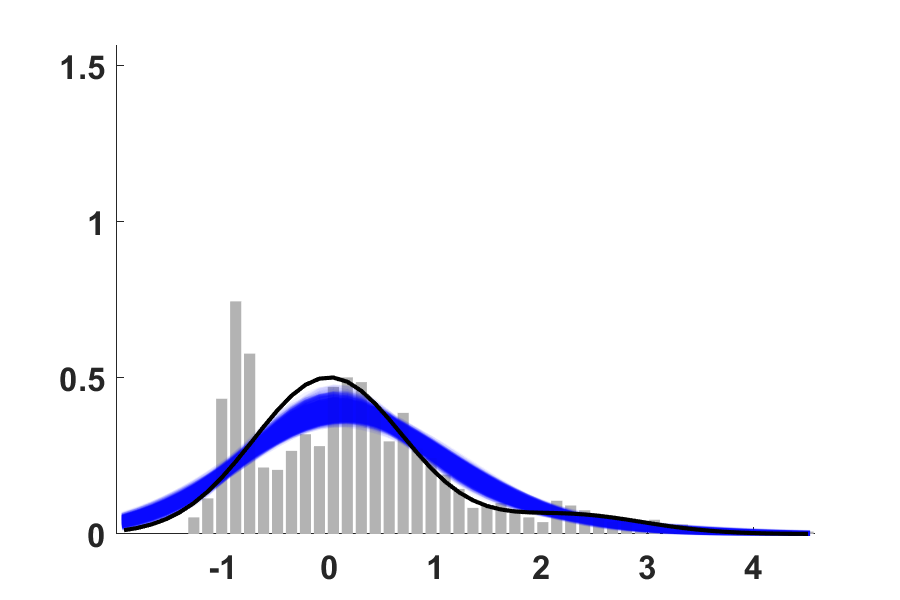}       &
                        \includegraphics[width=2in]{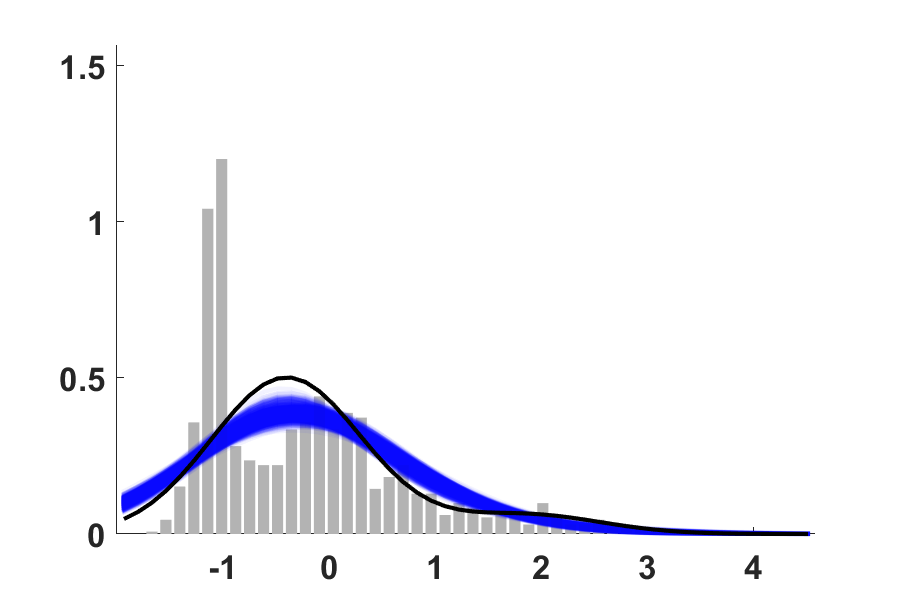}       &
                        \includegraphics[width=2in]{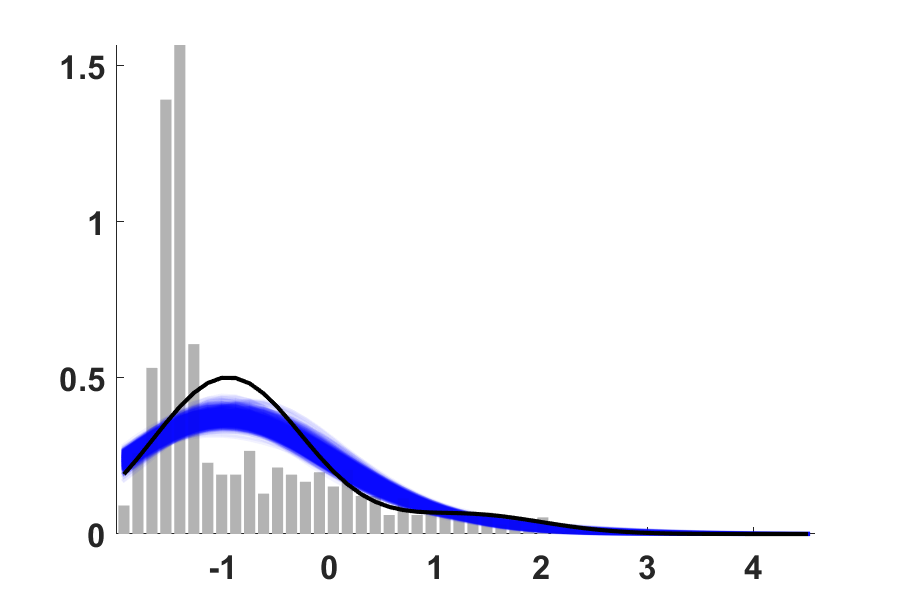}
                \end{tabular}
        \end{center}
        {\footnotesize {\em Notes:} The histograms depict $\mathbb{E}[\lambda_i|Y_{1:N,0:T}]$, $i=1,\ldots,N$. The shaded areas are hairlines obtained by generating draws from the posterior distribution of $\xi$ and plotting the corresponding  random effects densities $p(\lambda|\xi)$.  The black lines represent the true $p(\lambda)$. }\setlength{\baselineskip}{4mm}                                     
\end{figure}

In Figure~\ref{f_MC123results_phatlambda} we plot the cross-sectional distribution of posterior means of $\lambda_i$ as well as the estimated and ``true'' RE distribution. The left panel of the figure reproduces the left panel of Figure~1 in the main paper. By construction, the ``true'' distribution of the $\lambda_i$s shifts to the left for the other two designs (center and right panel of Figure~\ref{f_MC123results_phatlambda}). The spike in the empirical distribution of $\mathbb{E}[\lambda_i|Y_{1:N,0:T}]$ shifts to the left and increases in height because the estimated model needs to reproduce the number of zeros in the sample, which is done by lower estimates for $\lambda_i$. We are using a proper prior for the RE distribution to reduce the chance that draws of $\lambda_i$ take very large negative values. This contributes to the stability of the MCMC. 

\clearpage
\section{Data Set}
\label{appsec:data}

\noindent {\bf Charge-off rates.} The raw data are obtained from the website of the {\em Federal Reserve Bank
        of Chicago}.\footnote{https://www.chicagofed.org/banking/financial-institution-reports/commercial-bank-data} 
The raw data are available at a quarterly frequency. 
The charge-off rates are defined as charge-offs divided by the stock of loans and constructed in a similar manner as in Tables A-1 and A-2 of \cite{CovasRumpZakrajsek2014}. However, the construction differs in the following dimensions: (i) We focus on charge-off rates instead of net charge-off rates. (ii) We divide the charge-offs by the lagged stock of loans instead of the current stock of loans to reduce the timing issue.\footnote{According to bank report forms (e.g. \href{https://www.ffiec.gov/pdf/FFIEC_forms/FFIEC041_201709_f.pdf\#https://www.ffiec.gov/pdf/FFIEC_forms/FFIEC041_201709_f.pdf}{FFIEC 041}), the stocks of loans are given by quarterly averages. ``For all items, banks have the option of reporting either (1) an average of DAILY figures for the quarter, or (2) an average of WEEKLY figures (i.e., the Wednesday of each week of the quarter)."} (iii) For banks with domestic offices only (Form FFIEC 041), RIAD4645 (numerator for commercial and industrial loans) is not reported, so we switch to its domestic counterpart, RIAD4638. 

The charge-offs are reported as year-to-date values. Thus, in order to obtain quarterly data, we take differences: $Q1 \mapsto Q1$, $(Q2-Q1) \mapsto Q2$, $(Q3-Q2) \mapsto Q3$, and $(Q4-Q3) \mapsto Q4$.
The loans are stock variables and no further transformation is needed. We multiply the charge-off rates by 400 to convert them into annualized percentages. 
We construct charge-off rates for the following types of loans:
\begin{itemize}
        \item CI = commercial \& industrial; 
        \item CLD = construction \& land development;
        \item MF = multifamily real estate; 
        \item CRE = (nonfarm) nonresidential commercial real estate;
        \item HLC = home equity lines of credit (HELOCs); 
        \item RRE = residential real estate, excluding
        HELOCs; 
        \item CC = credit card; 
        \item CON = consumer, excluding credit card loans.                 
\end{itemize}

We focus on ``small" banks and relate the charge-off rates to local economic conditions. We include a bank
in the sample if its assets are below one billion dollars.  
The raw data set contains missing observations and outliers that we are unable to explain with
our econometric model. Thus, we proceed as follows to select a subset of observations from the raw data. For each rolling sample: 
\begin{enumerate}
        \item Eliminate banks for which domestic total assets are missing for all time periods in the sample.
        \item Compute average non-missing domestic total assets and eliminate banks with average assets above 1 billion dollars.
        \item For each loan category, eliminate banks for which the target charge-off rate is missing for at least one period of the sample.
        \item For each loan category, eliminate banks for which the target charge-off rate is  negative or greater than 400\% for at least one period of the sample.
        \item For each loan category proceed as follows: First, for each bank, drop the two largest observations $y_{it},\;t=0,\cdots,T+1$, and calculate the standard deviation (stdd) of the remaining observations. Then, eliminate a bank if any successive change $|y_{it}-y_{it-1}|+|y_{it+1}-y_{it}|>10\text{stdd}$. For $t=0$ and $t=T+1$, we only have one of the two terms and we set the other term in this selection criterion to zero.
\end{enumerate}
The remaining sample sizes after each of these steps  as well as some summary statistics for loan charge-off rates are reported in
Table~\ref{apptab:sampleselection_cc}.

\begin{table}[t!]
        \caption{Sample Sizes After Selection Steps and Summary Statistics for Charge-Off Rates}
        \label{apptab:sampleselection_cc}
        \begin{center}
                \scalebox{0.9}{
                        \begin{tabular}{llrrrrrrrrrr} \hline \hline
                                & & \multicolumn{5}{c}{Sample Sizes} & \multicolumn{4}{c}{Cross-sectional Statistics} \\
                                Loan    &       $t_0$   &       Initial &       Step1   &       Step2   &       Step3   &       Step4&       Step5   &       \% 0s      &       Mean    &   75\% &     Max     \\
                                \hline
                                CLD     &       2007Q3  &       7,903   &       7,903   &       7,299   &       3,290   &       3,146   &       1,304   &       77      &       1.5     &       0.0     &       106.8   \\
                                CLD     &       2007Q4  &       7,835   &       7,835   &       7,219   &       3,244   &       3,088   &       1,264   &       74      &       1.9     &       0.1     &       106.8   \\
                                CLD     &       2008Q1  &       7,692   &       7,692   &       7,084   &       3,204   &       3,032   &       1,257   &       71      &       2.2     &       0.5     &       180.2   \\
                                RRE     &       2007Q1  &       7,991   &       7,991   &       7,393   &       6,260   &       5,993   &       2,654   &       77      &       0.2     &       0.0     &       33.1    \\
                                RRE     &       2007Q2  &       7,993   &       7,993   &       7,383   &       6,152   &       5,894   &       2,576   &       76      &       0.3     &       0.0     &       33.1    \\
                                RRE     &       2007Q3  &       7,903   &       7,903   &       7,299   &       6,193   &       5,920   &       2,606   &       73      &       0.3     &       0.0     &       35.9    \\
                                RRE     &       2007Q4  &       7,835   &       7,835   &       7,219   &       6,146   &       5,859   &       2,581   &       70      &       0.4     &       0.1     &       69.2    \\
                                RRE     &       2008Q1  &       7,692   &       7,692   &       7,084   &       6,106   &       5,792   &       2,561   &       68      &       0.4     &       0.2     &       45.6    \\
                                RRE     &       2008Q2  &       7,701   &       7,701   &       7,080   &       6,029   &       5,721   &       2,492   &       67      &       0.4     &       0.2     &       63.6    \\
                                RRE     &       2008Q3  &       7,631   &       7,631   &       7,008   &       6,052   &       5,743   &       2,577   &       65      &       0.5     &       0.3     &       39.2    \\
                                RRE     &       2008Q4  &       7,559   &       7,559   &       6,938   &       6,005   &       5,679   &       2,600   &       63      &       0.5     &       0.3     &       45.6    \\
                                RRE     &       2009Q1  &       7,480   &       7,480   &       6,849   &       5,971   &       5,634   &       2,588   &       62      &       0.5     &       0.3     &       45.0    \\
                                RRE     &       2009Q2  &       8,103   &       8,103   &       7,381   &       5,895   &       5,564   &       2,536   &       62      &       0.5     &       0.3     &       45.0    \\
                                RRE     &       2009Q3  &       8,016   &       8,016   &       7,302   &       5,899   &       5,568   &       2,563   &       61      &       0.5     &       0.4     &       47.6    \\
                                RRE     &       2009Q4  &       7,940   &       7,940   &       7,229   &       5,846   &       5,508   &       2,553   &       60      &       0.5     &       0.4     &       45.0    \\
                                RRE     &       2010Q1  &       7,770   &       7,770   &       7,077   &       5,765   &       5,426   &       2,494   &       61      &       0.5     &       0.4     &       45.0    \\
                                RRE     &       2010Q2  &       7,770   &       7,770   &       7,072   &       5,635   &       5,308   &       2,420   &       61      &       0.5     &       0.4     &       45.0    \\
                                RRE     &       2010Q3  &       7,707   &       7,707   &       7,013   &       5,632   &       5,298   &       2,441   &       61      &       0.5     &       0.4     &       45.6    \\
                                RRE     &       2010Q4  &       7,608   &       7,608   &       6,910   &       5,583   &       5,255   &       2,443   &       61      &       0.5     &       0.3     &       38.2    \\
                                RRE     &       2011Q1  &       7,469   &       7,469   &       6,784   &       5,520   &       5,220   &       2,437   &       62      &       0.4     &       0.3     &       38.2    \\
                                RRE     &       2011Q2  &       7,472   &       7,472   &       6,783   &       5,398   &       5,110   &       2,385   &       62      &       0.4     &       0.3     &       38.2    \\
                                RRE     &       2011Q3  &       7,402   &       7,402   &       6,716   &       5,395   &       5,110   &       2,397   &       64      &       0.4     &       0.2     &       38.2    \\
                                RRE     &       2011Q4  &       7,334   &       7,334   &       6,649   &       5,341   &       5,059   &       2,395   &       65      &       0.3     &       0.2     &       38.2    \\
                                RRE     &       2012Q1  &       7,236   &       7,236   &       6,546   &       5,284   &       5,008   &       2,349   &       67      &       0.3     &       0.2     &       38.2    \\
                                RRE     &       2012Q2  &       7,234   &       7,234   &       6,534   &       5,584   &       5,283   &       2,430   &       66      &       0.3     &       0.2     &       38.2    \\
                                RRE     &       2012Q3  &       7,170   &       7,170   &       6,465   &       5,576   &       5,267   &       2,416   &       67      &       0.2     &       0.1     &       28.4    \\
                                RRE     &       2012Q4  &       7,073   &       7,073   &       6,358   &       5,495   &       5,197   &       2,362   &       69      &       0.2     &       0.1     &       22.2    \\
                                RRE     &       2013Q1  &       6,931   &       6,849   &       6,212   &       5,420   &       5,121   &       2,341   &       71      &       0.2     &       0.1     &       28.7    \\
                                RRE     &       2013Q2  &       6,934   &       6,857   &       6,200   &       5,296   &       5,008   &       2,298   &       71      &       0.2     &       0.1     &       28.7    \\
                                RRE     &       2013Q3  &       6,884   &       6,807   &       6,144   &       5,291   &       4,999   &       2,307   &       72      &       0.2     &       0.0     &       28.7    \\
                                RRE     &       2013Q4  &       6,803   &       6,726   &       6,061   &       5,212   &       4,932   &       2,271   &       74      &       0.1     &       0.0     &       28.7    \\
                                RRE     &       2014Q1  &       6,650   &       6,576   &       5,913   &       5,144   &       4,870   &       2,258   &       75      &       0.1     &       0.0     &       27.2    \\
                                RRE     &       2014Q2  &       6,650   &       6,578   &       5,897   &       5,012   &       4,746   &       2,190   &       76      &       0.1     &       0.0     &       16.9    \\
                                RRE     &       2014Q3  &       6,582   &       6,510   &       5,821   &       5,004   &       4,742   &       2,178   &       77      &       0.1     &       0.0     &       22.2    \\
                                RRE     &       2014Q4  &       6,502   &       6,431   &       5,729   &       4,945   &       4,691   &       2,210   &       78      &       0.1     &       0.0     &       16.9    \\
                                RRE     &       2015Q1  &       6,342   &       6,271   &       5,564   &       4,874   &       4,611   &       2,200   &       79      &       0.1     &       0.0     &       11.1    \\
                                RRE     &       2015Q2  &       6,348   &       6,278   &       5,560   &       4,751   &       4,500   &       2,134   &       79      &       0.1     &       0.0     &       11.1    \\
                                CC      &       2001Q2  &       9,031   &       9,031   &       8,532   &       1,691   &       1,540   &       875     &       33      &       3.4     &       4.7     &       162.5   \\
                                CC      &       2001Q3  &       8,995   &       8,995   &       8,491   &       1,666   &       1,515   &       844     &       33      &       3.4     &       4.8     &       88.9    \\
                                CC      &       2001Q4  &       8,887   &       8,887   &       8,382   &       1,636   &       1,489   &       836     &       34      &       3.3     &       4.6     &       88.9    \\

                                \hline
                        \end{tabular}  
                }       
        \end{center}
        {\footnotesize {\em Notes:} This table provides summary statistics for samples with cross-sectional dimension $N>400$ and percentage of zeros less than 80\%. 
                The date assigned to each panel refers to $t=t_0$, which is the conditioning information
                used to initialize the lag in the dynamic Tobit. We assume that $T=10$, which means that each sample has 12 time periods. The descriptive statistics are computed across $N$ and $T$ dimension of each panel.}\setlength{\baselineskip}{4mm} 
\end{table} 

\addtocounter{table}{-1}
\begin{table}[t!]
        \caption{Sample Sizes After Selection Steps and Summary Statistics for Charge-Off Rates (cont.)}
        \begin{center}
                \scalebox{0.9}{
                        \begin{tabular}{llrrrrrrrrrr} \hline \hline
                                & & \multicolumn{5}{c}{Sample Sizes} & \multicolumn{4}{c}{Cross-sectional Statistics} \\
                                Loan    &       $t_0$   &       Initial &       Step1   &       Step2   &       Step3   &       Step4&       Step5   &       \% 0s      &       Mean    & 75\% &      Max     \\
                                \hline
                                CC      &       2002Q1  &       8,723   &       8,723   &       8,228   &       1,612   &       1,466   &       814     &       35      &       3.3     &       4.4     &       400.0   \\
                                CC      &       2002Q2  &       8,823   &       8,823   &       8,312   &       1,670   &       1,519   &       817     &       38      &       3.2     &       4.3     &       88.9    \\
                                CC      &       2002Q3  &       8,805   &       8,805   &       8,286   &       1,631   &       1,488   &       821     &       38      &       3.2     &       4.3     &       88.9    \\
                                CC      &       2002Q4  &       8,728   &       8,728   &       8,199   &       1,606   &       1,468   &       813     &       39      &       3.1     &       4.1     &       88.9    \\
                                CC      &       2003Q1  &       8,611   &       8,611   &       8,077   &       1,573   &       1,445   &       811     &       40      &       3.0     &       4.0     &       128.5   \\
                                CC      &       2003Q2  &       8,754   &       8,754   &       8,203   &       1,544   &       1,422   &       787     &       40      &       3.0     &       3.9     &       136.1   \\
                                CC      &       2003Q3  &       8,755   &       8,755   &       8,198   &       1,513   &       1,395   &       754     &       41      &       2.9     &       3.8     &       136.1   \\
                                CC      &       2003Q4  &       8,671   &       8,671   &       8,120   &       1,500   &       1,387   &       724     &       42      &       2.8     &       3.6     &       136.1   \\
                                CC      &       2004Q1  &       8,526   &       8,526   &       7,989   &       1,468   &       1,355   &       707     &       43      &       2.7     &       3.6     &       136.1   \\
                                CC      &       2004Q2  &       8,662   &       8,662   &       8,108   &       1,440   &       1,331   &       677     &       42      &       2.8     &       3.6     &       136.1   \\
                                CC      &       2004Q3  &       8,626   &       8,626   &       8,067   &       1,411   &       1,308   &       664     &       43      &       2.7     &       3.5     &       136.1   \\
                                CC      &       2004Q4  &       8,552   &       8,552   &       7,989   &       1,391   &       1,284   &       657     &       44      &       2.6     &       3.3     &       140.9   \\
                                CC      &       2005Q1  &       8,384   &       8,384   &       7,829   &       1,369   &       1,271   &       639     &       44      &       2.5     &       3.2     &       151.3   \\
                                CC      &       2005Q2  &       8,507   &       8,507   &       7,938   &       1,332   &       1,236   &       611     &       44      &       2.6     &       3.2     &       175.0   \\
                                CC      &       2005Q3  &       8,482   &       8,482   &       7,897   &       1,315   &       1,218   &       596     &       45      &       2.6     &       3.2     &       175.0   \\
                                CC      &       2005Q4  &       8,404   &       8,404   &       7,816   &       1,290   &       1,203   &       604     &       46      &       2.6     &       3.2     &       210.5   \\
                                CC      &       2006Q1  &       8,263   &       8,263   &       7,674   &       1,275   &       1,188   &       614     &       47      &       2.6     &       3.1     &       175.0   \\
                                CC      &       2006Q2  &       8,307   &       8,307   &       7,708   &       1,247   &       1,164   &       594     &       47      &       2.7     &       3.2     &       269.2   \\
                                CC      &       2006Q3  &       8,240   &       8,240   &       7,639   &       1,231   &       1,156   &       594     &       46      &       2.8     &       3.4     &       269.2   \\
                                CC      &       2006Q4  &       8,137   &       8,137   &       7,537   &       1,211   &       1,139   &       595     &       45      &       3.0     &       3.6     &       269.2   \\
                                CC      &       2007Q1  &       7,991   &       7,991   &       7,393   &       1,197   &       1,129   &       574     &       44      &       3.2     &       3.9     &       269.2   \\
                                CC      &       2007Q2  &       7,993   &       7,993   &       7,383   &       1,173   &       1,107   &       561     &       43      &       3.3     &       4.1     &       269.2   \\
                                CC      &       2007Q3  &       7,903   &       7,903   &       7,299   &       1,159   &       1,091   &       544     &       44      &       3.2     &       4.2     &       175.0   \\
                                CC      &       2007Q4  &       7,835   &       7,835   &       7,219   &       1,133   &       1,066   &       534     &       43      &       3.3     &       4.2     &       175.0   \\
                                CC      &       2008Q1  &       7,692   &       7,692   &       7,084   &       1,123   &       1,056   &       527     &       44      &       3.3     &       4.2     &       175.0   \\
                                CC      &       2008Q2  &       7,701   &       7,701   &       7,080   &       1,101   &       1,035   &       512     &       45      &       3.2     &       4.1     &       158.3   \\
                                CC      &       2008Q3  &       7,631   &       7,631   &       7,008   &       1,096   &       1,036   &       509     &       44      &       3.1     &       4.0     &       158.3   \\
                                CC      &       2008Q4  &       7,559   &       7,559   &       6,938   &       1,082   &       1,020   &       506     &       45      &       3.1     &       3.9     &       149.4   \\
                                CC      &       2009Q1  &       7,480   &       7,480   &       6,849   &       1,059   &       999     &       498     &       46      &       3.0     &       3.7     &       147.3   \\
                                CC      &       2009Q2  &       8,103   &       8,103   &       7,381   &       1,045   &       989     &       492     &       45      &       2.8     &       3.7     &       78.5    \\
                                CC      &       2009Q3  &       8,016   &       8,016   &       7,302   &       1,042   &       988     &       492     &       47      &       2.7     &       3.5     &       77.6    \\
                                CC      &       2009Q4  &       7,940   &       7,940   &       7,229   &       1,032   &       978     &       479     &       49      &       2.7     &       3.3     &       400.0   \\
                                CC      &       2010Q1  &       7,770   &       7,770   &       7,077   &       1,020   &       963     &       459     &       49      &       2.5     &       3.2     &       100.0   \\
                                CC      &       2010Q2  &       7,770   &       7,770   &       7,072   &       997     &       940     &       454     &       50      &       2.3     &       3.0     &       62.0    \\
                                CC      &       2010Q3  &       7,707   &       7,707   &       7,013   &       994     &       940     &       450     &       50      &       2.2     &       2.8     &       62.0    \\
                                CC      &       2010Q4  &       7,608   &       7,608   &       6,910   &       976     &       920     &       454     &       51      &       2.1     &       2.6     &       56.3    \\
                                CC      &       2011Q1  &       7,469   &       7,469   &       6,784   &       961     &       906     &       451     &       52      &       2.0     &       2.5     &       68.6    \\
                                CC      &       2011Q2  &       7,472   &       7,472   &       6,783   &       941     &       889     &       450     &       53      &       1.9     &       2.4     &       67.9    \\
                                CC      &       2011Q3  &       7,402   &       7,402   &       6,716   &       933     &       879     &       443     &       54      &       1.9     &       2.3     &       67.9    \\
                                CC      &       2011Q4  &       7,334   &       7,334   &       6,649   &       920     &       869     &       430     &       55      &       1.8     &       2.2     &       67.9    \\

                                \hline
                        \end{tabular}  
                }       
        \end{center}
        {\footnotesize {\em Notes:} This table provides summary statistics for samples with cross-sectional dimension $N>400$ and percentage of zeros less than 80\%. 
                The date assigned to each panel refers to $t=t_0$, which is the conditioning information
                used to initialize the lag in the dynamic Tobit. We assume that $T=10$, which means that each sample has 12 time periods. The descriptive statistics are computed across $N$ and $T$ dimension of each panel.}\setlength{\baselineskip}{4mm} 
\end{table} 

\addtocounter{table}{-1}
\begin{table}[t!]
        \caption{Sample Sizes After Selection Steps and Summary Statistics for Charge-Off Rates (cont.)}
        \begin{center}
                \scalebox{0.9}{
                        \begin{tabular}{llrrrrrrrrrr} \hline \hline
                                & & \multicolumn{5}{c}{Sample Sizes} & \multicolumn{4}{c}{Cross-sectional Statistics} \\
                                Loan    &       $t_0$   &       Initial &       Step1   &       Step2   &       Step3 &       Step4  &       Step5   &       \% 0s      &       Mean    & 75\% &      Max     \\
                                \hline
                                CC      &       2012Q1  &       7,236   &       7,236   &       6,546   &       913     &       862     &       438     &       56      &       1.7     &       2.1     &       67.9    \\
                                CC      &       2012Q2  &       7,234   &       7,234   &       6,534   &       916     &       862     &       430     &       54      &       1.8     &       2.2     &       67.9    \\
                                CC      &       2012Q3  &       7,170   &       7,170   &       6,465   &       907     &       853     &       409     &       55      &       1.7     &       2.1     &       67.9    \\
                                CON     &       2009Q2  &       8,103   &       8,103   &       7,381   &       5,837   &       5,698   &       2,600   &       77      &       0.4     &       0.0     &       77.4    \\
                                CON     &       2009Q3  &       8,016   &       8,016   &       7,302   &       5,872   &       5,693   &       2,672   &       71      &       0.5     &       0.2     &       202.2   \\
                                CON     &       2009Q4  &       7,940   &       7,940   &       7,229   &       5,814   &       5,584   &       2,723   &       65      &       0.5     &       0.5     &       202.2   \\
                                CON     &       2010Q1  &       7,770   &       7,770   &       7,077   &       5,735   &       5,461   &       2,680   &       58      &       0.7     &       0.7     &       202.2   \\
                                CON     &       2010Q2  &       7,770   &       7,770   &       7,072   &       5,602   &       5,339   &       2,600   &       53      &       0.7     &       0.8     &       202.2   \\
                                CON     &       2010Q3  &       7,707   &       7,707   &       7,013   &       5,596   &       5,311   &       2,555   &       47      &       0.8     &       0.9     &       202.2   \\
                                CON     &       2010Q4  &       7,608   &       7,608   &       6,910   &       5,545   &       5,227   &       2,473   &       42      &       0.9     &       1.0     &       202.2   \\
                                CON     &       2011Q1  &       7,469   &       7,469   &       6,784   &       5,482   &       5,133   &       2,427   &       36      &       1.0     &       1.1     &       202.2   \\
                                CON     &       2011Q2  &       7,472   &       7,472   &       6,783   &       5,361   &       5,026   &       2,328   &       37      &       1.0     &       1.1     &       202.2   \\
                                CON     &       2011Q3  &       7,402   &       7,402   &       6,716   &       5,377   &       5,028   &       2,333   &       38      &       1.0     &       1.1     &       202.2   \\
                                CON     &       2011Q4  &       7,334   &       7,334   &       6,649   &       5,324   &       4,979   &       2,377   &       38      &       0.9     &       1.0     &       202.2   \\
                                CON     &       2012Q1  &       7,236   &       7,236   &       6,546   &       5,266   &       4,932   &       2,403   &       39      &       0.9     &       1.0     &       202.2   \\
                                CON     &       2012Q2  &       7,234   &       7,234   &       6,534   &       5,544   &       5,195   &       2,530   &       42      &       0.8     &       1.0     &       76.0    \\
                                CON     &       2012Q3  &       7,170   &       7,170   &       6,465   &       5,536   &       5,184   &       2,541   &       43      &       0.8     &       0.9     &       76.0    \\
                                CON     &       2012Q4  &       7,073   &       7,073   &       6,358   &       5,457   &       5,117   &       2,526   &       43      &       0.8     &       0.9     &       44.7    \\
                                CON     &       2013Q1  &       6,931   &       6,849   &       6,212   &       5,379   &       5,042   &       2,548   &       44      &       0.8     &       0.9     &       100.0   \\
                                CON     &       2013Q2  &       6,934   &       6,857   &       6,200   &       5,254   &       4,932   &       2,465   &       43      &       0.8     &       0.9     &       100.0   \\
                                CON     &       2013Q3  &       6,884   &       6,807   &       6,144   &       5,246   &       4,917   &       2,512   &       44      &       0.7     &       0.9     &       76.0    \\
                                CON     &       2013Q4  &       6,803   &       6,726   &       6,061   &       5,165   &       4,843   &       2,470   &       44      &       0.7     &       0.9     &       76.0    \\
                                CON     &       2014Q1  &       6,650   &       6,576   &       5,913   &       5,094   &       4,767   &       2,415   &       44      &       0.7     &       0.9     &       76.0    \\
                                CON     &       2014Q2  &       6,650   &       6,578   &       5,897   &       4,969   &       4,651   &       2,326   &       44      &       0.7     &       0.9     &       35.7    \\
                                CON     &       2014Q3  &       6,582   &       6,510   &       5,821   &       4,954   &       4,638   &       2,297   &       43      &       0.7     &       0.9     &       35.7    \\
                                CON     &       2014Q4  &       6,502   &       6,431   &       5,729   &       4,894   &       4,585   &       2,324   &       43      &       0.7     &       0.9     &       76.0    \\
                                CON     &       2015Q1  &       6,342   &       6,271   &       5,564   &       4,827   &       4,515   &       2,298   &       43      &       0.7     &       0.9     &       35.7    \\
                                CON     &       2015Q2  &       6,348   &       6,278   &       5,560   &       4,704   &       4,406   &       2,214   &       43      &       0.7     &       0.9     &       52.9    \\
                                CON     &       2015Q3  &       6,271   &       6,204   &       5,479   &       4,689   &       4,402   &       2,209   &       43      &       0.7     &       0.9     &       52.9    \\
                                CON     &       2015Q4  &       6,183   &       6,117   &       5,395   &       4,625   &       4,337   &       2,222   &       42      &       0.8     &       0.9     &       113.8   \\
                                CON     &       2016Q1  &       6,059   &       5,993   &       5,256   &       4,538   &       4,252   &       2,179   &       43      &       0.7     &       0.9     &       52.9    \\

                                \hline
                        \end{tabular}  
                }       
        \end{center}
        {\footnotesize {\em Notes:} This table provides summary statistics for samples with cross-sectional dimension $N>400$ and percentage of zeros less than 80\%. 
                The date assigned to each panel refers to $t=t_0$, which is the conditioning information
                used to initialize the lag in the dynamic Tobit. We assume that $T=10$, which means that each sample has 12 time periods. The descriptive statistics are computed across $N$ and $T$ dimension of each panel.}\setlength{\baselineskip}{4mm} 
\end{table} 

\clearpage

\noindent {\bf Local Market.} We use the annual {\em Summary of Deposits} data from the {\em Federal Deposit
        Insurance Corporation} to determine the local market for each bank. This data set contains information 
about the locations (at ZIP code level) in which deposits were made. Based on this information, for each bank in the charge-off data set we compute the amount of deposits received by state. We then associate each bank with the state from which it received the largest amount of deposits. 

\noindent {\bf Unemployment Rate ($\mbox{UR}_{it}$).} Obtained from the {\em Bureau of Labor Statistics}. We use seasonally adjusted monthly data, time-aggregated to quarterly frequency by simple averaging. 

\noindent {\bf Housing Price Index ($\mbox{HPI}_{it}$).} Obtained from the {\em Federal Housing Finance Agency} on all transactions, not seasonally adjusted. The index is available at a quarterly frequency.

\noindent {\bf Personal Income ($\mbox{INC}_{it}$).} Raw data are obtained from the {\em Bureau of Labor Statistics}. All quarterly series are seasonally adjusted. We first construct quarterly state-level personal income per capita, which is only available after 2010Q1. Before 2010Q1, there is no quarterly state-level population series available. We interpolate the annual population to quarterly frequency by assuming constant population growth rate within a year, and then divide the quarterly personal income by the imputed quarterly population.\footnote{To check whether this interpolation is reasonable, we also experimented with the same interpolation after 2010Q1, and the resulting time series are comparable to the available data.} Then, we deflate the personal income per capita by the personal consumption expenditure price index.

\noindent {\bf Geo Coding.} The annual {\em Summary of Deposits} data from the {\em Federal Deposit
        Insurance Corporation} also contains the state and county FIPS code associated with the headquarter location of each bank. Based on this information we can link the banks to counties and compute average forecasts for each county which are displayed in Figure~\ref{fig:tailprob.spatial} in the main text. 

\noindent {\bf Bank Characteristics.} Quarterly raw data are obtained from the website of the {\em Federal Reserve Bank
        of Chicago} (see above). We construct bank-characteristics variables as follows:
\begin{itemize}
        \item Log Assets = log(RCON2170); 
        \item Loan Fraction = specific loan stock / sum of all loan stocks;
        \item Capital-To-Asset Ratio= RCON3210/RCON2170; 
        \item Loan-To-Asset Ratio = RCON3360/RCON2170;
        \item ALLL-To-Loan Ratio = RCON3123/RCON3360; 
        \item Diversification = RIAD4079/(RIAD4079+RIAD4107); 
        \item Return on Assets = RIAD4340/RCON2170; 
        \item Overhead Costs-To-Asset Ratio (OCA) = RIAD4093/RCON2170.                 
\end{itemize} 
The unit of the balance sheet variables is thousand dollars. Except for log assets and loan fraction, the variables are similar to \cite{Ghosh2017}. The RIAD variables are year-to-date, so we take differences to obtain quarterly data.  
The RCON variables are stock quantities, so we use lagged values instead of current values to overcome the timing issue in ratios. The regressions in Table~\ref{tab:lambda.regressions} are based on period $t=0$ bank characteristics, to reduce concerns about simultaneity. Summary statistics for the variables are provided in Table~\ref{tab:bank.characteristics.stats}.

\begin{table}[t!]
        \caption{Summary Statistics for Bank Characteristics, RRE and CC 2007Q2}
        \label{tab:bank.characteristics.stats}
        \begin{center}
                \scalebox{0.76}{
                        \begin{tabular}{lccccccccccccc} \\ \hline \hline
                                & \multicolumn{6}{c}{RRE} & & \multicolumn{6}{c}{CC} \\
                                & \multicolumn{2}{c}{Low} & \multicolumn{2}{c}{High} & \multicolumn{2}{c}{All} & & \multicolumn{2}{c}{Low} & \multicolumn{2}{c}{High} & \multicolumn{2}{c}{All} \\ 
                                & Mean & StdD & Mean & StdD & Mean & StdD & & Mean & StdD & Mean & StdD & Mean & StdD\\
                                \cline{2-7} \cline{9-14}
                                Log Assets   &       11.607  &       0.865   &       12.427  &       0.719   &       12.088  &       0.880   &       &       12.105  &       0.726   &       12.501  &       0.674   &       12.440  &       0.696   \\
                                Loan Fraction    &       0.193   &       0.163   &       0.285   &       0.154   &       0.247   &       0.164   &       &       0.001   &       0.002   &       0.013   &       0.040   &       0.012   &       0.037   \\
                                Capital-Asset   &       0.104   &       0.037   &       0.095   &       0.023   &       0.099   &       0.030   &       &       0.099   &       0.040   &       0.095   &       0.021   &       0.095   &       0.025   \\
                                Loan-Asset      &       0.642   &       0.147   &       0.712   &       0.099   &       0.683   &       0.126   &       &       0.699   &       0.102   &       0.684   &       0.103   &       0.686   &       0.103   \\
                                ALLL-Loan     &       0.013   &       0.005   &       0.012   &       0.005   &       0.013   &       0.005   &       &       0.012   &       0.007   &       0.013   &       0.006   &       0.013   &       0.006   \\
                                Diversification &       0.099   &       0.093   &       0.099   &       0.178   &       0.099   &       0.149   &       &       0.102   &       0.054   &       0.126   &       0.081   &       0.122   &       0.078   \\
                                Ret.\ on Assets    &       0.003   &       0.003   &       0.003   &       0.002   &       0.003   &       0.003   &       &       0.003   &       0.002   &       0.003   &       0.002   &       0.003   &       0.002   \\
                                OCA     &       0.008   &       0.003   &       0.008   &       0.003   &       0.008   &       0.003   &       &       0.008   &       0.002   &       0.008   &       0.003   &       0.008   &       0.003   \\[1ex]
                                \textit{Sample Size}     &       \textit{515}     &               &       \textit{731}     &               &       \textit{1246}    &               &       &       \textit{61}      &               &       \textit{333}     &               &       \textit{394}     &               \\
                                
                                \hline      
                        \end{tabular}
                }
        \end{center}
        {\footnotesize {\em Notes:} Bank characteristics are the values observed at  2007Q2 ($t=0$). Low (High) refers to small (large) $\widehat{\lambda_i/\sigma_i}$ group of banks (cutoff is approx -2 for RRE and -1 for CC); see red and blue dots in Figure~\ref{fig:lambdasigma.vs.logassets}. The samples sizes of the ``All'' groups are smaller than those in Table~\ref{tab:baseline.summarystats} because the regression samples in Table~\ref{tab:lambda.regressions} only include banks with a full set of covariates. }\setlength{\baselineskip}{4mm}
\end{table}

\section{Additional Empirical Results}
\label{appsec:additionalresults}

\subsection{Tuning the CRE Prior}

In order to tune the prior for the CRE distribution, we recommend visualizing  certain characteristics of these distributions, such as moments and number of modes. In this subsection we consider two choices of the tuning constants, summarized in Table~\ref{apptab:tuning.constants}. We refer to the first choice of $\tau$ as ``initial,'' and the second choice as ``adjusted,'' based on the examination of the prior and posterior distribution resulting from the ``initial'' choice of $\tau$.

\begin{table}[h!]
        \caption{Tuning Constants for Prior Distribution, CC Sample}
        \label{apptab:tuning.constants}
        \begin{center}
                \begin{tabular}{lccccc} \hline \hline
                        & $\tau_\theta$ & $\tau_\nu$ & $\tau_\phi$ & $\tau_{\sigma}^{\lambda}$ & $\tau_{\sigma}^y$ \\ \hline      
                        Initial & 5.0 & 1.0 & 5.0 & 1.0 & 1.0 \\
                        Adjusted & 5.0 & 1.0 & 20.0 & 1.0 & 4.0  \\ \hline
                \end{tabular}
        \end{center}
\end{table}

\begin{figure}[h!]
        \caption{Prior and Posterior for CRE Distribution, CC} 
        \label{fig:cre.prior.posterior.cc.y0star}
        \begin{center}
                \begin{tabular}{ccccc}
                        & \multicolumn{2}{c}{Initial Tuning} & \multicolumn{2}{c}{Adjusted Tuning} \\
            & Prior & Posterior & Prior 2 & Posterior \\
            \rotatebox{90}{\hspace*{0.5cm} $y_{i0}^*$ Marginal} &
                        \includegraphics[width=1.4in]{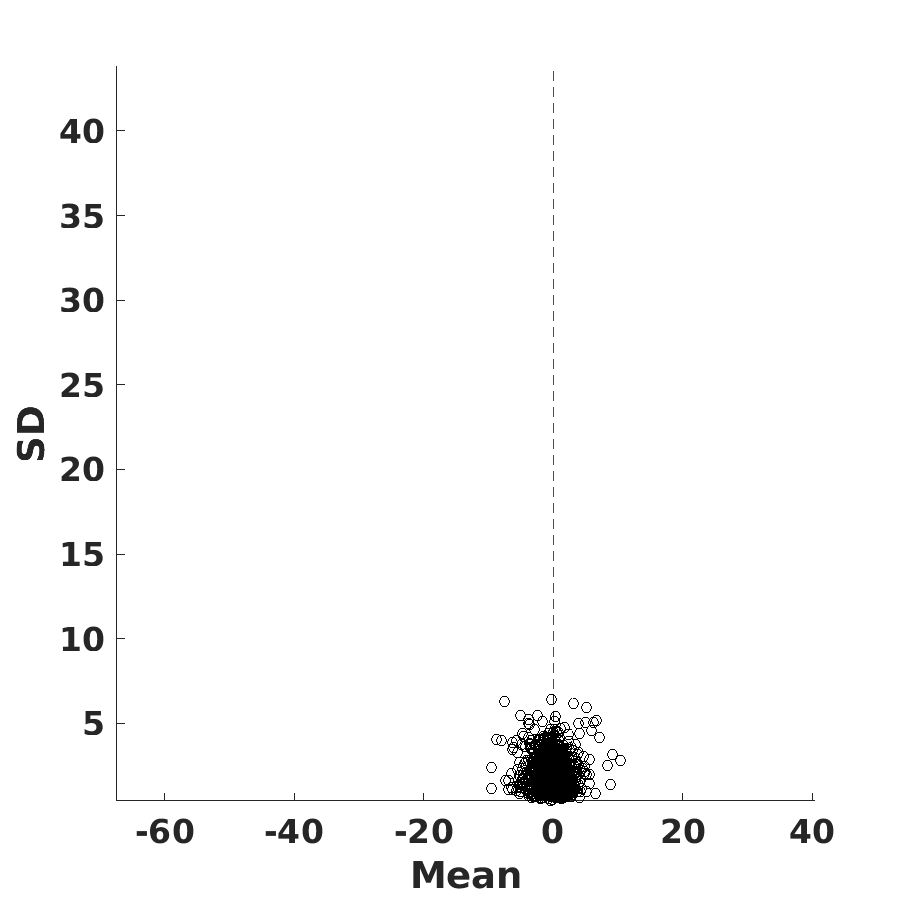}      & 
                        \includegraphics[width=1.4in]{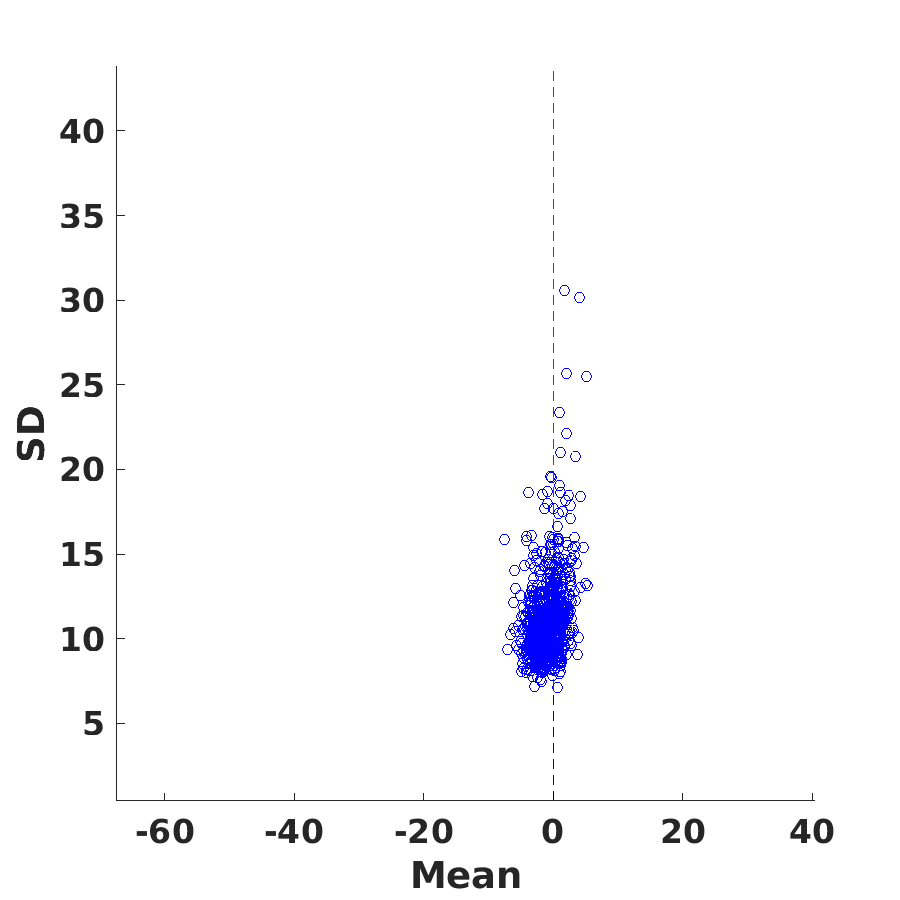} &
                        \includegraphics[width=1.4in]{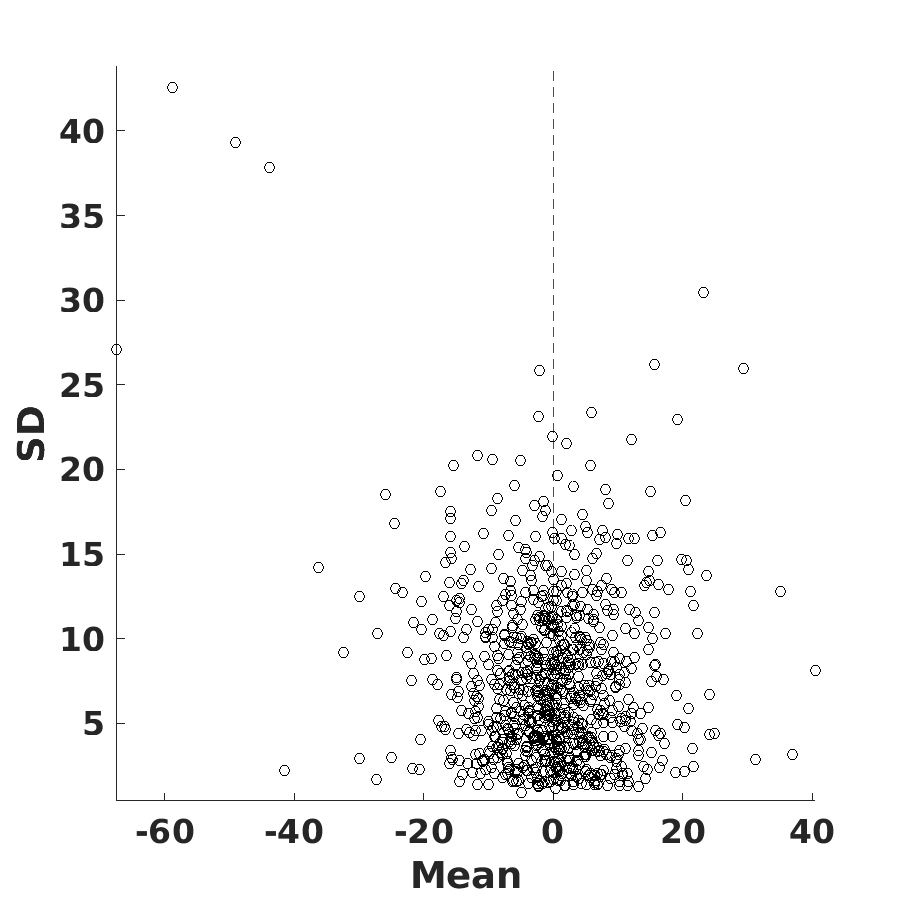} & 
                        \includegraphics[width=1.4in]{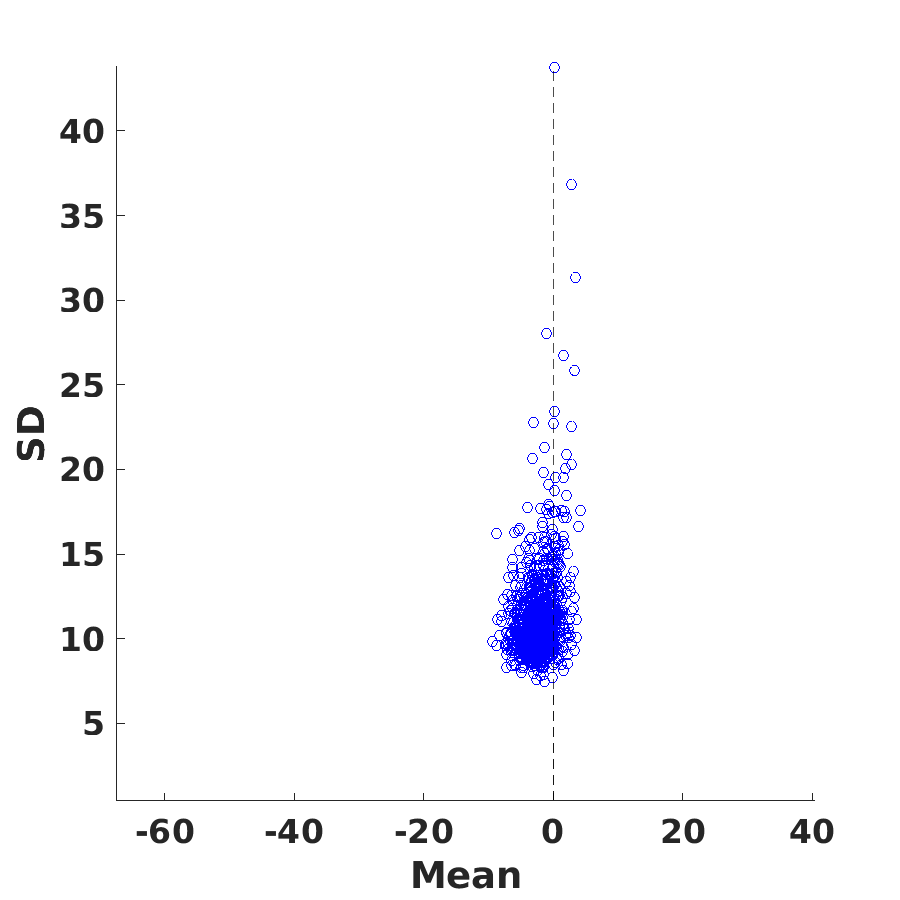} \\[2ex]
            \rotatebox{90}{\hspace*{0.5cm} $y_{i0}^*$ Marginal} &
                        \includegraphics[width=1.4in]{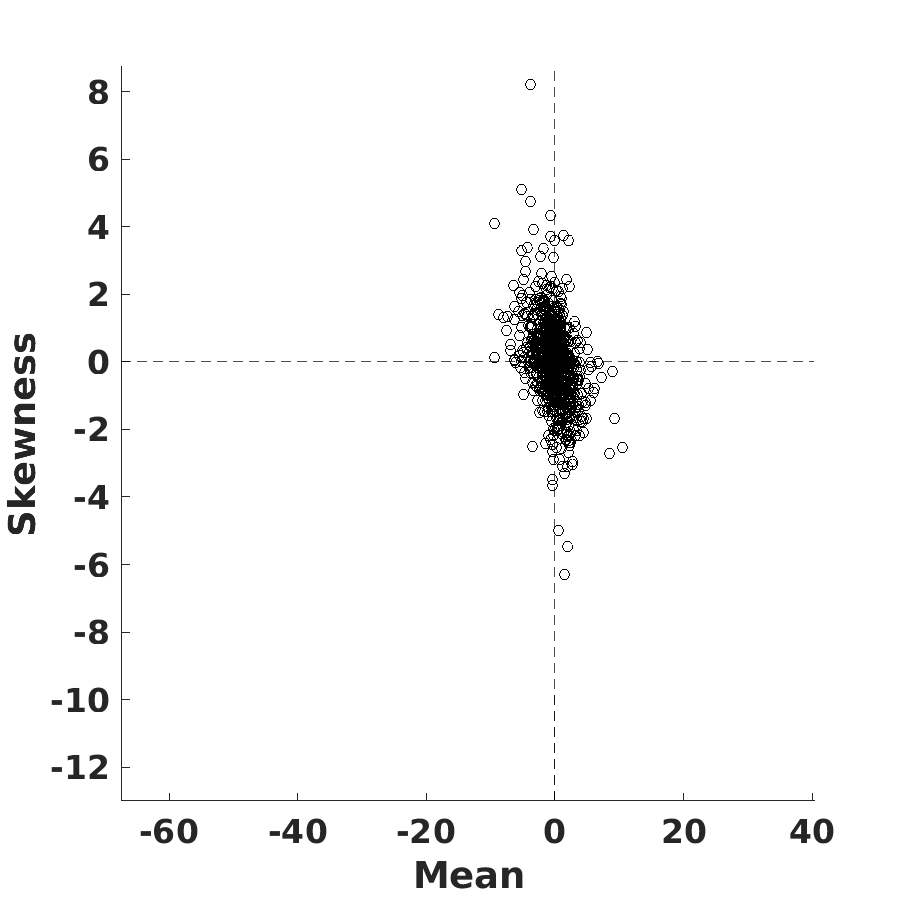}      &
                        \includegraphics[width=1.4in]{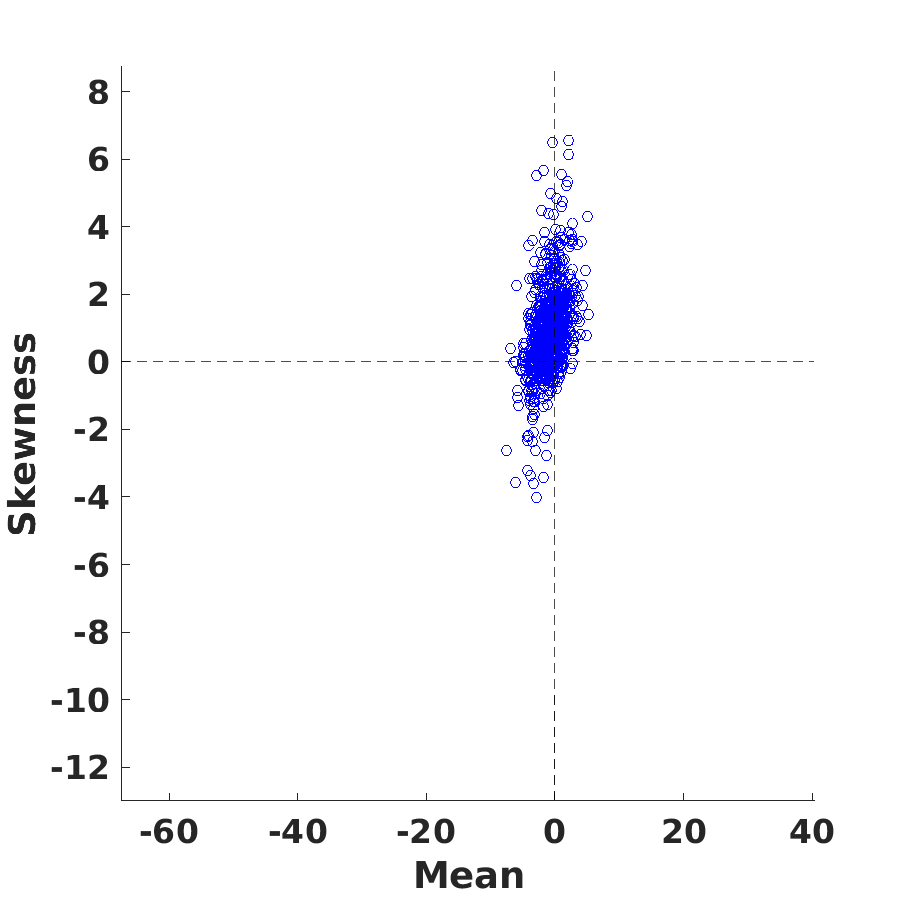} &
                        \includegraphics[width=1.4in]{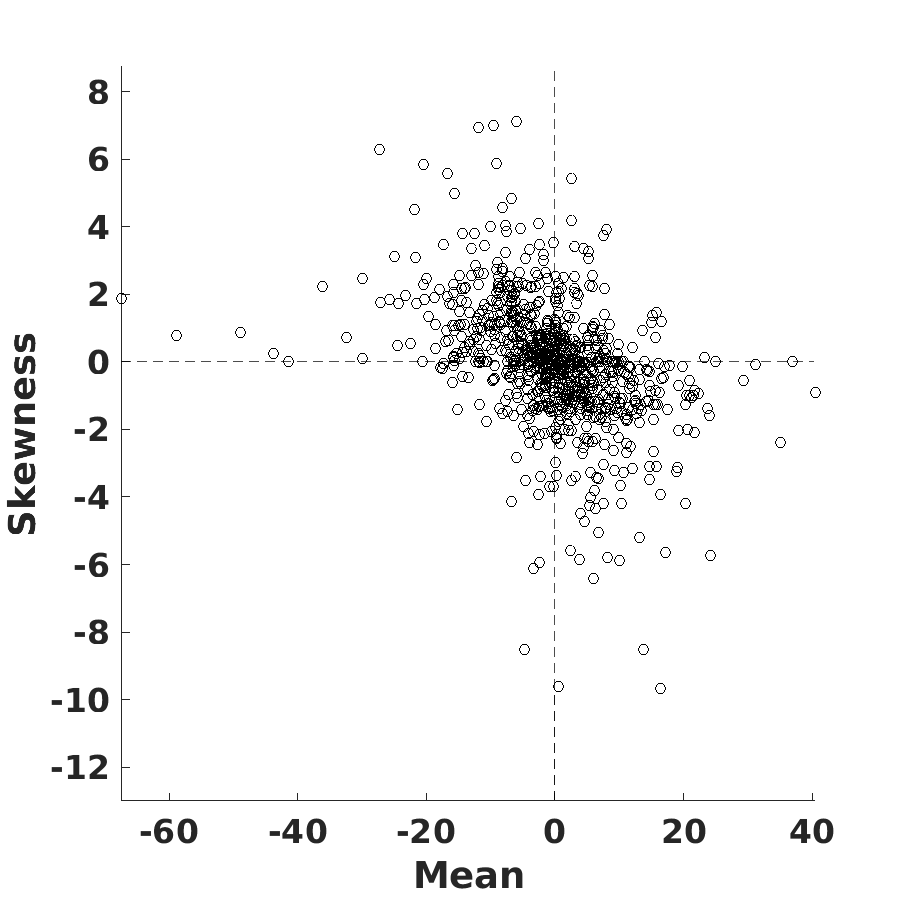} & 
                        \includegraphics[width=1.4in]{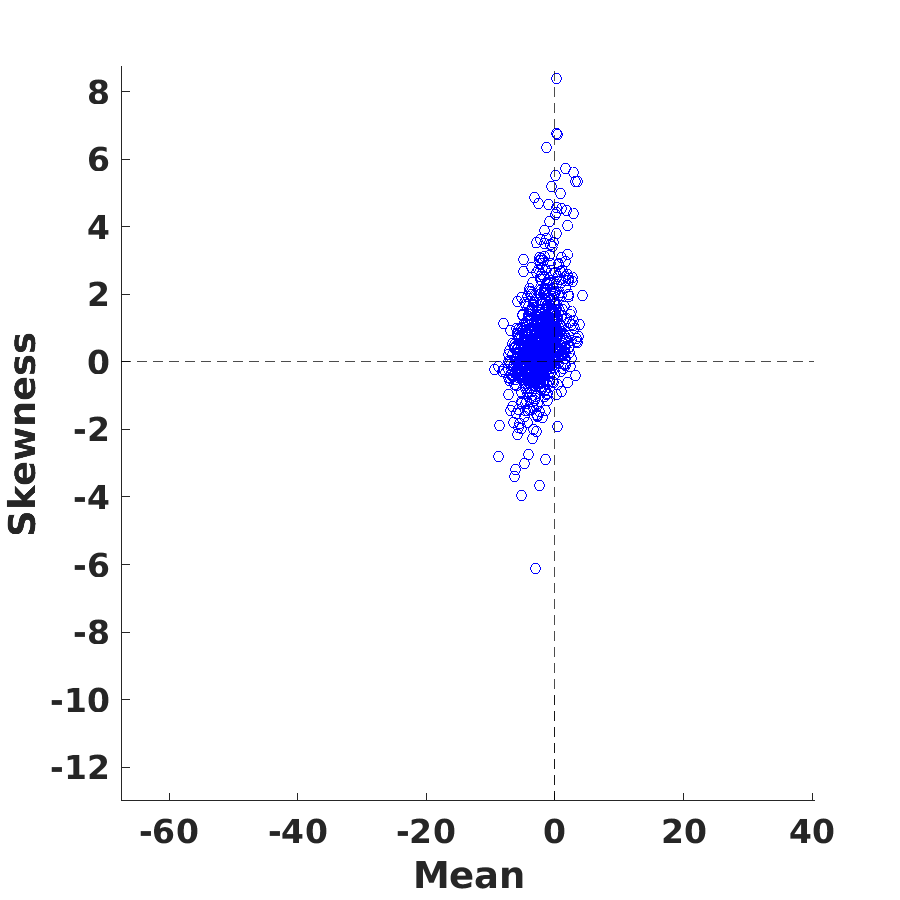} \\[2ex]
            \rotatebox{90}{\hspace*{0.5cm} $y_{i0}^*$ Marginal} &
                        \includegraphics[width=1.4in]{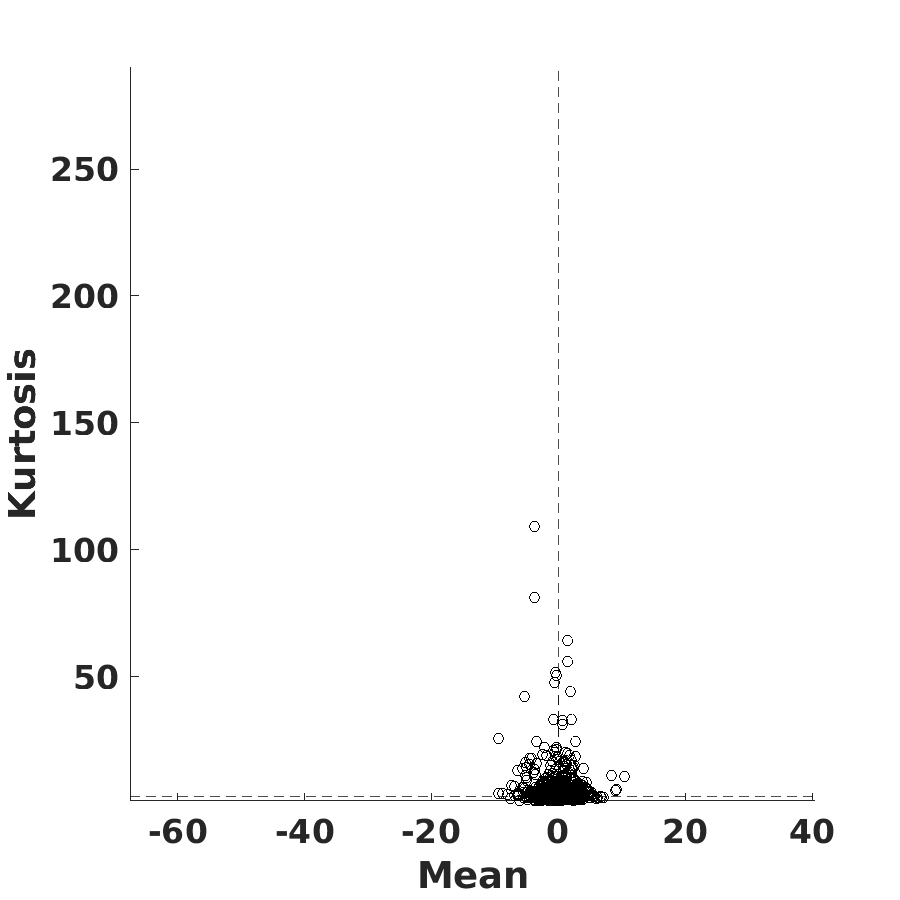}      & 
                        \includegraphics[width=1.4in]{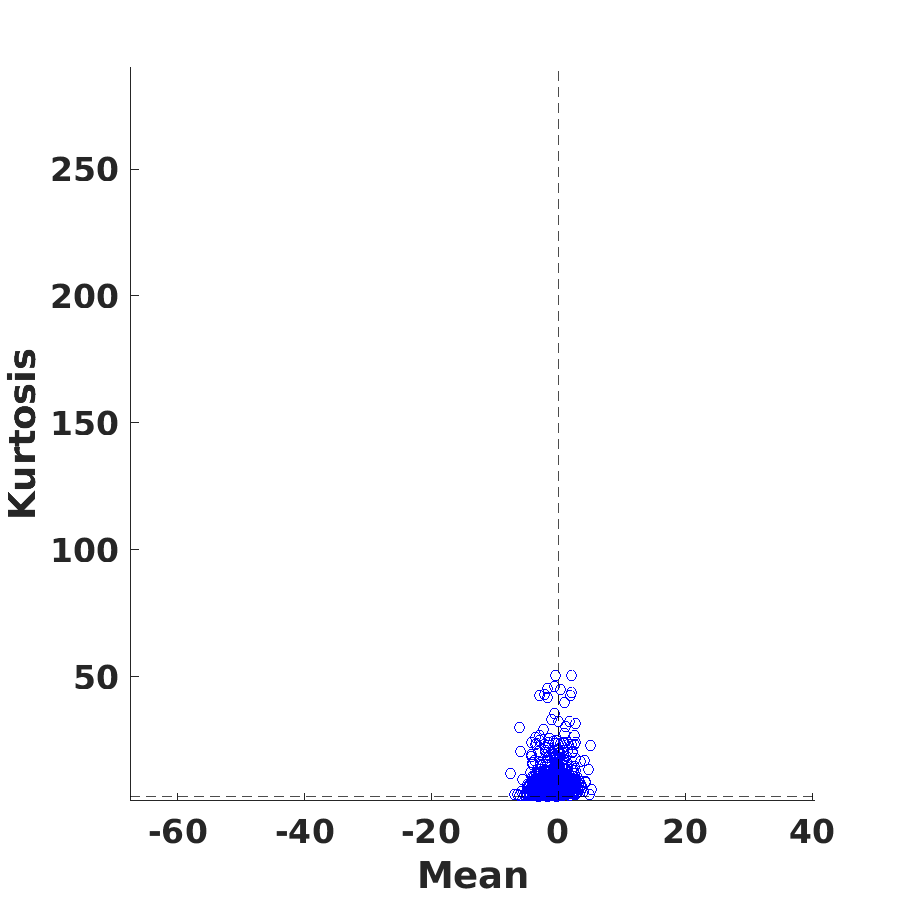} &
                        \includegraphics[width=1.4in]{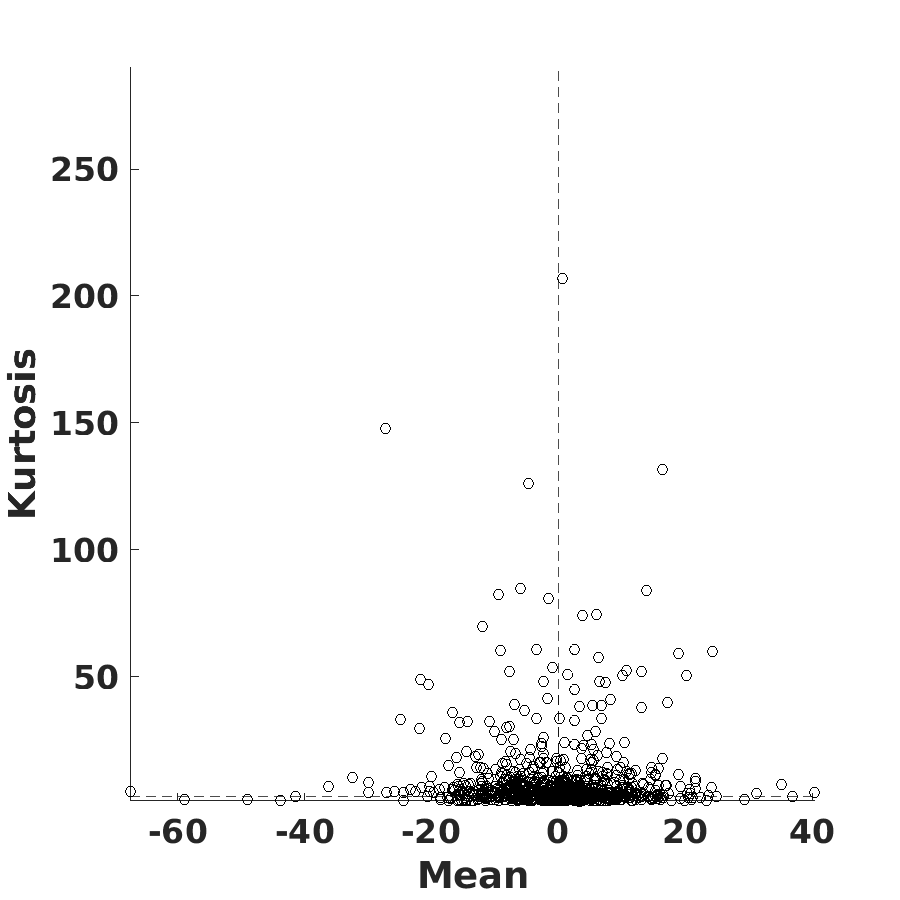} & 
                        \includegraphics[width=1.4in]{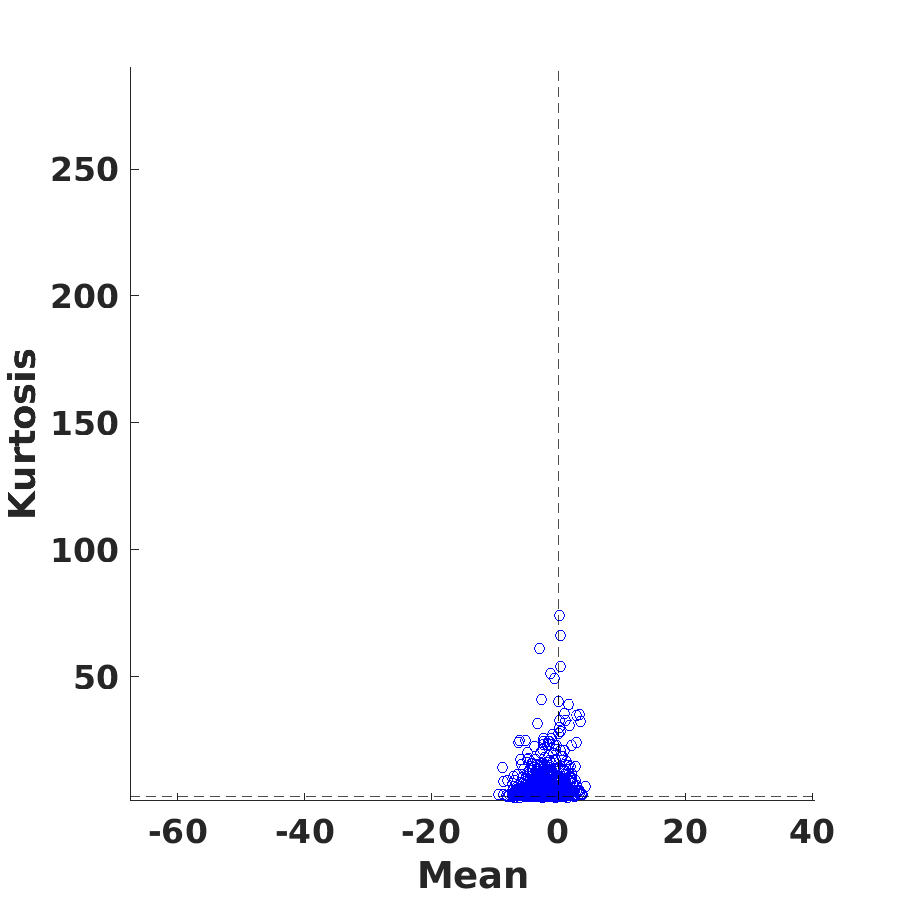} \\[2ex]
            \rotatebox{90}{\hspace*{0.5cm} $(\lambda_i,y_{i0}^*)$ Joint} &                       
                        \includegraphics[width=1.4in]{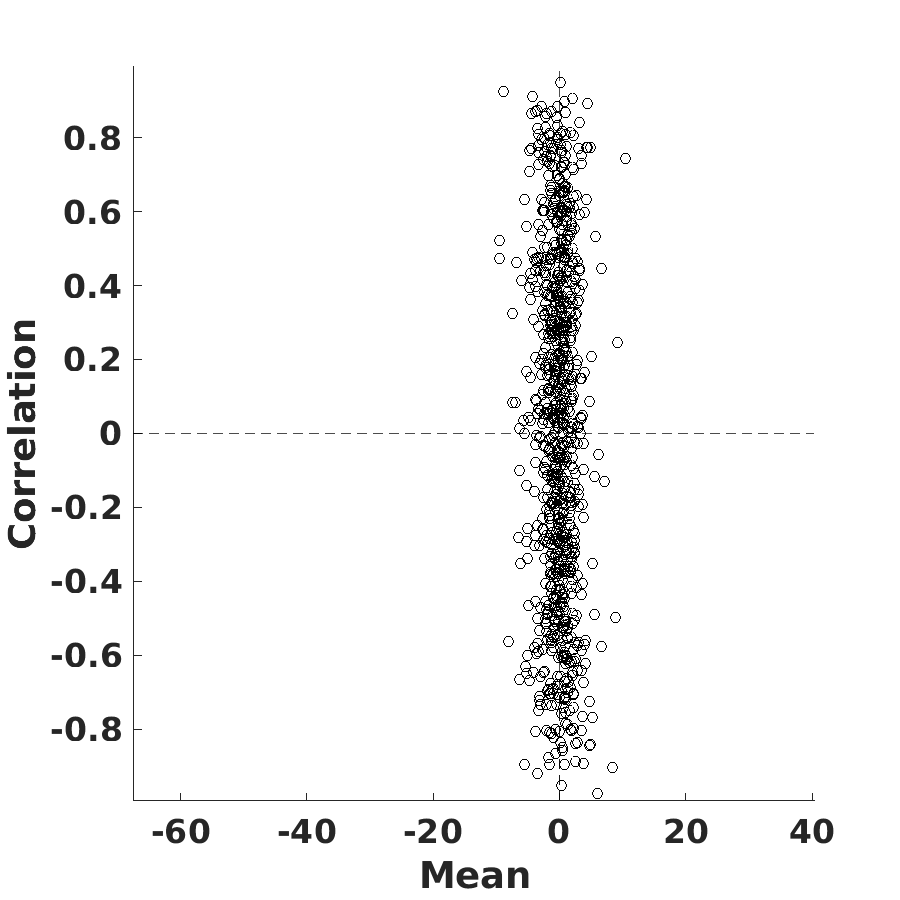}      & 
                        \includegraphics[width=1.4in]{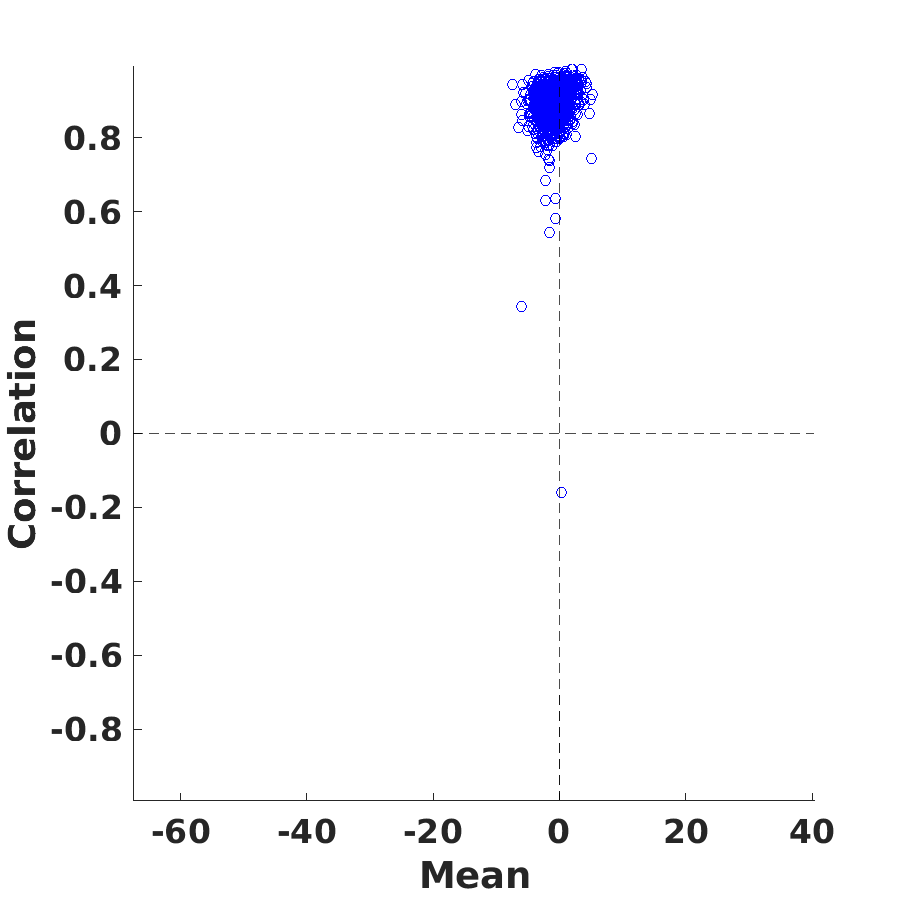} &
                        \includegraphics[width=1.4in]{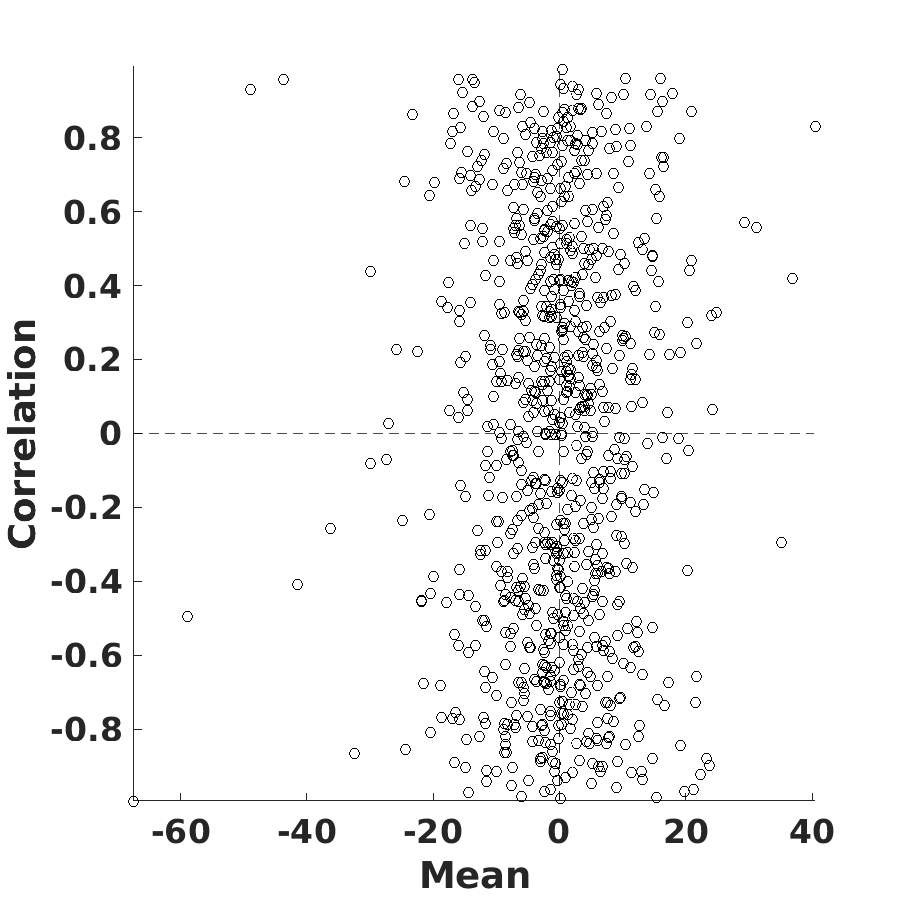} & 
                        \includegraphics[width=1.4in]{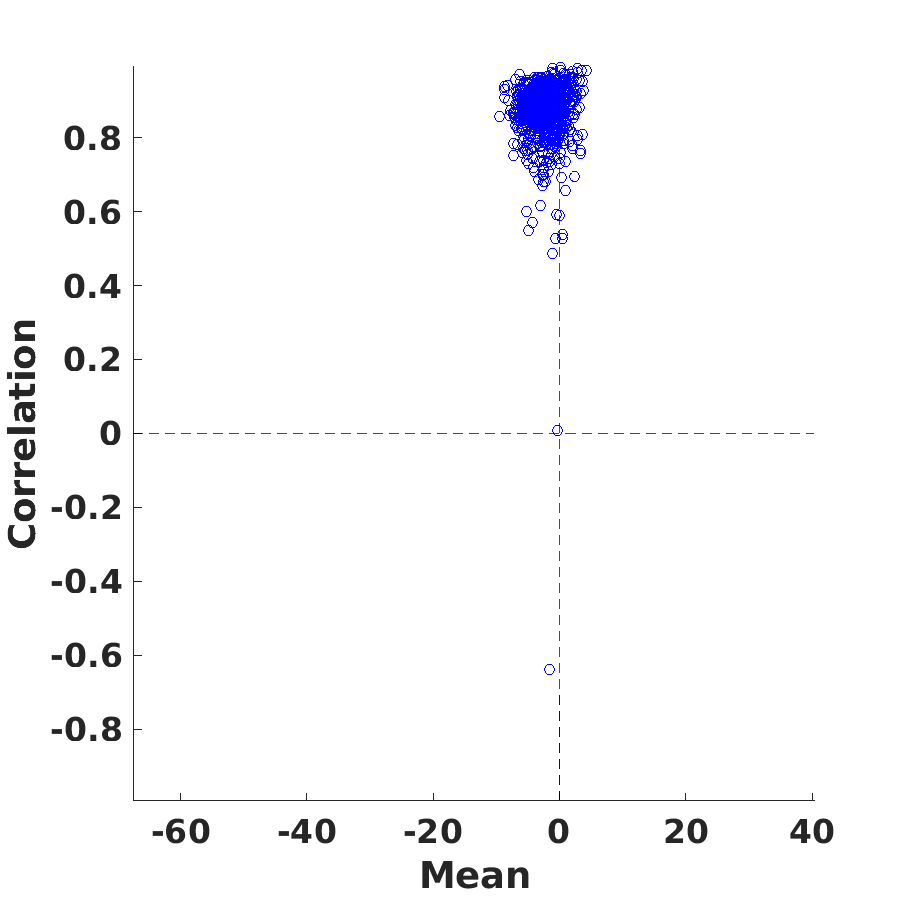} \\[2ex]                                
            \rotatebox{90}{$(\lambda_i,y_{i0}^*)$ Joint} &                      
                        \includegraphics[width=1.4in]{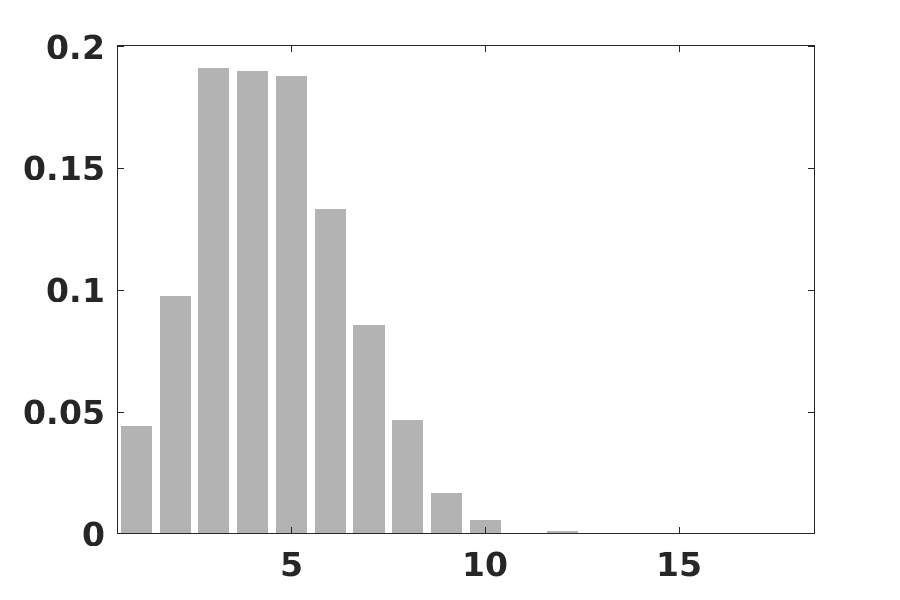}      & 
            \includegraphics[width=1.4in]{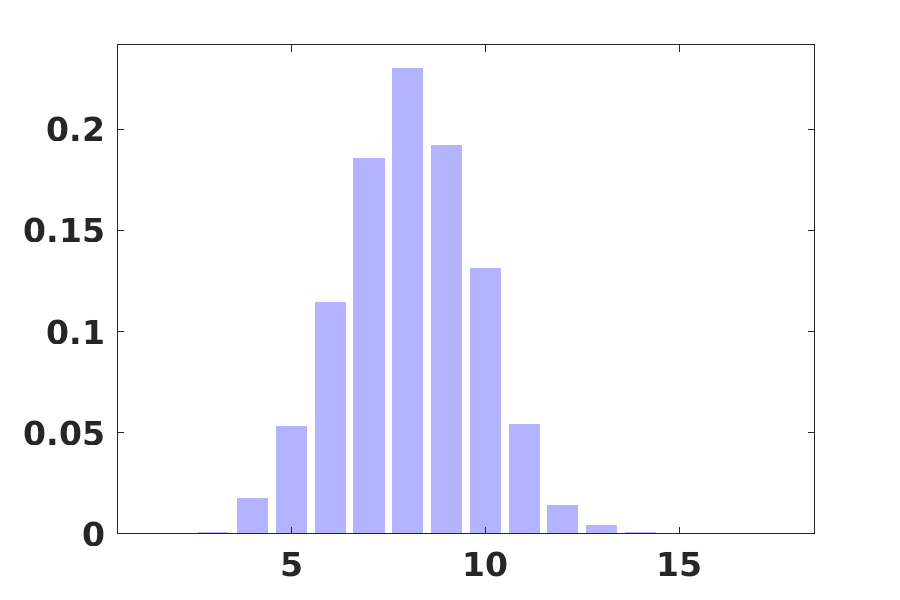} &
            \includegraphics[width=1.4in]{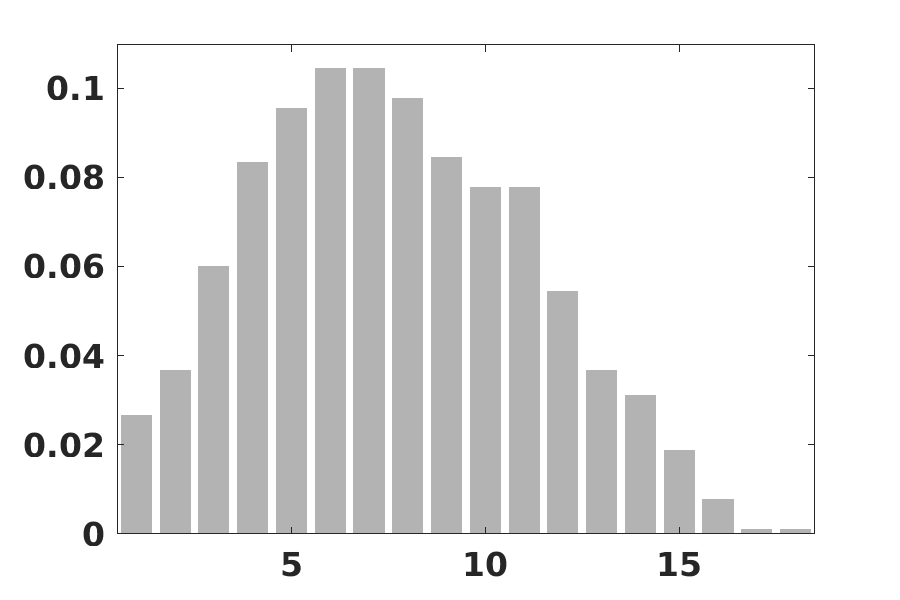} & 
            \includegraphics[width=1.4in]{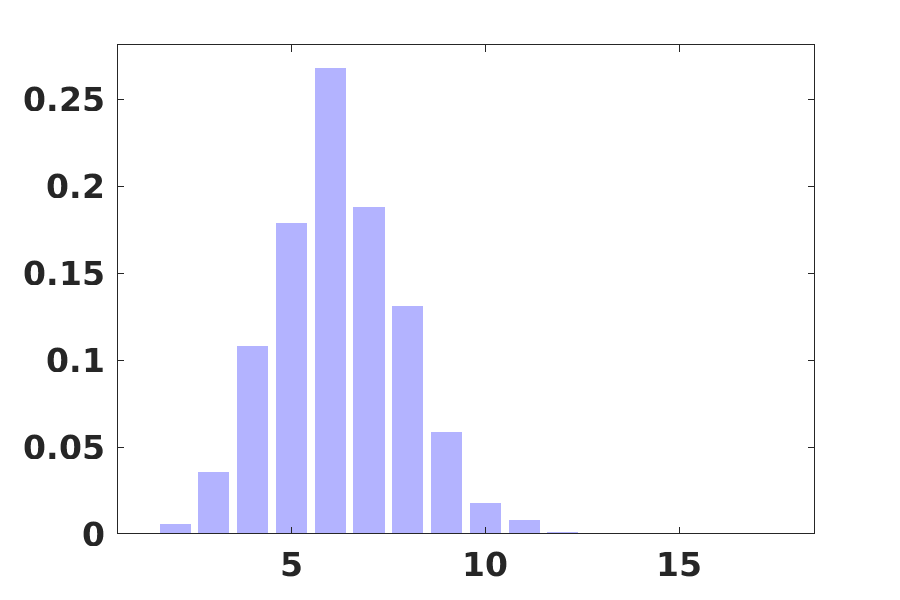}                        
                \end{tabular}
        \end{center}
        {\footnotesize {\em Notes:} The dots in the scatter plots in rows 1 to 4 correspond to draws from the prior or posterior distribution of the hyperparameter $\xi$ that indexes the CRE distribution. For each $\xi$ draw we compute the implied moments of the mixture of Normals CRE distribution. SD is the standard deviation and Correlation is the correlation between $\lambda_i$ and $y_{i0}^*$. The last row show the distribution of the number of modes. }\setlength{\baselineskip}{4mm}                                     
\end{figure}

For each draw of the hyperparameter vector $\xi$ from either the prior or posterior distribution, one can evaluate the moments of the CRE distribution, which is a mixture of Normals. The evaluation of a moment maps an infinite-dimensional object into a one-dimensional object whose distribution can be more easily visualized. Features of the prior for the CRE distribution for the CC sample are summarized in Figure~\ref{fig:cre.prior.posterior.cc.y0star}. To generate the figure, we need to choose values for the regressors $x_{it-1}$. Recall that the regressors are standardized to have mean zero and variance one. We set $x_{it-1} = \tilde{x}_{it-1} = \kappa [1,1]$ and choose $\kappa$ such that $\tilde{x}_{it-1}$ lies on the boundary of a 50\% coverage set constructed from a bivariate Normal distribution with mean zero, variances one, and a correlation that matches the correlation of $x_{it-1}$ in the sample. 

The dots in the scatter plots of the first three rows of Figure~\ref{fig:cre.prior.posterior.cc.y0star} represent moments of the marginal distribution of $y_{i0}^*$. The initial prior covers a wide range of distributions: the mean can range from -10 to 10, the standard deviation from close to 0 to 7, the distributions can be left-skewed or right-skewed, they may have a kurtosis similar to a Normal distribution or they may be very fat-tailed. The fourth row of the figure shows scatter plots of the correlation between $\lambda_i$ and $y_{i0}$, which can range from -1 to 1. 

A comparison of the prior and posterior plots under the initial tuning of the prior raises two concerns. First, the posterior location and scale of the distribution of means appear to be very similar to the prior location and scale. This could mean that the likelihood function does not contain any information about the mean of the distribution of $y_{i0}^*$. Second, the posterior location of the distribution of standard deviations appears to be very different from the prior location. Moreover, the posterior seems to be more spread out than the prior. Thus, in this particular dimension the prior seems to assign essentially no mass in an area of the parameter space that is favored by the likelihood function which could bias the posterior estimates in a way that may not be intended by the researcher. 

In view of these findings, we modify the choice of $\tau$ by raising $\tau_\phi$ from 5 to 20 and $\tau_\sigma^y$ from 1 to 4. A comparison of the first and third columns of Figure~\ref{fig:cre.prior.posterior.cc.y0star} indicates that the change in $\tau$ has the desired effect: the distribution of moments exhibits a larger variance. We proceed by computing the posterior distribution for the adjusted prior. Now the prior of the means is substantially more diffuse than the posterior of the means, and the posterior of the standard deviation does no longer lie in the far tail of the prior distribution. Comparing the posterior under the initial prior to the posterior under the adjusted prior, we find that the location of the posterior distributions is quite similar. The variance of the posterior increases slightly after the adjustment of $\tau$, but much less than the variance of the prior, so we see that the posterior is anchored by the information in the likelihood function.

The last row of Figure~\ref{fig:cre.prior.posterior.cc.y0star} shows histograms of the number of modes of the CRE distribution. Recall that the CRE distribution is a mixture of Normal distribution with $K=20$ components. This means that it could have up to 20 modes. Under the initial tuning, the prior distribution assigns probability close to one to the number of modes being between 0 to 10. The highest probability mass is associated with 3 to 5 modes. The posterior has a similar scale as the prior but is shifted to the right and peaks at 8 modes. Under the adjusted tuning, the prior distribution is more spread out which makes the posterior appear to be more concentrated relative to the prior.

\clearpage

\subsection{Density Forecasts} 

\begin{figure}[h!]
        \caption{Log Predictive Density Scores -- All Samples}
        \label{appfig:lps.scatters}
        \begin{center}
                \begin{tabular}{cc}
                        CRE \& Heteroskedastic & Flexible \& Heteroskedastic\\
             \includegraphics[width=2in]{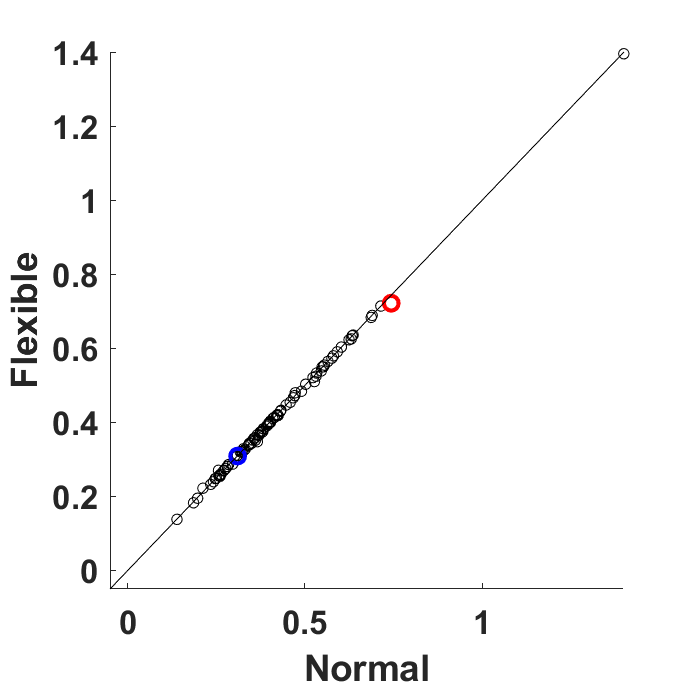} & 
                        \includegraphics[width=2in]{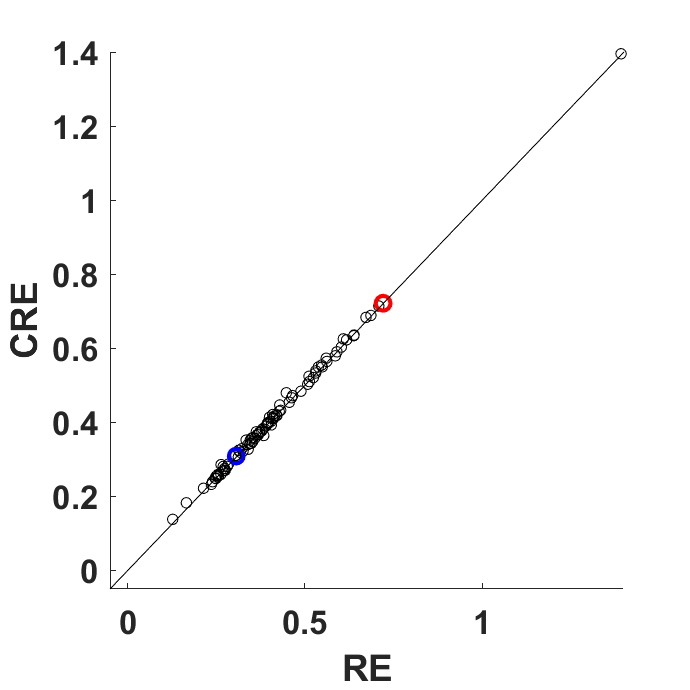} 
                \end{tabular}
        \end{center}
        {\footnotesize {\em Notes:} The panels provide pairwise comparisons of log predictive scores. We also show the 45-degree line. Log probability scores are depicted as differentials relative to pooled Tobit. The blue (red) circle corresponds to RRE (CC). We use $x_{it} = [\Delta \ln \mbox{HPI}_{it}, \Delta \mbox{UR}_{it}]'$.}\setlength{\baselineskip}{4mm}                                     
\end{figure}

Figure~\ref{appfig:lps.scatters} resembles Figure~\ref{fig:lps.scatters} in the main text and shows that accuracy differentials of Normal versus flexible CREs and of CREs versus REs are small. 

\clearpage

\begin{figure}[h!]
        \caption{CC Charge-Off Rate Predictive Tail Probabilities, Spatial Dimension}
        \label{fig:tailprob.spatial.CC}
        \begin{center}
                \begin{tabular}{cc}
                        2010Q1 & 2015Q2 \\
                        \includegraphics[width=3in,trim=0in 0in 0in 0in, clip]{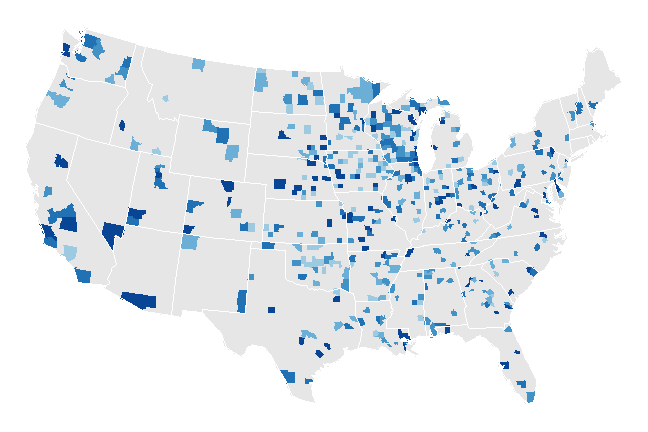} &
                        \includegraphics[width=3in,trim=0in 0in 0in 0in, clip]{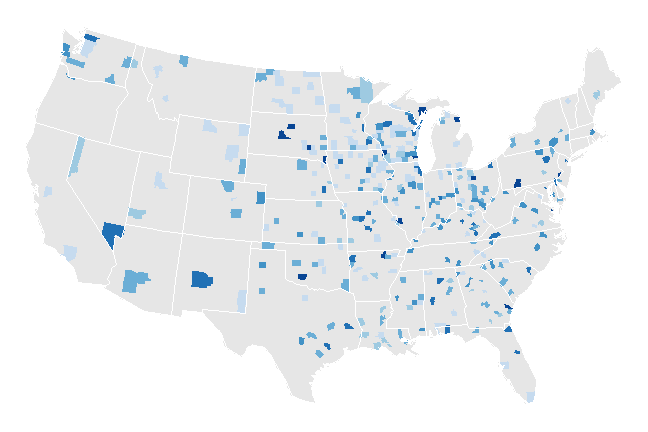} \\
                        \multicolumn{2}{c}{\includegraphics[width=6.4in,trim=0in 0in 0in 0in, clip]{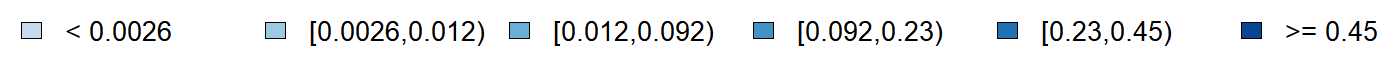}}
                \end{tabular}
        \end{center}
        {\footnotesize {\em Notes:} Predictive tail probabilities are defined as $\mathbb{P}\{y_{iT+1} \ge c |Y_{1:N,0:T},X_{1:N,-1:T}\}$, where $c=5$\%. Flexible CRE specification with heteroskedasticity. The estimation samples range from 2007Q2 ($t=0$) to 2009Q4 ($t=T=10$) and 2012Q3 ($t=0$) to 2015Q1 ($t=T=10$).  }\setlength{\baselineskip}{4mm}                                     
\end{figure}

Figure~\ref{fig:tailprob.spatial.CC} resembles Figure~\ref{fig:tailprob.spatial} in the main text and shows the spatial distribution of CC charge-off rate forecasts. Averaging across banks in each county contained in our sample we report predictive tail probabilities $\mathbb{P}\{y_{iT+1} \ge 5\% |Y_{1:N,0:T},X_{1:N,-1:T}\}$.

\clearpage
\subsection{Parameter Estimates}

\begin{figure}[h!]
        \caption{Heterogeneous Coefficient Estimates, CC Charge-Off Rates}
        \label{fig:lambda.sigma.CC}
        \begin{center}
                \begin{tabular}{ccc}
                        $\mathbb{E}[\lambda_i/(1-\rho)|\cdot]$ & $\mathbb{E}\big[\ln (\sigma_i/\sqrt{1-\rho^2})|\cdot\big]$ & Scatter \\[1ex] 
                        \includegraphics[width=2in]{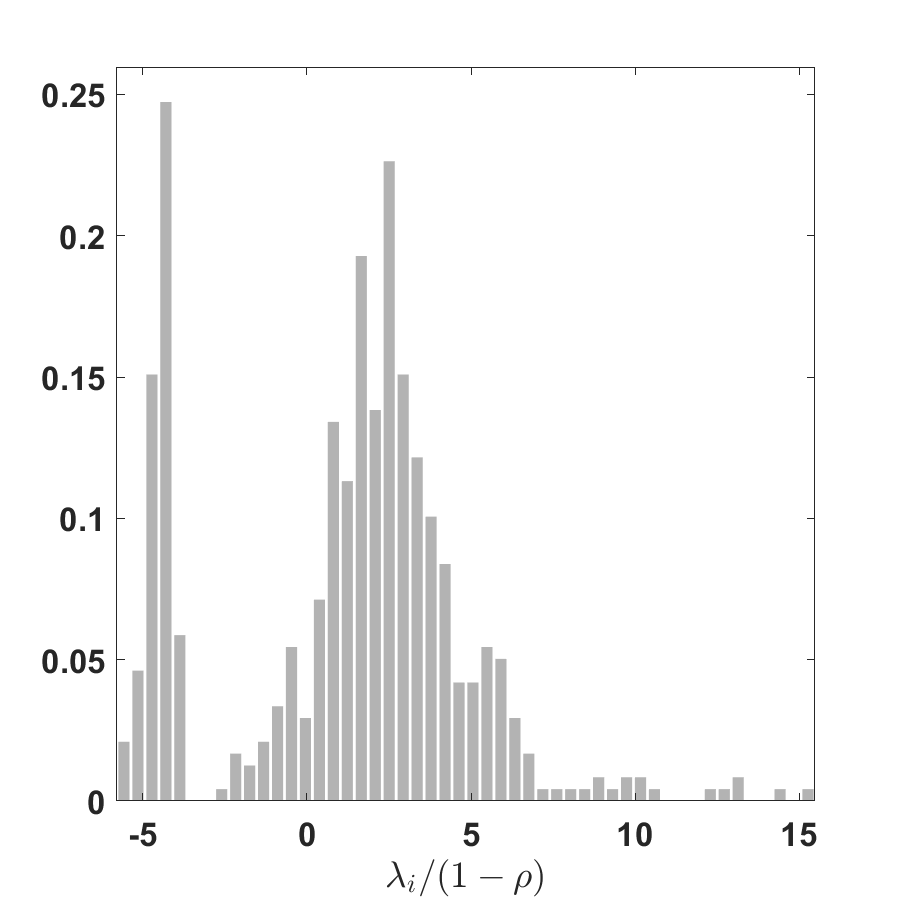}      & \includegraphics[width=2in]{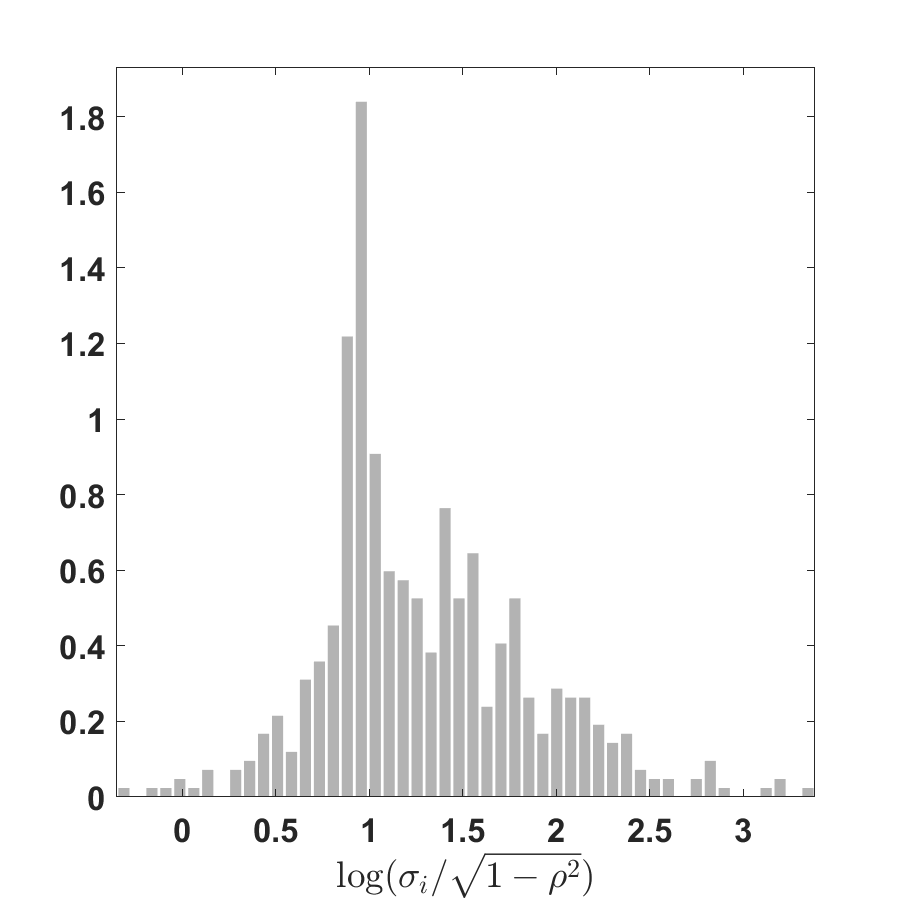} & 
                        \includegraphics[width=2in]{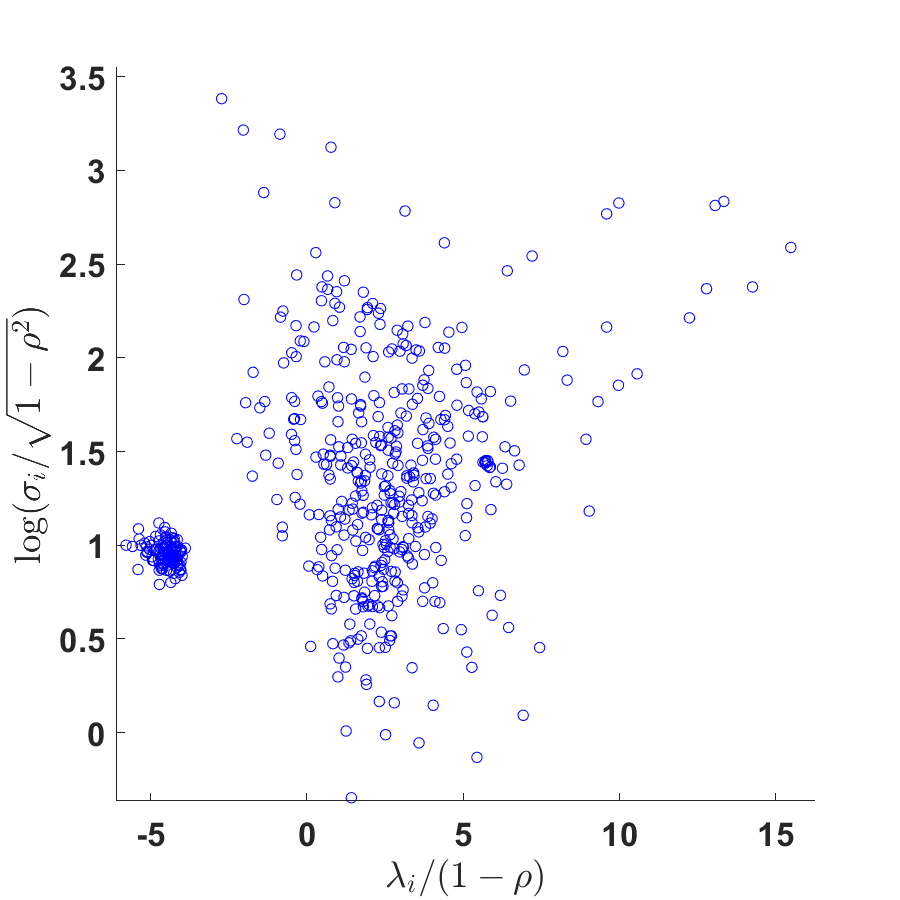}                                   
                \end{tabular}
        \end{center}
        {\footnotesize {\em Notes:} Heteroskedastic flexible CRE specification. The estimation sample ranges from 2007Q2 ($t=0$) to 2009Q4 ($t=T=10$). A few extreme observations are not visible in the plots. The conditioning set is $(Y_{1:N,0:T},X_{1:N,-1:T})$.}\setlength{\baselineskip}{4mm}                                     
\end{figure}

\begin{figure}[h!]
        \caption{Effects (Terms I and II) of HPI and UR on RRE Charge-Off Rates}
        \label{fig:xit.marginal.effects.RRE}
        \begin{center}
                \begin{tabular}{cc}
                        HPI Fall & Unemployment Increase \\
                        \includegraphics[width=3in]{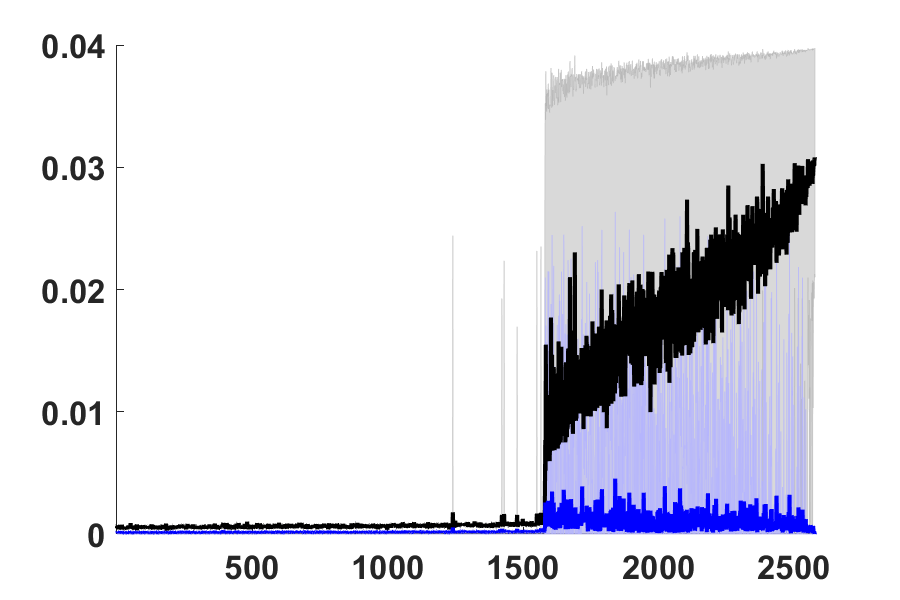} & 
                        \includegraphics[width=3in]{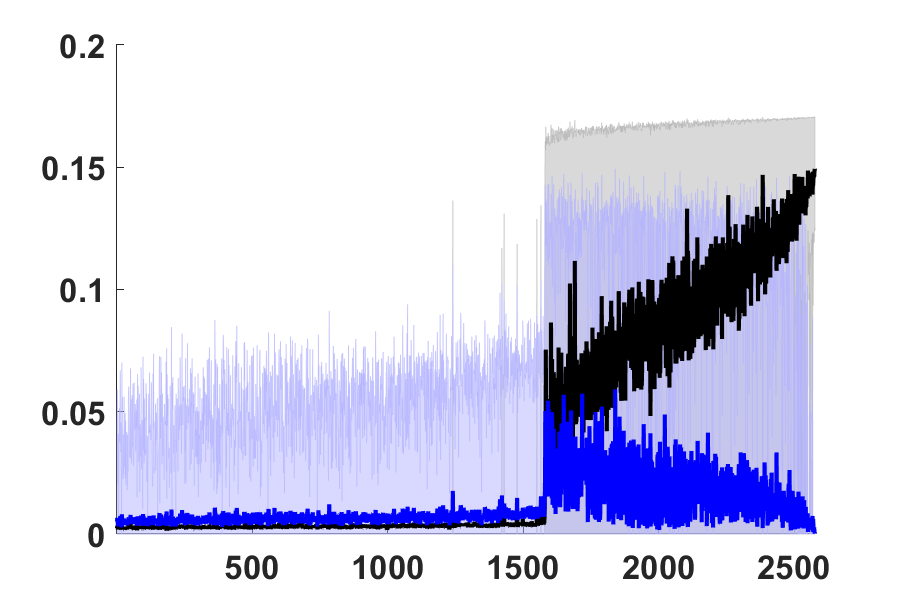} 
                \end{tabular}
        \end{center}
        {\footnotesize {\em Notes:} Heteroskedastic flexible CRE specification. The estimation sample ranges from 2007Q2 ($t=0$) to 2009Q4 ($t=T=10$). The banks $i=1,\ldots,N$ along the $x$-axis are sorted based on the posterior means  $\widehat{\lambda_i/\sigma_i}$. Terms $I_i$ are shown in black/grey and terms $II_i$ in dark/light blue. The units on the $y$-axis are in percent. The solid lines indicate the posterior means of the treatment effect components and the shaded areas delimit 90\% credible bands. }\setlength{\baselineskip}{4mm}                                     
\end{figure}

Figures~\ref{fig:lambda.sigma.CC} (CC charge-off rates) and~\ref{fig:xit.marginal.effects.RRE} (RRE charge-off rates) resemble Figures~\ref{fig:lambda.sigma} and~\ref{fig:xit.marginal.effects.CC} in the main text.

\clearpage
\subsection{Predictive Checks}

\begin{figure}[h!]
        \caption{Additional Posterior Predictive Checks: Cross-sectional Distribution of Sample Statistics}
        \label{appfig:predictive.checks}
        \begin{center}
                \begin{tabular}{cccc}
                        & & & Robust \\
                        & Mean of $Y_{iT+1}$  & Mean of $Y_{iT+1}$  & Correlation of $(y_{it},y_{it-1})$ \\
                        & After Obs. Zero     & Before Obs. Zero    & All Observations \\[1ex]    
                        \rotatebox{90}{\phantom{abcde} RRE} &
                        \includegraphics[width=2in]{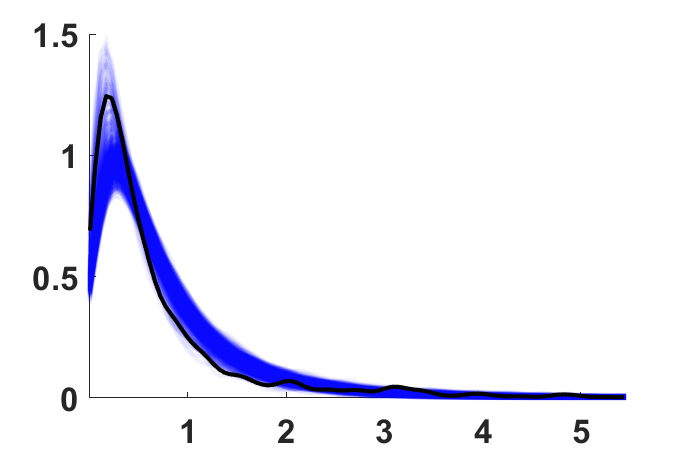}      & \includegraphics[width=2in]{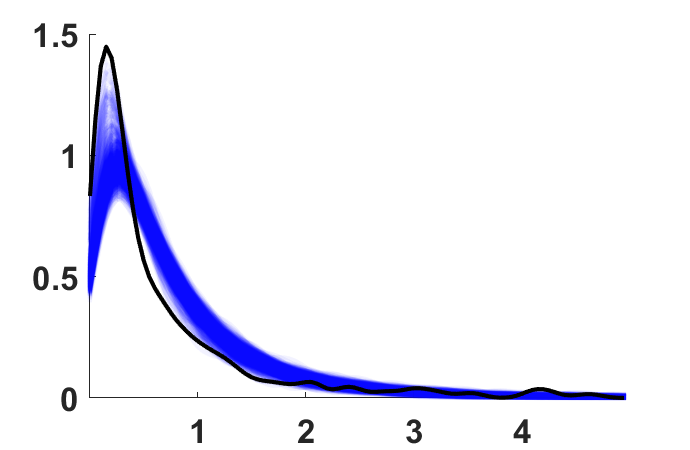} & 
                        \includegraphics[width=2in]{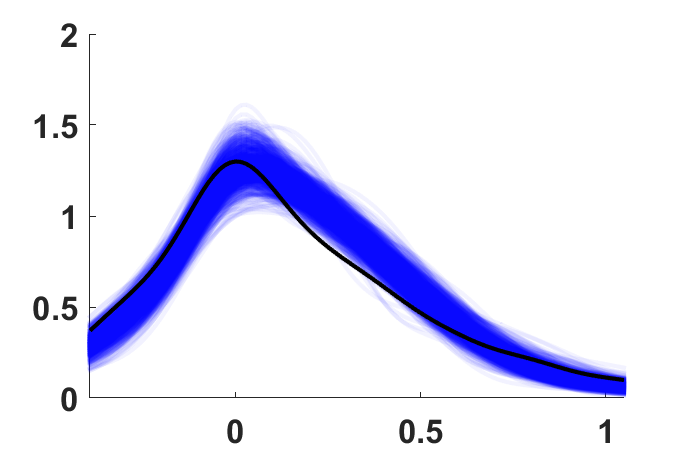} \\
                        \rotatebox{90}{\phantom{abcdef} CC} &                                 
                        \includegraphics[width=2in]{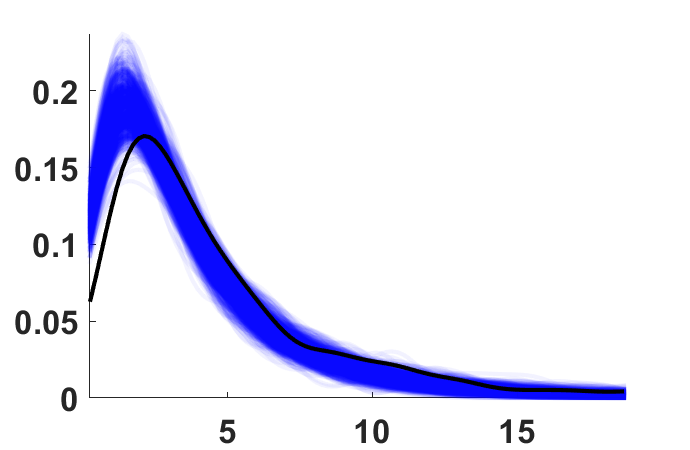}      & \includegraphics[width=2in]{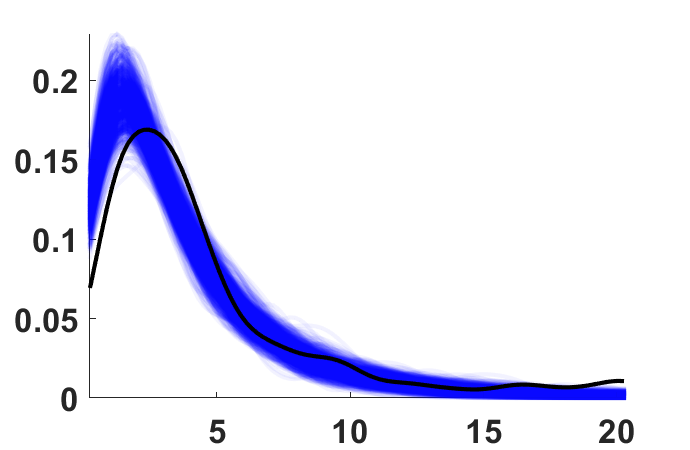} & 
                        \includegraphics[width=2in]{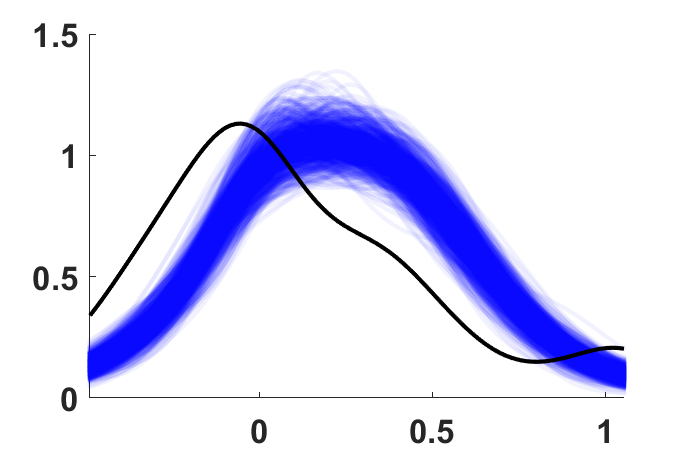} \\
                \end{tabular}
        \end{center}
        {\footnotesize {\em Notes:} Heteroskedastic flexible CRE specification. The estimation sample ranges from 2007Q2 ($t=0$) to 2009Q4 ($t=T=10$). The black lines  are computed from the actual data. Each hairline corresponds to a simulation of a sample $\tilde{Y}_{1:N,0:T+1}$ of the panel Tobit model based on a parameter draw from the posterior distribution. Robust autocorrelations are computed using the MM estimator in \cite{ChangPolitis2016}.}  \setlength{\baselineskip}{4mm}                                     
\end{figure}

\end{document}